\documentclass[
  ,draft            
  ,numberedheadings 
  ,a4paper                 
]
  {aipproc}

\layoutstyle{6x9}

\usepackage{amssymb}
\usepackage{amsmath}
\usepackage{natbib}

\newcommand{\myvec}[1]{\mathbf{#1}}

\newcommand{\vnhat}{\hat{\myvec{n}}}
\newcommand{\vkhat}{\hat{\myvec{k}}}
\newcommand{\vx}{\myvec{x}}
\newcommand{\vy}{\myvec{y}}
\newcommand{\vz}{\myvec{z}}
\newcommand{\vk}{\myvec{k}}
\newcommand{\vm}{\myvec{m}}
\newcommand{\ve}{\myvec{e}}
\newcommand{\vv}{\myvec{v}}
\newcommand{\vzero}{\myvec{0}}
\newcommand{\vV}{\myvec{V}}
\newcommand{\vvb}{\myvec{v}_{\mathrm{b}}}
\newcommand{\velb}{v_{\mathrm{b}}}
\newcommand{\vtheta}{\myvec{\theta}}
\newcommand{\vphi}{\myvec{\phi}}
\newcommand{\vnabla}{\myvec{\nabla}}

\newcommand{\sigT}{\sigma_{\mathrm{T}}}
\newcommand{\clp}{\mathcal{P}}
\newcommand{\clr}{\mathcal{R}}

\newcommand{\D}{\mathrm{d}}
\newcommand{\E}{\mathrm{e}}

\newcommand{\threej}[6]{\left(\begin{array}{ccc} #1 & #2 & #3 \\
                                                 #4 & #5 & #6
                        \end{array}\right)}

\newcommand{\refjnl}[1]{#1}

\def\aj{\refjnl{AJ}}                   
\def\araa{\refjnl{ARA\&A}}             
\def\apj{\refjnl{ApJ}}                 
\def\apjl{\refjnl{ApJ}}                
\def\apjs{\refjnl{ApJS}}               
\def\aap{\refjnl{A\&A}}                

\def\mnras{\refjnl{MNRAS}}             
\def\prd{\refjnl{Phys.~Rev.~D}}        
\def\prl{\refjnl{Phys.~Rev.~Lett.}}    
\def\nat{\refjnl{Nature}}              

\def\physrep{\refjnl{Phys.~Rep.}}   

\renewcommand{\beth}{\bar{\eth}}

\begin{document}

\title{Lecture notes on the physics of cosmic microwave background anisotropies}


\author{Anthony Challinor}{
  address={Institute of Astronomy and Kavli Institute for Cosmology Cambridge, Madingley Road, Cambridge, CB3 0HA, UK}
,altaddress={DAMTP, Centre for Mathematical Sciences, Wilberforce Road,
Cambridge, CB3 0WA, UK}}

\author{Hiranya Peiris}{
  address={Institute of Astronomy and Kavli Institute for Cosmology Cambridge, Madingley Road, Cambridge, CB3 0HA, UK}
}

\begin{abstract}
We review the theory of the temperature anisotropy and polarization of the
cosmic microwave background (CMB) radiation, and describe what we have learned
from current CMB observations. In particular, we
discuss how the CMB is being used to
provide precise measurements of the composition and geometry of the
observable universe, and to constrain the physics of the early universe.
We also briefly review the physics of the small-scale CMB fluctuations
generated during and after the epoch of reionization, and which are the
target of a new breed of arcminute-resolution instruments.
\end{abstract}

\maketitle


\section{Introduction}

The last two decades have seen remarkable advances in observational
cosmology and quantitative cosmological constraints now come from a number of
complementary probes. Amongst these, the cosmic microwave background
temperature and polarization anisotropies have played a particularly
important role. The physics is very well understood -- the primary
contribution to the anisotropies is accurately described by linear
perturbation theory -- and so theoretical predictions are believed
to be accurate to better than 1\%. Furthermore, the CMB currently
provides the most
direct observational link to the physics of the early universe since the
fluctuations were mostly imprinted around the time of recombination.

The physics of the CMB anisotropies is now a textbook
subject~\cite{DurrerCMBbook}. The basic picture is that the angular variations
in temperature that we observe today are a snapshot of the local
properties (density, peculiar velocity and the total gravitational potential)
of the gas of CMB photons at the time the primordial plasma recombined
at redshift $z \approx 1100$. These quantities are related to the
primordial perturbations, plausibly generated during an epoch of
cosmological inflation, by the rather simple acoustic physics of the
pre-recombination plasma. The CMB anisotropies therefore encode information
on the primordial perturbation itself, as well as the matter composition
and geometry of the universe. By mapping the microwave sky, cosmologists thus
hope to answer some of the biggest questions in physics.

\begin{figure}[t!]
\centerline{\includegraphics[width=0.7\textwidth,angle=0]{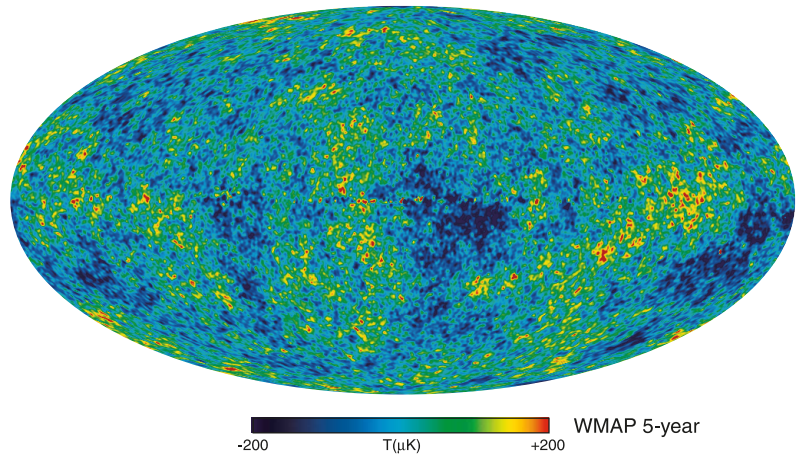}}
\caption{Fluctuations in the CMB temperature, as determined from five years
of WMAP data, about the average temperature of $2.725\,\mathrm{K}$.
Reproduced from~\cite{2009ApJS..180..225H}.
}
\label{fig:wmap5}
\end{figure}

Large-angle temperature anisotropies were first discovered by COBE-DMR in
1992~\cite{1992ApJ...396L...1S}. 
Since then, tens of instruments have mapped the anisotropies and we
now have high quality measurements of the statistics of the anisotropies
over three decades of angular scales.
The current state-of-the-art dataset is provided by COBE's successor,
the WMAP satellite, which released its five-year results in March
2008~\cite{2009ApJS..180..225H}. A map of the temperature fluctuations from WMAP,
after removing the effect of microwave emission from our galaxy, is shown
in Fig.~\ref{fig:wmap5}
The WMAP results are beautifully fit by a simple cosmological
model with flat space, baryons and cold dark matter making up
4\%\ and 21\%\ respectively of the energy density with the remaining
75\%\ in some form of dark energy, and Gaussian-distributed, adiabatic
primordial density perturbations with
an almost, but not quite, scale-invariant spectrum.
The accuracy of these numbers is quite remarkable by cosmological standards;
for example, the baryon density is
measured to better than 3\%\ and space is flat to better than 1\%.
Inflation continues
to stand up to exacting comparison with observations. Indeed, the quality
of the data is now good enough to rule out several popular inflation models
from the plethora that theorists have proposed.
The community now eagerly awaits the launch of Planck, the third-generation
CMB satellite, in mid-2009. This will image the sky in nine wavelength bands
with a best resolution below 5 arcmin.
Forecasts
indicate that Planck should measure several important parameters to better
than 1\%, and place tight constraints on inflation
models~\cite{2006astro.ph..4069T}.

%

The CMB temperature fluctuations are complemented by its linear
polarization~\cite{1968ApJ...153L...1R}. Polarization is
generated by scattering of anisotropic radiation and for the
CMB this occurs at two epochs: around recombination and, later, from
the epoch of reionization~\cite{1997PhRvD..55.1822Z}.
The small polarization signal (r.m.s. $\sim 5\,\mu\mathrm{K}$) was
first detected in 2002 by the DASI instrument~\cite{2002Natur.420..772K},
and has subsequently been measured by several groups. The signal-to-noise
of these measurements is still poor compared to the temperature
anisotropies, but already important constraints have emerged on
the epoch of reionization from WMAP~\cite{2007ApJS..170..335P,2009ApJS..180..306D}.
Polarization also provides the only viable route in the near-future for
detecting the background of gravitational waves predicted from
inflation in models with large, i.e.\ super-Planckian, field
values~\cite{1997PhRvL..78.2054S,1997PhRvL..78.2058K}.
A positive detection would be particularly interesting for attempts to realise inflation
in string theory, and would rule out some alternatives to inflation such as the
cyclic model~\cite{2002Sci...296.1436S}.
Since Planck may not have the sensitivity to make a
definitive test of gravitational waves,
a new generation of ambitious sub-orbital instruments
that will deploy hundreds or even thousands of superconducting detectors to beat down the
noise are under construction.
These exciting instruments aim to probe the highly uncertain
physics of energies near $10^{16}$~GeV.

In addition to these \emph{primary} anisotropies, there are a number
of \emph{secondary} processes that add further structure to the CMB at late
times (see Ref.~\cite{2008RPPh...71f6902A} for a recent review).
These include gravitational redshifts induced by the linear evolution of the
gravitational potentials once dark energy becomes dynamically
important~\cite{1968Natur.217..511R}, gravitational lensing of CMB photons by the large-scale
distribution of matter~\cite{1987A&A...184....1B,1989MNRAS.239..195C,2006PhR...429....1L},
and various scattering effects that
become operative after the onset of
reionization~\cite{1972CoASP...4..173S,1980MNRAS.190..413S,1986ApJ...306L..51O}.
While the
observation and interpretation of the primary anisotropies is now a mature
field, for the secondary processes, in which the CMB is used as a back-light,
this is just beginning. A new breed of instruments with arcminute resolution
are now surveying these secondary effects with the hope of
improving our understanding of the epoch of reionization and the growth of
structure over a wide range of redshifts.
If these new observations
are to fulfill their scientific potential, accurate modelling of secondary
anisotropies is required. This properly requires large numerical
simulations and is currently a very active area of CMB theory.

In these lecture notes we outline the basic physics of the primary
temperature and polarization anisotropies of the CMB, and discuss how
observations of these are being used to constrain cosmological models
and the physics of the early universe. 
For earlier reviews that are similar in spirit, see for
example~\cite{2002ARA&A..40..171H,2003AnPhy.303..203H,2005LNP...653...71C}.
We also briefly review secondary anisotropies and
look forward to what we may learn from the new high-resolution
surveys.

\section{Introduction to CMB temperature anisotropies}

The CMB radiation has an exquisite
blackbody (thermal) spectrum, with very nearly the same temperature of
2.725~K in all directions on the
sky~\cite{1994ApJ...420..439M,2002ApJ...581..817F}. The largest
temperature variation is a dipole due to our motion relative to the
CMB. Initially, ignore the small effect of the temperature fluctuations
due to the primordial perturbations. Then the CMB is isotropic in its rest
frame with a Lorentz-invariant distribution function\footnote{The
one-particle distribution function gives the number of photons per
proper phase-space volume.}
$\bar{f}(p^\mu) \propto 1/ (e^{E_{\mathrm{CMB}}/T_{\mathrm{CMB}}}-1)$
where $E_{\mathrm{CMB}}$ is the photon energy in the CMB frame and
$T_{\mathrm{CMB}}=\bar{T}$ is the isotropic temperature. If we observe
a photon with energy $E$ and direction $\ve$, then the relativistic Doppler
shift gives $E_{\mathrm{CMB}} = E \gamma (1+\vv\cdot \ve)$,
where $\vv$ is our velocity relative to the CMB.
In terms of the
energy and direction in our reference frame,
\begin{equation}
\bar{f}(p^\mu) \propto \frac{1}{e^{E\gamma(1+\vv\cdot
\ve)/T_{\mathrm{CMB}}}-1} \, .
\label{eq:1}
\end{equation}
This still looks like a blackbody along any direction but the observed
temperature varies over the sky as $T(\ve) =
T_{\mathrm{CMB}}/[\gamma(1+\vv\cdot \ve)] \approx
T_{\mathrm{CMB}} (1-\vv \cdot \ve)$ for $|\vv| \ll 1$, i.e.\
a dipole anisotropy. The measured dipole~\cite{1993ApJ...419....1K} implies
the solar system barycentre has speed $3.68 \times 10^5\,\mathrm{m\,s}^{-1}$
relative to the CMB. 
It is clear from Eq.~(\ref{eq:1}) that relative motion
also produces quadrupole anisotropies at $O(|\vv|^2)$ and CMB data
is routinely corrected for this (small) effect. 
Reinstating the contributions of the primordial perturbation to the temperature
anisotropies, these are at the level of one part in $10^5$ and dominate
over the relative-velocity effects for the observed quadrupole and
higher multipoles.

\subsection{CMB observables}

It is convenient to
expand the temperature fluctuation in spherical harmonics,
\begin{equation}
\Delta T(\vnhat) / \bar{T} = \sum_{lm} T_{lm} Y_{lm}(\vnhat) \; ,
\label{eq:2}
\end{equation}
with $T_{lm}^* = (-1)^m T_{l-m}$ since the temperature is a real field.
The sum in Eq.~(\ref{eq:2}) runs over $l \geq 1$, but the dipole
($l=1$) is
usually removed explicitly when analysing data
since it depends linearly on
the velocity of the observer. Multipoles at $l$ encode angular information
with characteristic scale (i.e.\ period) $\sim 2\pi/l$.

The statistical properties of the fluctuations in a perturbed cosmology
can be expected to respect the symmetries of the background
model. In the case of Robertson-Walker models, this means the
\emph{statistics} should be independent of spatial position (homogeneous) and
invariant under rotations of the field. Under a rotation specified
by Euler angles $\alpha$, $\beta$ and $\gamma$, the $T_{lm}$ transform
to $\sum_{m'} D^l_{mm'}(\alpha,\beta,\gamma) T_{l m'}$ where
$D^l_{mm'}$ is a Wigner rotation matrix.
Demanding
invariance under rotations fixes the second-order statistics to be of the form
\begin{equation}
\langle T_{lm} T^*_{l'm'} \rangle = C_l^T \delta_{ll'} \delta_{mm'} \; ,
\label{eq:3}
\end{equation}
which defines the power spectrum $C_l^T$. The angle brackets in this equation
denote the average over an ensemble of realisations of the fluctuations.
The simplest models of inflation predict that the fluctuations should also be
Gaussian at early times, and this is preserved by linear evolution of the
small fluctuations. Equation~(\ref{eq:3}) tells us that the $T_{lm}$
are uncorrelated for different $l$ and $m$ and, if Gaussian, they are
also independent. In this case, the 
power spectrum provides the complete statistical description of the
temperature anisotropies. For this reason, measuring the anisotropy
power spectrum has, so far, been the main goal of observational CMB research.

If we were given ideal (i.e.\ noise-free) measurements of the temperature
over the full sky, the CMB power spectrum could be estimated by
\begin{equation}
\hat{C}_l^T = \frac{1}{2l+1} \sum_m |T_{lm}|^2 \;  .
\label{eq:4}
\end{equation}
This is an unbiased estimator of the true ensemble $C_l^T$, but has
an irremovable \emph{cosmic variance} due to the finite number ($2l+1$) of
modes that we can observe. If the temperature anisotropies are Gaussian
distributed, $\hat{C}_l^T$ has a $\chi^2$ distribution with $2l+1$ degrees of
freedom and the cosmic variance is~\cite{1995PhRvD..52.4307K}
\begin{equation}
\mathrm{var}(C_l^T) = \frac{2}{2l+1} (C_l^{T})^2 \; .
\label{eq:5}
\end{equation}
In practice, estimating the power spectrum is complicated by a number
of real-world complexities such as partial sky coverage and instrumental
noise. The critical review by Efstathiou~\cite{2004MNRAS.349..603E}
provides a useful
entry point to the large literature on practical methods for
estimating the temperature power spectrum.

\begin{figure}[t!]
\centerline{\includegraphics[width=0.7\textwidth,angle=0]{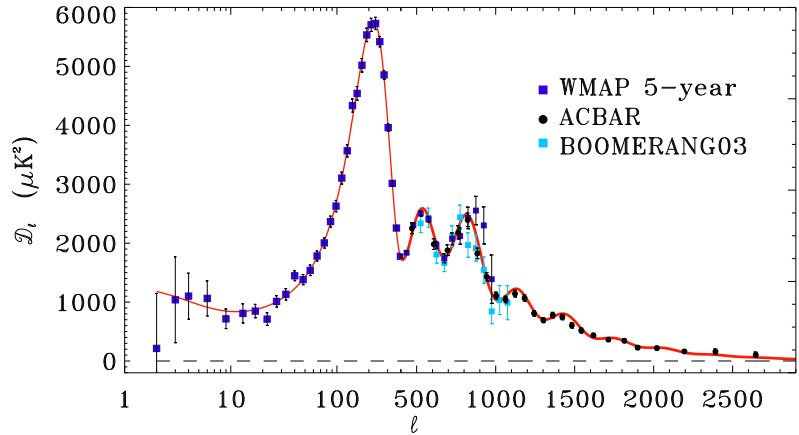}}
\caption{The temperature anisotropy power spectrum, $\mathcal{D}_l
\equiv l(l+1)C_l^T/(2\pi)$,
as measured
by WMAP5 (blue;~\cite{2009ApJS..180..296N}), ACBAR (black;~\cite{2008arXiv0801.1491R}) and
BOOMERanG (cyan;~\cite{2006ApJ...647..823J}). The line is the best-fit
flat, $\Lambda$CDM model to the WMAP5 and ACBAR data.
Reproduced from~\cite{2008arXiv0801.1491R}.
}
\label{fig:Tdata}
\end{figure}
%

We show in Fig.~\ref{fig:Tdata} a selection of
recent measurements of $C_l^T$ taken from~\cite{2008arXiv0801.1491R}.
WMAP and ACBAR are currently the
most constraining of all measurements of $C_l^T$ on large
and small scales, respectively, but will be improved upon significantly
by Planck for $l > 500$. (WMAP5 is cosmic-variance
limited to $l = 530$~\cite{2009ApJS..180..296N}.)
Also shown in Fig.~\ref{fig:Tdata} is the
theoretical power spectrum for the best-fit flat, $\Lambda$CDM model to
the WMAP5 and ACBAR data. The agreement is striking, with five acoustic
peaks clearly delimited.
Moreover, the agreement between different experiments (and technologies)
is excellent where they overlap in $l$ coverage. Understanding the main
features in the $C_l^T$ spectrum is one of the principal aims of these lecture
notes.

The correlation between the temperature anisotropies along two directions
evaluates to
\begin{equation}
\langle \Delta T(\vnhat_1) \Delta T(\vnhat_2) \rangle
= \bar{T}^2 \sum_l \frac{2l+1}{4\pi} C_l^T P_l(\cos\theta)\;,
\label{adc:eq3}
\end{equation}
which depends only on the angular separation $\theta$ as required by rotational
invariance. Here, $P_l(x)$ are the Legendre polynomials. The mean-square
temperature anisotropy is
\begin{equation}
\langle \Delta T ^2 \rangle
= \bar{T}^2 \sum_l \frac{2l+1}{4\pi} C_l^T
\approx \bar{T}^2 \int \frac{l(l+1)}{2\pi} C_l^T \, \D \ln l\; ,
\label{adc:eq4}
\end{equation}
so that the quantity $l(l+1)C_l^T /2\pi$, which is conventionally plotted,
is approximately the power per decade in $l$ of the temperature anisotropies.

\paragraph{Higher-Order Statistics}
\label{subsec:higherorderstatistics}

In Gaussian theories, all higher-order correlations of the $T_{lm}$ can
be reduced to the power spectrum by Wick's theorem. 
In non-Gaussian theories, the form of the correlators is restricted by
rotational invariance. Here, we shall just consider the three-point
function which must take the form
\begin{equation}
\langle T_{l_1 m_1} T_{l_2 m_2} T_{l_3 m_3} \rangle =
B_{l_1 l_2 l_3} \left( \begin{array}{ccc} l_1 & l_2 & l_3 \\
                                          m_1 & m_2 & m_3 
                       \end{array}\right) \; , 
\label{eq:5a}
\end{equation}
where the right-hand side involves the Wigner 3$j$ symbol (which
enforces $m_1 + m_2 +m_3 = 0$) and the
\emph{angular bispectrum} $B_{l_1 l_2 l_3}$. If we further
assume that the generation and propagation of cosmological perturbations
respects parity symmetry, the bispectrum must
then vanish for $l_1 + l_2 + l_3$ odd. Parity invariance further
ensures that the bispectrum is real and invariant under arbitrary
perturbations of its indices. There is, as yet, no strong evidence for a non-zero
CMB bispectrum; see Sec.~\ref{subsubsec:nonGauss}. For a
discussion of the CMB four-point
function, or trispectrum, see~\cite{2001PhRvD..64h3005H}.

\subsection{Thermal history}
\label{subsec:thermal}

The high temperature of the early universe maintained a low
equilibrium fraction of neutral atoms, and a correspondingly high
number density of free electrons. Coulomb scattering between the ions
and electrons kept them in local kinetic equilibrium, and Thomson scattering
of photons tended to maintain the isotropy of the CMB in the baryon rest frame.
As the universe expanded and cooled, the dominant element hydrogen started to
recombine when the temperature fell below $\sim$ 4000~K. This is a factor of
40 lower than might be anticipated from the
13.6-eV ionization potential of hydrogen, and is
due to the large ratio of the number of photons to baryons.
Through recombination, the mean-free path for Thomson scattering grew to the
horizon size and CMB photons ``decoupled'' from matter. More precisely,
the probability density that photons last scattered at some time
defines the \emph{visibility function}.
This is now known to peak 380~kyr after the big bang~\cite{2009ApJS..180..225H}, with a
width $\sim 120\,\mathrm{kyr}$. Since then, CMB photons have
propagated relatively unimpeded for $13.7~\mathrm{Gyr}$, covering a
comoving distance $\chi_* = 14.1\,\mathrm{Gpc}$. The distribution
of their energies carries the imprint of 
fluctuations in the radiation temperature,
the gravitational potentials, and the peculiar velocity of the radiation
where they last scattered, as the temperature anisotropies that we observe
today.
Only a small fraction of CMB photons -- current results from CMB polarization
measurements~\cite{2009ApJS..180..306D} indicate around 10 per cent -- underwent further
scattering once the universe reionized, due to the ionizing radiation from
the first stars.

The details
of hydrogen recombination are complicated since it occurs out of
quasi-equilibrium,
but the broad details are well
understood~\cite{1968ApJ...153....1P,1968ZhETF..55..278Z}. Most hydrogen
atoms recombined through the (forbidden) two-photon transition between
the $2s$ and $1s$ states. Direct recombination to $1s$, or via
allowed transitions from excited states, would have produced a large excess
of photons more energetic than Ly-$\alpha$ (i.e.\ the
$2p$--$1s$ transition) over those in the Wien tail of
an equilibrium blackbody spectrum, and would therefore have been rapidly
re-absorbed.

Understanding hydrogen recombination at the level required for sub-percent
accuracy in CMB observables is both a challenging and ongoing activity.
A major step forward was made in~\cite{2000ApJS..128..407S}
where a number of approximations made in the original calculations
were improved upon. However, this improved calculation still misses
a number of effects that are thought to influence the CMB power
spectra by more than 1\% for $l>1500$ and so there has been a flurry of recent
activity in this area driven by the imminent launch of the Planck satellite.
For a recent review of some of the issues involved, see~\cite{2008arXiv0807.2577F}. 

\subsection{Linear anisotropy generation: gravity}

We shall mostly be concerned with linearised scalar (density)
perturbations about
spatially-flat models in general relativity.
It is convenient to work in the conformal
Newtonian gauge in which the 
spacetime metric is
\begin{equation}
\D s^2 = a^2(\eta)[(1+2\psi)\D\eta^2 - (1-2\phi)\D\vx^2]\; ,
\label{eq:6}
\end{equation}
where $\eta$ is conformal time (related to proper time
$t$ by $\D t = a \D\eta$), $a$ is the scale factor in the background model
and $\vx$ is comoving position. The physics of the CMB is
particularly intuitive in this gauge.
The two
scalar potentials $\phi$ and $\psi$ represent the scalar perturbation to the
metric, and play a similar role to the Newtonian gravitational
potential. In the absence of anisotropic stress in matter and fields,
$\phi=\psi$.

We parameterise a CMB photon by its energy $\epsilon/a$
as seen by an observer at rest in our coordinate system, where
$\epsilon$ is the \emph{comoving energy},
and the direction cosines
$\ve = (e^1,e^2, e^3)$, with $\ve^2 = 1$, relative to an orthonormal
triad of spatial vectors, $a^{-1}(1+\phi) \partial_i$.
The comoving energy is constant in the absence of metric perturbations.
The photon four-momentum is then
\begin{equation}
p^\mu = a^{-2} \epsilon [1-\psi,(1+\phi)\ve]\; .
\label{eq:7}
\end{equation}
We write the photon distribution function, $f(x^\mu,p^\mu)$, as
the sum of a background part $\bar{f}(\epsilon)$, which is a Planck function
with $E/ \bar{T}(\eta) = \epsilon / (a \bar{T})$ and so has no explicit time
dependence as $a \bar{T}(\eta) = \mathrm{const.}$,
and a linear perturbation. Denoting small changes in the 
temperature of the radiation for different directions by 
$\bar{T}(\eta) \Theta(\ve)$, the perturbation to the distribution function
can be written as $-\Theta(\ve) \epsilon \D \bar{f} / \D \epsilon$
so that
\begin{equation}
f(\eta,\vx,\epsilon,\ve) = \bar{f}(\epsilon)[1-\Theta(\eta,\vx,\ve)
\D \ln \bar{f}(\epsilon) / \D \ln \epsilon] \; .
\label{eq:8}
\end{equation}
The quadrupole and higher multipoles of the fractional temperature
fluctuation $\Theta$ are gauge-invariant.

The distribution function is conserved along the photon path in phase space
when there is no scattering. Parameterising the path with $\eta$, we
have $\D \vx / \D \eta = (1+\phi + \psi)\ve$ from $p^\mu = \D x^u / \D
\lambda$ (with $\lambda$ an affine parameter). The evolution of
$\epsilon$ and $\ve$ follows from the geodesic equation; at linear order
\begin{eqnarray}
\D\epsilon / \D \eta &=& - \epsilon \D \psi / \D \eta
+ \epsilon (\dot{\phi} + \dot{\psi}) \; , \label{eq:9} \\
\D \ve / \D \eta &=& - (\vnabla - \ve \ve \cdot \vnabla) (\phi + \psi) \; ,
\label{eq:10}
\end{eqnarray}
where overdots denote differentiation with respect to $\eta$ and
$\D \psi / \D \eta = \dot{\psi} + \ve \cdot \vnabla \psi$ is the
derivative of $\psi$ along the photon path. The comoving energy
thus evolves due to variation in the gravitational potentials
experienced by the photon, and the direction is deflected by
the gradient of $\phi + \psi$ perpendicular to $\ve$ (i.e.\ gravitational
lensing). The derivative of the distribution function along the spacetime
path \emph{at fixed $\epsilon$ and $\ve$} is sourced by the gravitational
redshifting in Eq.~(\ref{eq:9}), so that
\begin{equation}
\left. \frac{\D \Theta (\eta,\vx,\ve)}{\D \eta} \right|_{\mathrm{grav}}
= \left. \left(\frac{\partial}{\partial \eta} + \ve \cdot \vnabla \right)
\Theta \right|_{\mathrm{grav}} = 
- \frac{\D \psi}{\D \eta} + \dot{\phi} + \dot{\psi} \; .
\label{eq:11}
\end{equation}
The gravitational lensing effect does not contribute at linear order
since the zero-order distribution function $\bar{f}$ is isotropic.
If our observation event has coordinates $(\eta_R,\vx_R)$, a photon
would have last scattered around recombination at the event $(\eta_*,
\vx_*)$, where $\eta_* = \eta_R - \chi_*$ and $\vx_* = \vx_R - \ve \chi_*$.
On integrating Eq.~(\ref{eq:11}) between these events, the first term
on the right produces a temperature perturbation given by the difference
between $\psi$ at last scattering and at reception. A photon suffers
a net redshift if it climbs out of a deeper potential well than it is
received in and this corresponds to a negative temperature fluctuation.
The second term on the right of Eq.~(\ref{eq:11})
gives the integrated Sachs-Wolfe contribution to $\Theta$,
\begin{equation}
\Theta_{\mathrm{ISW}} =  \int_*^R (\dot{\phi}+\dot{\psi}) \, \D \eta \; ,
\label{eq:12}
\end{equation}
where the integral is taken along the line of sight. At linear order,
the only contributions to this integral are from the evolution of the
potentials around recombination, due to the duration of the
matter-radiation transition (the early ISW effect), and the late-time ISW
effect, when the potentials decay as dark energy becomes dynamically
important. For the latter, in traversing a potential well that is getting
shallower in time (i.e.\ $\dot{\phi} +\dot{\psi} > 0$),
photons receive a net blueshift and a positive contribution
to $\Theta$. 

\subsection{Linear anisotropy generation: scattering}
\label{subsec:scattering}

The dominant scattering mechanism to affect CMB anisotropies is classical
Thomson scattering off non-relativistic, free electrons (with proper
number density $n_{\mathrm{e}}$). The evolution
of the temperature fluctuation due to scattering takes the form
\begin{equation}
\left. \frac{\D \Theta}{\D \eta}\right|_{\mathrm{scattering}} =
- a n_{\mathrm{e}} \sigma_{\mathrm{T}} \Theta + \frac{3a n_{\mathrm{e}} \sigma_{\mathrm{T}}}{16\pi}
\int \D \hat{\vm}\, \Theta(\epsilon,\hat{\vm})\left[1+(\ve\cdot \hat{\vm})^2
\right]
+ a n_{\mathrm{e}} \sigma_{\mathrm{T}} \ve \cdot \vv_{\mathrm{b}} \; ,
\label{eq:13}
\end{equation}
where $\sigma_{\mathrm{T}}$ is the Thomson cross-section. The first term
on the right describes scattering out of the beam; the second,
scattering into the beam and involves the familiar $1+\cos^2\theta$
dependence where $\theta$ is the scattering angle; the final term
is a Doppler effect that arises from the peculiar (bulk) velocity of the
electrons, $\vv_{\mathrm{b}} = \D \vx / \D \eta$.
For the latter, scattering from a moving electron boosts the energy of
a photon scattered in the direction that the electron is moving.
Note that the effect of scattering vanishes for an isotropic distribution
function and $\vv_{\mathrm{b}} = 0$. If we neglect the anisotropic
nature of Thomson scattering, replacing $1+\cos^2\theta$ by its
angular average $4/3$, Eq.~(\ref{eq:13}) reduces to
\begin{equation}
\left. \frac{\D \Theta}{\D \eta}\right|_{\mathrm{scattering}} \approx
- a n_{\mathrm{e}} \sigma_{\mathrm{T}} (\Theta - \Theta_0 - \ve \cdot \vv_{\mathrm{b}})
\; ,
\label{eq:14}
\end{equation}
where $\Theta_0 = \int \Theta \, \D \ve / (4\pi)$ is the monopole of
the temperature fluctuation (which equals $\delta_\gamma / 4$ with
$\delta_\gamma$ the photon energy density contrast).
As expected, scattering tends to isotropise the photons in the
baryon rest frame, $\Theta \rightarrow \Theta_0 + \ve \cdot \vv_{\mathrm{b}}$.

Combining the gravitational and scattering contributions gives the
Boltzmann equation for $\Theta$, which is most conveniently written as
\begin{eqnarray}
\frac{\D}{\D \eta} \left[ \E^{-\tau}(\Theta + \psi)\right]
&=& - \dot{\tau} \E^{-\tau} \left( \psi + \ve\cdot \vv_{\mathrm{b}}
+ \frac{3}{16\pi} \int \D \hat{\vm}\, \Theta(\epsilon,\hat{\vm})
\left[1+(\ve\cdot \hat{\vm})^2
\right]\right) \nonumber \\
&& \mbox{} + \E^{-\tau} (\dot{\phi} + \dot{\psi}) \; ,
\label{eq:15}
\end{eqnarray}
where $\tau(\eta) \equiv \int_{\eta}^{\eta_R} a n_{\mathrm{e}}
\sigma_{\mathrm{T}}\, \D \eta'$ is the optical depth to Thomson scattering.
The quantity $- \dot{\tau} \E^{-\tau}$ is the (conformal) visibility function,
introduced in Sec.~\ref{subsec:thermal}. 

A formal solution to Eq.~(\ref{eq:15}) can be found by integrating
along the line of sight from some early time, when $\tau \rightarrow \infty$,
to the reception event $R$ where $\tau = 0$:
\begin{equation}
[\Theta(\ve) + \psi]_R = \int_0^R S\, \D \eta \; ,
\label{eq:16}
\end{equation}
with $S$ the source term on the right-hand side of Eq.~(\ref{eq:15}).
The standard Boltzmann codes, CMBFAST~\cite{1996ApJ...469..437S} and CAMB~\cite{2000ApJ...538..473L}
evaluate this integral directly and efficiently by expanding the source term
in Fourier modes. For our purposes here, we gain useful insight into the
physics of anisotropy formation by approximating last scattering as
instantaneous, in which case the visibility function is a delta-function,
and ignoring the effect of reionization and the anisotropic nature of
Thomson scattering; Eq.~(\ref{eq:16}) then approximates to
\begin{equation}
[\Theta(\ve) + \psi]_R = \Theta_0 |_* + \psi_* + \ve \cdot \vv_{\mathrm{b}}|_*
+ \int_*^R (\dot{\phi} + \dot{\psi})\, \D \eta \; .
\label{eq:17}
\end{equation}
The various terms have a simple physical interpretation.
The temperature received along $\ve$ is the isotropic
temperature of the CMB at the last scattering event on the line of sight,
$\Theta_0 |_*$, corrected for the gravitational redshift due to the
difference in the potential $\psi_* - \psi_R$,
and the Doppler shift
$\ve \cdot \vv_{\mathrm{b}}|_*$ resulting from scattering off moving electrons.
Finally, there is the integrated Sachs-Wolfe contribution from evolution
of the potentials along the line of sight.

\subsubsection{Spatial-to-angular projection}
\label{subsubsection:spatang}

The main contribution to the CMB temperature anisotropies comes from the
projection of $\Theta_0 + \psi$ at the time of last scattering onto
the sphere of radius $\chi_*$ centred on $\vx_R$. For a general,
statistically homogeneous and isotropic field $f(\vx)$, with
a Fourier representation
\begin{equation}
f(\vx) = \int \frac{\D^3 \vk}{(2\pi)^{3/2}}\,
f(\vk) \E^{i \vk \cdot \vx} \; ,
\label{eq:18}
\end{equation}
the spherical multipoles of the projection of $f$ on the
sphere $|\vx - \vx_R | = \chi_*$ are
\begin{equation}
f_{lm} = 4\pi i^l \int \frac{\D^3 \vk}{(2\pi)^{3/2}}
f(\vk) j_l(k\chi_*) Y_{lm}^*(\hat{\vk}) \E^{i \vk \cdot \vx_R} \; .
\label{eq:19}
\end{equation}
Here, $j_l(k\chi_*)$ are spherical Bessel functions. For large $l$ they peak
sharply around $k \chi_* = l$ so that the multipoles $f_{lm}$ receive their
majority contribution from Fourier modes of wavenumber $k \sim l /\chi_*$.
For any line of sight, such wavevectors aligned perpendicular to the line of
sight give angular structure on the sky with characteristic period
$2\pi / (k \chi_*) \sim 2\pi / l$ matching that of $Y_{lm}(\vnhat)$.
The $j_l(k\chi_*)$ have oscillatory tails for $k \chi_* \gg l$ so that
modes with $k > l /\chi_*$ also contribute to the $f_{lm}$; this
corresponds to shorter wavelength Fourier modes not oriented
perpendicular to the line of sight. Introducing the dimensionless
power spectrum, $\mathcal{P}_f(k)$, of $f(\vx)$,
\begin{equation}
\langle f(\vk) f^*(\vk') \rangle = \frac{2\pi^2}{k^3} \mathcal{P}_f(k)
\delta(\vk-\vk') \; ,
\label{eq:20}
\end{equation}
the angular power spectrum of the projection is
\begin{equation}
C_l^f = 4\pi \int \mathcal{P}_f(k) j_l^2(k\chi_*) \, \D \ln\, k \; .
\label{eq:21}
\end{equation}
The contribution to $C_l^T$ from $\Theta_0 + \psi$ at last scattering takes
this form with $f = \Theta_0 + \psi$.

The arguments for the projection of a field that is locally a dipole, like
$\ve \cdot \vv_{\mathrm{b}}$, are a little different. For scalar perturbations,
the Fourier components of $\vv_{\mathrm{b}}$ take the form
$i \hat{\vk} v_{\mathrm{b}}(\vk)$ and so the projection of 
$\ve \cdot \vv_{\mathrm{b}}$ has multipoles
\begin{equation}
(\ve \cdot \vv_{\mathrm{b}})_{lm} = -4\pi i^l \int \frac{\D^3 \vk}{(2\pi)^{3/2}}
v_{\mathrm{b}}
(\vk) j_l'(k\chi_*) Y_{lm}^*(\hat{\vk}) \E^{i \vk \cdot \vx_R} \; .
\label{eq:21b}
\end{equation}
Here, $j_l'(k\chi_*)$ is the derivative of $j_l(k \chi_*)$ with respect to
its argument. It does not have the same prominent peak as $j_l(k \chi_*)$,
rather having positive and negative wings either side of $k \chi_* = l$.
The projection from $k$ to $l$ is therefore much less sharp for terms
that are locally a dipole compared to monopole terms. Physically,
this arises because there is no contribution to
the projection of $\ve \cdot \vv_{\mathrm{b}}$ from Fourier modes
with their wavevector perpendicular to the line of sight.

\subsection{Acoustic physics}
\label{sec:acosutic}

The photon isotropic temperature perturbation $\Theta_0 = \delta_\gamma/ 4$ and
baryon velocity $\vv_{\mathrm{b}}$ depend on the acoustic physics
of the pre-recombination plasma.
Before recombination, the comoving mean-free path of CMB photons is
\begin{equation}
l_p = (a n_{\mathrm{e}} \sigma_{\mathrm{T}})^{-1} = - \dot{\tau}^{-1}
\approx
4.9\times 10^4 (\Omega_\mathrm{b} h^2)^{-1}(1+z)^{-2} \,\, \mathrm{Mpc} \; .
\label{eq:22}
\end{equation}
On scales larger than this, Thomson scattering keeps the photons
isotropic in the baryon rest frame and the photon-baryon system
can be modelled as a tightly-coupled fluid characterised by a single
peculiar velocity $\vv_{\mathrm{b}}$. The Euler equation for this system,
in the Newtonian gauge, is 
\begin{equation}
\dot{\vv}_{\mathrm{b}} + \frac{R}{1+R} \mathcal{H} \vv_{\mathrm{b}}
+ \frac{1}{4(1+R)} \vnabla \delta_\gamma + \vnabla \psi = 0 \; ,
\label{eq:23}
\end{equation}
where $R \equiv \bar{\rho}_{\mathrm{b}}/(\bar{\rho}_\gamma + \bar{p}_\gamma)
= 3 \bar{\rho}_{\mathrm{b}} / (4 \bar{\rho}_\gamma) \propto a$ and
$\bar{\rho}_i$ and $\bar{p}_i$ are the unperturbed energy density and
pressure of species $i$. The drag term, involving the conformal
Hubble parameter $\mathcal{H} \equiv \dot{a}/a$, arises since the peculiar
velocity of non-relativistic massive particles (the baryons) decays as
$1/a$ in an expanding universe. The $\vnabla \delta_\gamma$ term
comes from the gradient of the photon pressure. This acts on the full inertia
of the photon-baryon system and so its effect reduces for increasing
$R$. The final term describes (universal) gravitational infall in th
potential gradient
$\vnabla \psi$. The energy density of the photons is not changed by scattering
off cold electrons and so $\delta_\gamma$ evolves as for a non-interacting
photon gas:
\begin{equation}
\dot{\delta}_\gamma + \frac{4}{3} \vnabla \cdot \vv_{\mathrm{b}}
- 4 \dot{\phi} =0 \; .
\label{eq:24}
\end{equation}
The second and third terms in this equation are the usual
$\bar{\rho} + \bar{p}$ times volume expansion rate, where
the perturbed expansion rate receives contributions from
$\vnabla \cdot \vv_{\mathrm{b}}$ and the perturbation $-3 \phi \D^3 \vx$ to the
volume element (see Eq.~(\ref{eq:6})).

Combining Eqs~(\ref{eq:23}) and~(\ref{eq:24}) gives the equation of 
a damped, harmonic oscillator driven by
gravity~\cite{1995ApJ...444..489H}:
\begin{equation}
\ddot{\delta}_\gamma + \frac{\mathcal{H}R}{1+R} \dot{\delta}_\gamma
- \frac{1}{3(1+R)}\vnabla^2 \delta_\gamma  = 4 \ddot{\phi} +
\frac{4 \mathcal{H}R}{1+R} \dot{\phi} + \frac{4}{3} \nabla^2  \psi\; .
\label{eq:25}
\end{equation}
First, consider the solutions to the homogeneous equation, ignoring the
effects of gravity. On sub-Hubble scales, $k \gg \mathcal{H}$, the
WKB approximation can be used to show that
\begin{equation}
\delta_\gamma = (1+R)^{-1/4} \cos k r_\mathrm{s}  \phantom{xx}
\mathrm{and} \phantom{xx} \delta_\gamma = (1+R)^{-1/4} \sin k r_\mathrm{s}\; ,
\label{eq:26}
\end{equation}
are independent solutions, where 
the \emph{sound horizon} $r_\mathrm{s} \equiv \int_0^\eta \D \eta' /
\sqrt{3(1+R)}$. These are acoustic oscillations at instantaneous frequency
$k / \sqrt{3(1+R)}$, where $1/\sqrt{3(1+R)}$ is the sound speed of the
photon-baryon system, with a slow decay in amplitude.
We shall show shortly that for \emph{adiabatic
initial conditions}, as expected in single-field models of inflation,
the system oscillates essentially in the cosine mode. All Fourier modes
of a given $k$, irrespective of the direction of their wavevector,
reach extrema of their acoustic oscillations at the same time. This
\emph{phase coherence} is an inevitable consequence of the dominance of
a single ``growing mode'' in the primordial perturbation at the end of
inflation, and is not present in models that continually
source perturbations such as those with topological defects~\cite{1997PhRvL..79.1611P}.
Phase coherence implies a series of acoustic peaks in the CMB angular
power spectrum (see Fig.~\ref{fig:Tdata})
at multipoles $l$ such that $l/\chi_*$ corresponds to
the wavenumber of modes that have reached extrema at $\eta_*$.
For adiabatic initial conditions, the acoustic peaks are located
approximately at $l = n \pi \chi_* / r_{\mathrm{s}}(\eta_*)$.

\subsubsection{Initial conditions and gravitational driving}

We can get the initial conditions for the oscillator deep in the radiation era
by considering the self-gravitating perturbations of a radiation fluid.
Such a treatment makes a number of assumptions. First, neutrinos
are treated as a fluid but, after decoupling, they actually free-stream
and so do not cluster appreciably on sub-Hubble scales. Moreover,
free-streaming generates anisotropic stress $\sim \bar{\rho}_\nu
(k\eta)^2 \phi$ in the neutrinos which produces a
time-independent 14\% offset between (the constant) $\phi$ and $\psi$ on
large scales.
Second,
we are assuming that the perturbations in the baryonic matter and
CDM have a negligible effect on the gravitational potentials. This is
appropriate for adiabatic initial conditions deep in the radiation
era, where 
all species are initially distributed homogeneously
on large scales on the same spacetime hypersurfaces, i.e.\ the
density perturbations of all species vanish on hypersurfaces over which
the total density perturbation vanishes.
In any other gauge,
$\bar{\rho}_i \delta_i / \dot{\bar{\rho}}_i$ is equal for all species
and there are therefore no relative perturbations in the number density of
particles. For adiabatic initial conditions, the total energy density
contrast in the radiation era is dominated by the radiation perturbations.
This should be compared to isocurvature initial conditions, where,
for example, the CDM density contrast is initially much larger than the
radiation contrast in such a way as to give no net curvature perturbation.

In a radiation fluid, $\mathcal{H} = 1 /\eta$ and
the trace of the $ij$ Einstein equation (see,
for example,~\cite{2003moco.book.....D}) gives
\begin{equation}
\ddot{\phi} + \frac{3}{\eta} \dot{\phi} - \frac{1}{\eta^2} \phi =
\frac{1}{2\eta^2} \delta_r \; ,
\label{eq:27}
\end{equation}
where the subscript $r$ denotes radiation (photons plus neutrinos).
The $00$ Einstein equation is
\begin{equation}
\vnabla^2 \phi = \frac{3}{\eta^2} \left(\eta \dot{\phi} + \phi\right)
+ \frac{3}{2\eta^2} \delta_r \; ,
\label{eq:28}
\end{equation}
so the potential evolves as
\begin{equation}
\ddot{\phi} + \frac{4}{\eta} \dot{\phi} - \frac{1}{3} \vnabla^2 \phi = 0 \; ,
\label{eq:29}
\end{equation}
with regular (growing-mode) solution $\phi \propto j_1(k\eta/\sqrt{3})
/ (k\eta)$. This is constant outside the sound horizon,
$k \eta \ll \sqrt{3}$, but undergoes decaying oscillations on small scales
with asymptotic behaviour\footnote{The
presence of the CDM, which does not undergo acoustic
oscillations, leads to a $k$-dependent offset, $\sim k^{-2} \ln k$,
in the oscillation as the matter-radiation transition is approached.}
$\phi \propto - \sqrt{3} \cos(k\eta/\sqrt{3}) / (k \eta)^2$.
Solving for $\delta_r$ from Eq.~(\ref{eq:28}) gives a solution
with $\delta_r = -2 \phi(0)$ on large scales (i.e.\ $\delta_r$ is
constant) and
constant amplitude oscillations,
$\delta_r = - 6 \cos(k\eta/ \sqrt{3}) \phi(0)$,  well inside the
sound horizon. As advertised above, adiabatic initial conditions
excite the cosine mode of the oscillator. Note also that the
radiation starts off over-dense in potential wells.

If we now return to Eq.~(\ref{eq:25}), we can make two important observations
that affect the height of the acoustic peaks in the CMB power spectrum.
First, for modes that enter the sound horizon between matter-radiation
equality and recombination, the oscillations occur in potentials
that approximately equal the primordial value $\phi(0)$. More carefully, the
potential decays by 10\% through the matter-radiation
transition on large scales  -- although this is not complete at
recombination -- and, by Eq.~(\ref{eq:24}), $\delta_\gamma$ is enhanced
by 20\%. The potential
shifts the midpoint of the oscillation to
$\delta_\gamma \approx -4 (1+R) \psi$ which corresponds to
the over-density needed for photon pressure to balance gravitational infall.
Since baryons contribute to the inertia
but not the pressure, they enhance $\delta_\gamma$ at the midpoint.
The source term for the temperature anisotropy, $\Theta_0 + \psi =
\delta_\gamma / 4 + \psi$, therefore oscillates about $- R \psi$
with an amplitude of $3\phi(0)/10$ (ignoring non-zero $R$ in the amplitude).
The dependence of the midpoint
on the baryon density makes the relative heights of the first
few acoustic peaks sensitive to the baryon density.
The second observation is that for modes shorter than the
sound horizon at matter-radiation equality, the decay of the potential
during the oscillations in the radiation renders the driving term
on the right of Eq.~(\ref{eq:25}) negligible during the matter era.
However, our treatment of the radiation fluid above shows that
the driving term acts resonantly around the time of sound-horizon crossing
in the radiation era and enhances the amplitude of the acoustic
oscillation of $\delta_\gamma /4$ towards the asymptotic
value $3\phi(0)/2$, five times the value on large scales.
Increasing the matter density limits the scales for which resonant
driving is effective to smaller scales.
In practice, diffusion damping (see Sec.~\ref{subsec:complications})
exponentially damps the amplitude of the oscillations on small scales
and the asymptotic amplitude is not attainable.

\subsection{$C_l^T$ for adiabatic initial conditions}

\begin{figure}[t!]
\centerline{\includegraphics[width=0.5\textwidth,angle=-90]{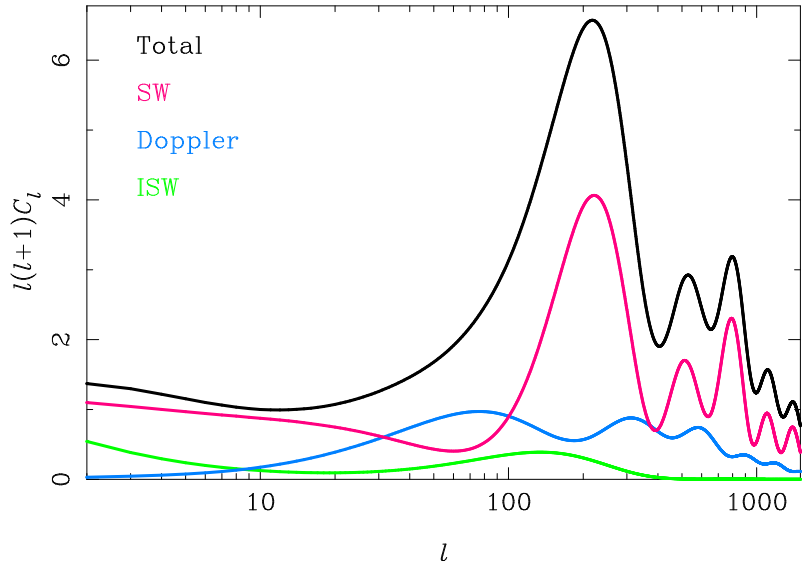}}
\caption{Contribution of the various terms in Eq.~(\ref{eq:17}) to the
temperature-anisotropy power spectrum from adiabatic initial conditions:
$\delta_\gamma/4 + \psi$ (denoted
SW for Sachs-Wolfe~\cite{1967ApJ...147...73S}; magenta);
Doppler effect from $v_\mathrm{b}$ (blue);
and the integrated Sachs-Wolfe effect (ISW; green)
coming from evolution of the potential along the line of sight.
The units of the spectrum are arbitrary.}
\label{fig:adiabatic_spectrum}
\end{figure}

The contributions of the various terms in Eq.~(\ref{eq:17})
to the temperature anisotropy power spectrum for adiabatic initial
conditions are shown in Fig.~\ref{fig:adiabatic_spectrum}. The main
contribution is from $\Theta_0 + \psi$. The plateau on large scales
comes from modes that have not had time to oscillate by recombination;
for such modes, $\Theta_0 + \psi \approx 3 \phi(0)/10 = \phi/3$.
It is convenient to express the initial value of the potential
in terms of the comoving-gauge curvature perturbation,
\begin{equation}
\mathcal{R} \equiv - \phi - \frac{\mathcal{H}(\dot{\phi} + \mathcal{H}
\psi)}{4\pi G a^2 (\bar{\rho} + \bar{p})} \; ,
\end{equation}
which is conserved on super-Hubble scales for adiabatic perturbations
independent of the equation of state relating the total
pressure $\bar{p}$ and energy density $\bar{\rho}$. During radiation
domination, $\mathcal{R} = - 3 \phi(0)/2$ ignoring neutrino anisotropic
stress, and so on large scales $\Theta_0 + \psi \approx - \mathcal{R}/5$.
This directly relates $C_l^T$ to the primordial power
spectrum $\mathcal{P}_\mathcal{R}(k)$ (which is calculable given a model
of inflation) on large scales. In particular, a nearly scale-invariant
$\mathcal{P}_\mathcal{R}(k)$ projects to give $l (l+1) C_l \approx
\mathrm{const}$. The prominent acoustic peaks in
Fig.~\ref{fig:adiabatic_spectrum} reflect the acoustic oscillations
of the photon-baryon system. The first peak corresponds to
modes that have just reached the first maximum of their oscillation
by last scattering. The resonant-driving arguments given earlier
suggest that the acoustic peaks should grow in amplitude with increasing
$l$. In practice, an exponential decline in power is seen because of
diffusion damping.

The late-time ISW effect adds power incoherently to the contribution
from recombination since the former is sourced at late times.
However, the early-ISW effect, which peaks close to the first
acoustic peak, adds power coherently~\cite{1995ApJ...444..489H}.
It enhances the first peak
considerably since $\dot{\phi} + \dot{\psi}$ has the same sign, i.e.\
positive in potential wells, as $\Theta_0 + \psi$. It is
less important for the higher peaks since the potential at last scattering
for these is suppressed by the acoustic oscillations during
radiation domination.

The baryon peculiar velocity oscillates $\pi/2$ out of phase with
$\delta_\gamma$ by Eq.~(\ref{eq:24}), and so the Doppler term tends to
fill in the troughs of the angular power spectrum. The oscillatory
structure would be destroyed completely were it not for two effects:
the geometric effect noted for the projection
of dipole-like fields in Sec.~\ref{subsec:scattering}, and the
effect of baryons. The latter
shift the midpoint of the oscillations in
$\Theta_0 + \psi$ from zero, so its zeroes do not coincide with the
extrema of the peculiar velocity. Furthermore, baryons reduce
the amplitude of the oscillations in $\vv_{\mathrm{b}}$ by a factor
of $\sqrt{(1+R)}$.

\subsection{Complications to the simple picture}
\label{subsec:complications}

\subsubsection{Photon diffusion}

On small scales, the CMB anisotropies are exponentially damped
due to photon diffusion. Before recombination, fluctuations with
wavelengths comparable to or smaller than the mean-free path of photons
to Thomson scattering are damped out, as photons can diffuse out of
over-dense regions into neighbouring under-densities. How far will
a photon have diffused by last scattering? To answer this, consider
the comoving mean-free
path of the photons, $l_p$, which is related to the Thomson cross section and the
number density of free electrons via
\begin{equation}
l_p = \frac{1}{a n_{\mathrm{e}} \sigma_{\mathrm{T}}} \, .
\label{eq:261}
\end{equation}
In some interval of conformal time, $d\eta$, a photon undergoes
$\D\eta/ l_p$ scatterings and random walks through a mean-squared
distance $l_p^2 \D\eta/ l_p = l_p \D\eta$. The total mean-squared distance
that a photon will have moved by such a random walk by the time
$\eta_*$ is therefore
\begin{equation}
\int_0^{\eta_*} \frac{\D\eta'}{a n_{\mathrm{e}} \sigma_{\mathrm{T}}} \sim \frac{1}{k_D^2}\, ,
\end{equation}
which defines a damping scale $k_D^{-1}$. The photon density and velocity
perturbations at wavenumber $k$ are damped exponentially by photon
diffusion, going like $e^{-k^2/k_D^2}$; this produces a similar damping in the
angular power spectrum $C_l$ (see Fig.~\ref{fig:adiabatic_spectrum}).
Evaluating $k_D$ around recombination gives a comoving damping scale
$\sim 30\,\mathrm{Mpc}$.

\subsubsection{Reionization}
\label{subsubsec:Treion}

Once structure formation had proceeded to produce the first sources of
ionizing radiation, neutral hydrogen and helium began to reionize. The
resulting free
electron density could then re-scatter CMB photons, and this affected
the observed CMB in several ways. The only significant
effect for the temperature anisotropies in linear theory is a uniform
screening by $\E^{-\tau_{\mathrm{re}}}$, where $\tau_{\mathrm{re}}$ is
the optical depth through reionization, on scales $l > 10$.
Anisotropies on such scales are generated from perturbations that
are sub-Hubble at the time of reionization. For these modes, the
radiation is already significantly anisotropic at reionization since
the perturbations vary significantly over the scattering electron's
own last-scattering surface. The net effect of in-scattering from
different lines of sight thus averages to zero, leaving a suppression
by $\E^{-\tau_{\mathrm{re}}}$ due to scattering out of the line of sight.
In the power spectrum, this becomes $\E^{-2\tau_{\mathrm{re}}}$.
However, Fourier modes that are still super-Hubble at
reionization produce negligible \emph{anisotropy} by the time of reionization
since the
perturbations over the scattering electron's own last scattering surface
are almost uniform. Scattering isotropic radiation has no effect, and
so reionization does not alter the temperature anisotropies from
recombination for $l < 10$. Scattering around
reionization also generates new large-angle polarization -- see
Sec.~\ref{subsec:pol_reion} -- and, at second-order, small-scale
temperature anisotropies; see Sec.~\ref{subsec:secondary_scattering}.

Note that screening at reionization makes the determination of the
amplitude and shape of the primordial power spectrum from temperature data
alone rather degenerate with $\tau_{\mathrm{re}}$.

\subsubsection{Gravitational waves}

Tensor modes, describing gravitational waves, represent the
transverse trace-free perturbations to the spatial metric:
\begin{equation}
\D s^2 = a^2(\eta) [\D \eta^2 - (\delta_{ij} + h_{ij})\D x^i \D x^j]\; , 
\label{adc:eq40}
\end{equation}
with $h_i^i=0$ and $\partial_i h^i_j = 0$. A convenient parameterisation of
the photon four-momentum in this case is
\begin{equation}
p^\mu = \frac{\epsilon}{a^2}\left[1, e^i - \frac{1}{2} h^i_j e^j\right]\; ,
\label{adc:eq41}
\end{equation}
where, as for scalar perturbations, $\ve^2=1$ and $\epsilon$ is $a$ times the 
energy of the photon as seen
by an observer at constant $\vx$. The components of $\ve$ are the
direction cosines of the photon direction for this observer on an orthonormal
spatial triad of vectors $a^{-1}(\partial_i - h_i^j \partial_j / 2)$,
and are constant in the unperturbed universe.
The evolution of the comoving energy, $\epsilon$, follows from the
geodesic equation:
\begin{equation}
\frac{1}{\epsilon} \frac{\D \epsilon}{\D \eta} + \frac{1}{2} \dot{h}_{ij}
e^i e^j = 0\; ,
\label{adc:eq42}
\end{equation}
and so the Boltzmann equation for the fractional temperature
anisotropies is
\begin{eqnarray}
\frac{\partial \Theta}{\partial \eta} + \ve \cdot \vnabla
\Theta  &=&
- a n_{\mathrm{e}} \sigma_{\mathrm{T}} \Theta + \frac{3a n_{\mathrm{e}} \sigma_{\mathrm{T}}}{16\pi}
\int \D \hat{\vm}\, \Theta(\epsilon,\hat{\vm})\left[1+(\ve\cdot \hat{\vm})^2
\right] \nonumber \\
&&\mbox{} - \frac{1}{2} \dot{h}_{ij} e^i e^j\; .
\label{adc:eq43}
\end{eqnarray}
All perturbed scalar and vector quantities vanish for tensor perturbations
so $\Theta(\ve)$ has only $l \geq 2$ moments.
Neglecting the anisotropic nature of Thomson scattering, or, equivalently,
the temperature quadrupole at last scattering,
the solution of Eq.~(\ref{adc:eq43}) is an integral along the unperturbed
line of sight:
\begin{equation}
[\Theta(\ve)]_R = - \frac{1}{2} \int_*^R \E^{-\tau} \dot{h}_{ij} e^i
e^j \, \D\eta\; .
\label{adc:eq44}
\end{equation}
The physics behind this solution is as follows.
The time derivative $\dot{h}_{ij}$ is the shear induced by the gravitational
waves. This quadrupole perturbation to the expansion rate
produces an anisotropic
redshifting of the CMB photons and an associated temperature anisotropy.

The shear source term for the tensor anisotropies is locally
a quadrupole since $\dot{h}_{ij}$ is trace-free. Projecting this
contribution at a given time onto angular multipoles on the sky is
a little more involved than for scalar perturbations~\cite{1997PhRvD..56..596H}.
We start by decomposing
$\dot{h}_{ij}$ into circularly-polarized Fourier modes,
\begin{equation}
h_{ij} = \sum_{\pm} \int \frac{\D^3 \vk}{(2\pi)^{3/2}}
h_{ij}^{(\pm 2)}(\vk) \E^{i\vk \cdot \vx} \, ,
\qquad
h_{ij}^{(\pm 2)}(\vk) =
\frac{1}{\sqrt{2}} m_{ij}^{(\pm 2)}(\vk) h^{(\pm 2)}(\vk)
\; .
\end{equation}
Here, the basis tensors $m_{ij}^{(\pm 2)}(\vk)$ are complex, symmetric
trace-free and orthogonal to $\vk$, and $h^{(\pm 2)}(\vk)$ are scalar
Fourier amplitudes.
For $\vk$ along the $z$-axis, we
choose $m_{ij}^{(\pm 2)} = (\hat{\vx} \pm i \hat{\vy})_i
(\hat{\vx} \pm i \hat{\vy})_j / 2$. For such Fourier modes
the shear source is
\begin{equation}
\dot{h}_{ij}^{(\pm 2)}(k\hat{\vz})e^ie^j \propto \frac{1}{2\sqrt{2}}
\left[(\hat{\vx}\pm i \hat{\vy})\cdot \ve\right]^2 = \frac{1}{2\sqrt{2}}
\sin^2\theta e^{\pm 2i\phi} = \sqrt{\frac{4\pi}{15}} Y_{2\, \pm 2}(\ve) \, .
\label{eq:shearsource}
\end{equation}
The projection of this at comoving distance $\chi$ is
\begin{eqnarray}
\sqrt{\frac{4\pi}{15}} Y_{2\, \pm 2}(\ve) \E^{-i k \chi \cos\theta} &=&
\frac{4\pi}{\sqrt{15}} \sum_L (-i)^L \sqrt{2L+1} j_L(k\chi) Y_{2\,\pm 2}(\ve)
Y_{L0}(\ve) \nonumber \\
&=& \sqrt{\frac{4\pi}{3}} \sum_L \Biggl[ (-i)^L (2L+1) j_L(k\chi)
\nonumber \\
&& \hspace{-30pt} \times \sum_l
\sqrt{2l+1} \threej{2}{L}{l}{\mp 2}{0}{\pm 2}
\threej{2}{L}{l}{0}{0}{0} Y_{l\pm 2}(\ve) \Biggr] \, ,
\end{eqnarray}
where we have used the Rayleigh plane-wave expansion. The $3j$ symbols arise
from coupling the angular dependence of the source to that of the plane
wave. Since they force $L$ and $l$ to have the same parity, the $l$th
multipoles of the projection involve the $l$ and $l\pm 2$ multipoles of the
plane wave. We can simplify further
by writing out the $3j$ symbols explicitly and
using the recursion relations for Bessel functions to express
$j_{l\pm 2}$ in terms of $j_l$. The final result is remarkably compact:
\begin{equation}
\dot{h}_{ij}^{(\pm 2)}(k\hat{\vz})e^ie^j \E^{-i k \chi \cos\theta}
= -\sqrt{\frac{\pi}{2}}\dot{h}^{(\pm 2)}(k\hat{\vz}) \sum_{l}
(-i)^l 
\sqrt{2l+1} \sqrt{\frac{(l+2)!}{(l-2)!}}
\frac{j_l(k\chi)}{(k\chi)^2} Y_{l\,\pm 2}(\ve) \, .
\label{eq:tensproj}
\end{equation}
The spatial-to-angular projection is now controlled by
$j_l(k\chi)/(k\chi)^2$. 
As for scalar perturbations, this is concentrated on multipoles
$l \sim k \chi$. The result for general $\vk$ can be obtained by a rotation
$\hat{D}(\phi_\vk,\theta_\vk,0)$, where $\theta_\vk$ and $\phi_\vk$ are
the polar and azimuthal angles of $\hat{\vk}$ and the third Euler angle
can be taken to be zero.
Pre-empting a little the discussion in Sec.~\ref{sec:pol}, we can relate
the rotation matrices to spin-weighted spherical harmonics,
$D^l_{m\, \pm 2}(\phi_\vk,\theta_\vk,0) = \sqrt{4\pi/(2l+1)} {}_{\mp 2}
Y_{lm}^*(\hat{\vk})$, so that
\begin{equation}
\dot{h}_{ij}^{(\pm 2)}(\vk)e^ie^j \E^{-i \chi \vk \cdot \ve}
= - \sqrt{2\pi^2} \dot{h}^{(\pm 2)}(\vk) \sum_{lm}
(-i)^l 
\sqrt{\frac{(l+2)!}{(l-2)!}}
\frac{j_l(k\chi)}{(k\chi)^2}{}_{\mp 2}Y_{lm}^*(\hat{\vk})
 Y_{l m}(\ve)  \, .
\end{equation}
This generalises the scalar result for which the multipoles were given
in Eq.~(\ref{eq:19}).

\begin{figure}[t!]
\centerline{\includegraphics[width=0.5\textwidth,angle=-90]{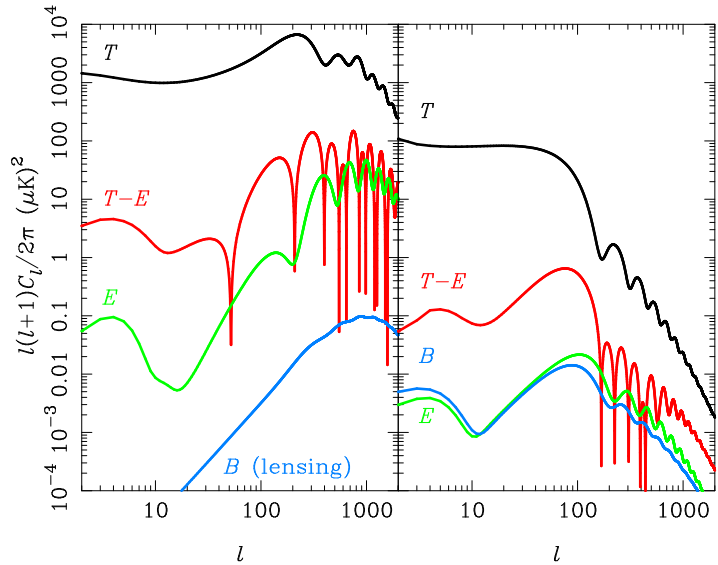}}
\caption{Temperature
(black), $E$-mode (green), $B$-mode (blue) and $T$-$E$ cross-correlation
(red) CMB power spectra from scalar perturbations (left) and
tensor perturbations (gravitational waves; right). The amplitude of
the tensor perturbations is shown at the maximum amplitude allowed
by current data ($r=0.22$~\cite{2009ApJS..180..330K}). The $B$-mode spectrum induced
by weak gravitational lensing is also shown in the left-hand panel (blue;
see Sec.~\ref{subsec:lensing}).}
\label{fig:scalarplustensor}
\end{figure}

Figure~\ref{fig:scalarplustensor} compares the power spectra due to
gravitational waves with those from scalar perturbations. The former
is shown at the maximum amplitude allowed by current data.
Note that the spectra from gravitational waves fall sharply on scales
that are sub-Hubble at recombination ($l > 60$) since the amplitude of
gravitational waves decays away as $1/a$ inside the Hubble radius. 
This limits the power of temperature anisotropies to constrain
gravitational waves since the sampling variance of the dominant
scalar perturbations is large at low $l$.
Fortunately, CMB polarization provides an alternative route to
detecting the effect of gravitational waves on the CMB which is not
limited by cosmic variance~\cite{1997PhRvD..55.1830Z,1997PhRvD..55.7368K};
see Sec.~\ref{sec:pol}.

\subsubsection{Isocurvature modes}
\label{subsubsec:iso}

Adiabatic fluctuations are a generic prediction of single-field inflation
models. However, multiple scalar fields typically arise in models
inspired by high-energy physics, such as the axion model~\cite{2002PhLB..543...14B},
curvaton~\cite{2003PhRvD..67b3503L} and multi-field inflation~\cite{1994PhRvD..50.6123P,2001PhRvD..63b3506G}.
In such models, if the fields decay asymmetrically
and the decay products are unable to reach chemical equilibrium with each other,
an isocurvature contribution to the primordial perturbation will result.
The simplest, and best-motivated, possibility is an isocurvature mode
where initially the dominant fractional over-density is
in the CDM, with a compensating (very small) fractional
fluctuation in the radiation and baryons~\cite{1986MNRAS.218..103E}.
The amplitude of the CDM isocurvature mode is quantified by the
gauge-invariant quantity $\mathcal{S} \equiv \delta_{\mathrm{c}} - 3 \delta_\gamma /4$,
where $\delta_c$ is the CDM fractional over-density. Generally,
$\mathcal{S}$ can be correlated with the curvature perturbation
$\mathcal{R}$, for example
in the curvaton and multi-field models.

In the CDM mode,
the photons are initially
unperturbed, as
is the geometry: $\delta_\gamma(0)=0=\phi(0)$ and $\vv_{\mathrm{b}} = 0$.
The different equations of state of the CDM and radiation lead to the
generation of a curvature perturbation. On large scales, $\mathcal{R}$
grows like $a$ in radiation domination and is constant in matter domination
with an asymptotic value $\approx \mathcal{S}/3$. The gravitational potential
follows suit and, for super-Hubble modes at last scattering,
$\phi_* = -3 \mathcal{R}_* / 5 = -\mathcal{S}/5$.
On large scales
$\delta_\gamma/4 = \phi$ is preserved, so that the source term for
the temperature anisotropies on large scales is $\Theta_0 + \psi \approx
-2 \mathcal{S}/5$. For a general mix of isocurvature and adiabatic
perturbations, the large-angle temperature anisotropies are related to
the primordial perturbations by $\Delta T / T \sim - \mathcal{R}/5
- 2 \mathcal{S}/5$. 

The evolution of the potential for isocurvature modes makes the driving term
in Eq.~(\ref{eq:25}) mimic the sine solution of the homogeneous
equation, and
so $\delta_\gamma$ oscillates as $\sin kr_\mathrm{s}$ inside the
sound horizon. This $\pi/2$ phase difference compared to adiabatic
initial conditions shifts the 
acoustic peaks to multipoles $l = (n+1/2)\pi \chi_*/r_{\mathrm{s}}(\eta_*)$.
The different peak positions for isocurvature
initial conditions allow the CMB to constrain their relative contribution
to the total fluctuations. The constraints on isocurvature modes
depend strongly on the assumptions made about the shape of the
primordial power spectrum for
$\mathcal{S}$ and any correlations with $\mathcal{R}$.
For uncorrelated,
axion-type models with scale-free
CDM isocurvature fluctuations, the ratio of the primordial spectra
for $\mathcal{S}$ to $\mathcal{R} < 0.2$ from CMB data alone~\cite{2009ApJS..180..330K}.
Recent constraints in more general models can be found in~\cite{2006PhRvD..74f3503B}.

\section{Introduction to CMB polarization}
\label{sec:pol}

As the universe recombines, the mean-free path to Thomson scattering
grows and temperature anisotropies start to develop. Subsequent scattering of
the CMB quadrupole generates partial linear polarization in the
CMB~\cite{1968ApJ...153L...1R}. The signal is small, with r.m.s.\
$\sim 5\,\mu\mathrm{K}$, but contains complementary information to the
temperature anisotropies and is therefore being actively pursued
by observers.

\subsection{Polarization observables}

Polarized specific brightness is most conveniently described in
terms of Stokes parameters $I(\vnhat;\nu)$, $Q(\vnhat;\nu)$, $U(\vnhat;\nu)$
and $V(\vnhat;\nu)$. The $I$ parameter is the total intensity, $Q$ measures
the difference in brightness between two orthogonal linear polarizations, and
$V$ measures the circular polarization. The $U$ parameter, like $Q$,
measures linear polarization and is defined as the difference
in brightness between two linear polarizations at $45^\circ$ to those
used to define $Q$. Thomson scattering of the CMB does not generate circular
polarization so we expect the primordial $V=0$. 
The Stokes parameters $Q$ and $U$
are functions of frequency as well as direction. In linear theory,
their frequency dependence follows that of the temperature anisotropies;
dividing by the derivative of the CMB blackbody (i.e.\ Planck) brightness
with respect to (log) temperature allows us to express the Stokes parameters as
frequency-independent fractional \emph{thermodynamic equivalent temperatures}.
We shall denote the equivalent temperatures by
$Q(\vnhat)$ and $U(\vnhat)$.

The linear Stokes parameters depend on the choice of basis.
For a line of sight $\vnhat$, we define the Stokes parameters on a local
$x$-$y$ basis defined by $-\hat{\vtheta},\hat{\vphi}$ of the spherical-polar
basis at $\vnhat$. Note that $-\hat{\vtheta},\hat{\vphi}$ and $-\vnhat = \ve$
-- the radiation propagation direction -- form a right-handed basis.
The reader is cautioned that this convention differs from some of the CMB
literature, most notably~\cite{1997PhRvL..78.2054S}. Under the
right-handed rotation about the propagation direction,
$\hat{\theta} \pm i \hat{\vphi} \rightarrow
\E^{\pm i \psi} (\hat{\theta} \pm i \hat{\vphi})$, the complex
polarization
\begin{equation}
Q \pm i U \rightarrow \E^{\mp 2 i \psi} (Q \pm i U) \; ,
\end{equation}
and so $Q \pm i U$ is said to be spin $\mp 2$. The complex polarization
can be expanded in terms of a real scalar field $\sum_{lm} E_{lm}
Y_{lm}(\vnhat)$ and a real pseudo-scalar field
$\sum_{lm} B_{lm} Y_{lm}(\vnhat)$, where the summations are over $l \geq 2$,
as~\cite{1997PhRvL..78.2054S}
\begin{equation}
(Q \pm i U)(\vnhat) = \sum_{lm} (E_{lm} \mp i B_{lm}) {}_{\mp 2}Y_{lm}(\vnhat)
\; . 
\end{equation}
Here, the ${}_{\mp 2}Y_{lm}(\vnhat)$ are spin-$\mp 2$ spherical
harmonics~\cite{1967JMP..8..2155G}. Spin-$s$ spherical harmonics
are defined by
\begin{equation}
{}_s Y_{lm} = \sqrt{\frac{(l-|s|)!}{(l+|s|)!}} \eth^s Y_{lm} \; ,
\end{equation}
where $\eth^{-|s|} \equiv (-1)^s \beth^{|s|}$. The spin-raising
and lowering operators, $\eth$ and $\beth$ have action
\begin{eqnarray}
\eth {}_s \eta &=& -\sin^s \theta (\partial_\theta + i \mathrm{cosec}
\partial_\phi)(\sin^{-s}\theta \, {}_s \eta) \label{pol:eq8b} \\
\beth {}_s \eta &=& -\sin^{-s} \theta (\partial_\theta - i \mathrm{cosec}
\partial_\phi)(\sin^{s}\theta \, {}_s \eta) \label{pol:eq8c}
\end{eqnarray}
when acting on a spin-$s$ quantity ${}_s \eta$.
They raise and lower the spin by one
unit respectively.

The decomposition of the polarization into $E$ and $B$ modes is
analogous to writing a vector on the 2-sphere, $V_a$,
as the sum of a gradient part, $\nabla_a V_E$, and a divergence-free
part, which can always be written as $\epsilon_{a}{}^b \nabla_b V_B$
where $\epsilon_{ab}$ is the alternating tensor. The components of
$V_a$ on the null basis $\vm_\pm \equiv (\hat{\vtheta} \pm i \hat{\vphi})
/\sqrt{2}$ are then
\begin{eqnarray}
V_+ &= \vV \cdot \vm_+ = \frac{1}{\sqrt{2}} (\partial_\theta + i \mathrm{cosec}\theta)V_E - \frac{i}{\sqrt{2}} (\partial_\theta + i \mathrm{cosec}\theta)V_B
= -\frac{1}{\sqrt{2}} \eth (V_E - i V_B) \; , \nonumber \\
V_- &= \vV \cdot \vm_- = \frac{1}{\sqrt{2}} (\partial_\theta - i \mathrm{cosec}\theta)V_E + \frac{i}{\sqrt{2}} (\partial_\theta - i \mathrm{cosec}\theta)V_B
= -\frac{1}{\sqrt{2}} \beth (V_E + i V_B) \; .
\end{eqnarray}
We see that the spin-1 $V_+$ is written as the action of
the spin-raising operator on the potential $-V_E + i V_B$ whose real and
imaginary parts are the (real) $E$ and $B$ potentials for $\vV$.
Things work similarly for polarization since $Q\pm i U$ are the
null components of a symmetric, trace-free tensor
$\mathcal{P}_{ab}$~\cite{2002PhRvD..65b3505L}.

\begin{figure}[t!]
\includegraphics[width=2.5cm,angle=-90]{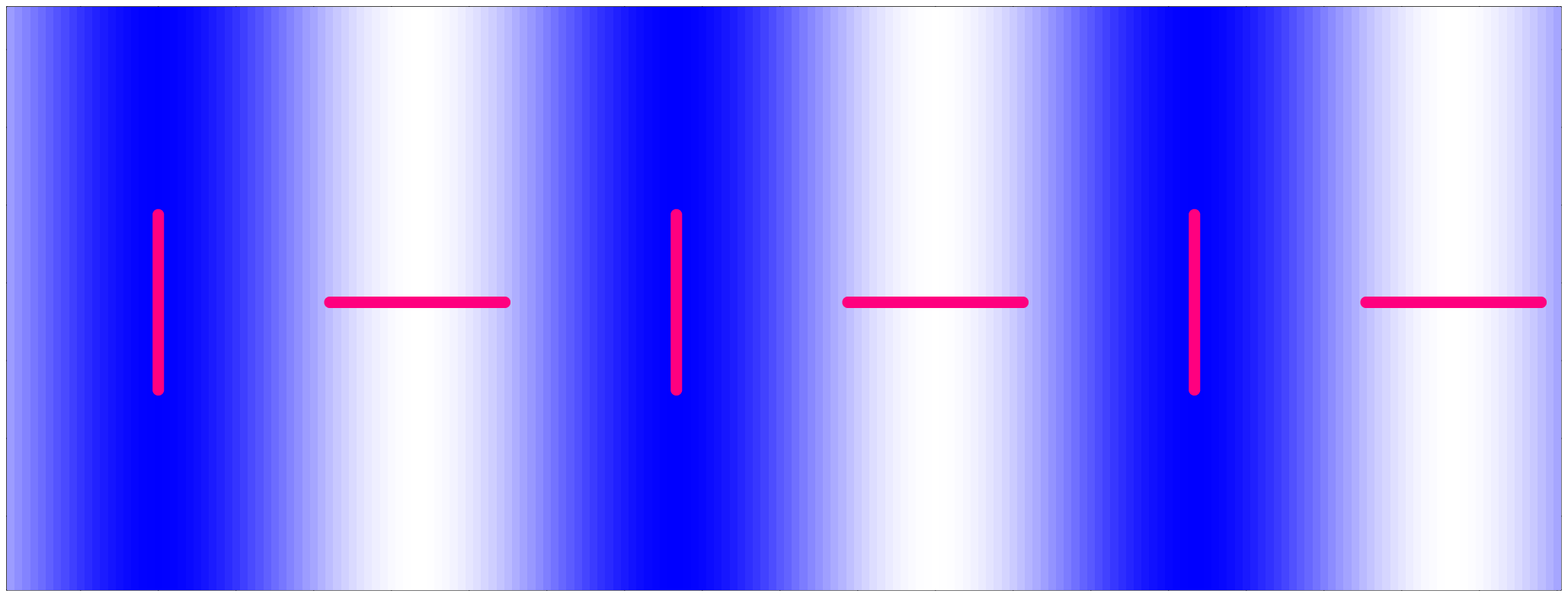}
\quad
\includegraphics[width=2.5cm,angle=-90]{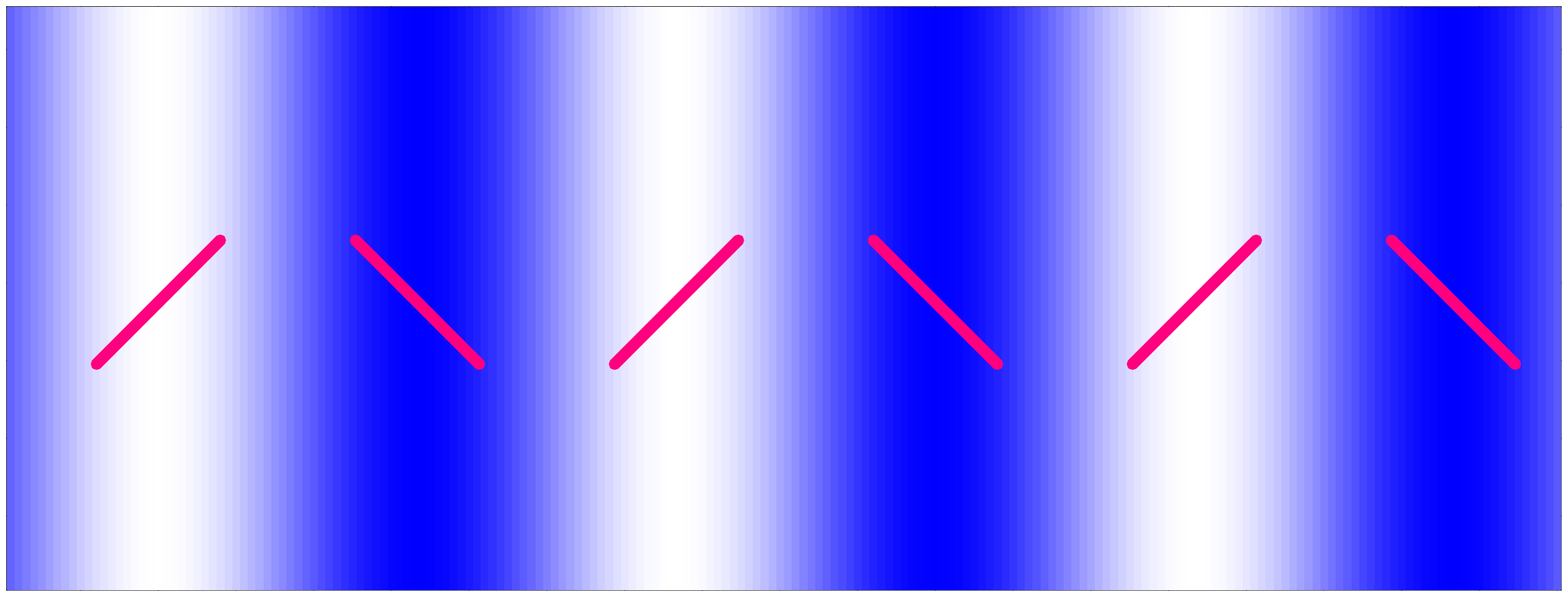}
\caption{Polarization patterns for an $E$ mode (left) and
$B$ mode (right) on a small patch of the sky for potentials that
are locally Fourier modes. The shading denotes the amplitude of the potential
and the headless vectors denote the unsigned direction of the polarization.
For the electric pattern the polarization is aligned with or perpendicular to
the Fourier wavevector depending on the sign of the potential; for the
magnetic pattern the polarization is at 45 degrees.}
\label{fig:pol_patterns}
\end{figure}

To gain further
intuition for $E$ modes and $B$ modes, consider the case where
the potentials $\sum_{lm} E_{lm} Y_{lm}(\vnhat)$ and
$\sum_{lm} B_{lm} Y_{lm}(\vnhat)$
behave locally like a plane wave across a small patch of the
sky (that can accurately be treated as flat). The $E$ and $B$
contributions to the polarization are depicted in Fig.~\ref{fig:pol_patterns}.
Quite generally, a given potential $\psi$ generates a $B$ mode
with $Q + i U = -i \beth^2 \psi$ and $Q-iU = i \eth^2 \psi$, and an
$E$ mode with $Q + i U = \beth^2 \psi$ and $Q-iU = \eth^2 \psi$.
Since $i= \E^{2i\pi/4}$, the $B$-mode polarization pattern is obtained from
the $E$ mode by rotating by $45^\circ$ about the line of sight.

\paragraph{Statistics of CMB polarization}

The decomposition of the polarization field into $E$ and $B$ parts
is invariant under rotations, and the $E$ and $B$
multipoles transform like those for the temperature
anisotropies, e.g.\ $E_{lm} \rightarrow \sum_{m'} D^l_{mm'} E_{lm'}$.
Under the operation of parity,
$(Q\pm i U)(\vnhat) \rightarrow (Q\mp i U)(-\vnhat)$ so that
$E_{lm} \rightarrow (-1)^l E_{lm}$ (electric parity) while
$B_{lm} \rightarrow (-1)^{l+1} B_{lm}$ (magnetic parity). 

Rotational and parity invariance in the mean limit the non-zero two-point
correlations between polarization (and temperature) multipoles
to
\begin{eqnarray}
\langle E_{lm} E^*_{l'm'} \rangle &=& \delta_{ll'} \delta _{mm'} C_l^E \\
\langle B_{lm} B^*_{l'm'} \rangle &=& \delta_{ll'} \delta _{mm'} C_l^B \\
\langle T_{lm} E^*_{l'm'} \rangle &=& \delta_{ll'} \delta _{mm'} C_l^{TE} \; ,
\end{eqnarray}
which define the power spectra $C_l^E$, $C_l^B$ and $C_l^{TE}$.
Note that there is no correlation between $B$ and either $\Delta T$ or
$E$.

Assuming Gaussian statistics, the cosmic variance in the polarization
(auto-)power spectra are
\begin{equation}
\mathrm{var}(\hat{C}_l^E ) = \frac{2}{2l+1} (C_l^E)^2 \; , \quad
\mathrm{var}(\hat{C}_l^B ) = \frac{2}{2l+1} (C_l^B)^2 \; .
\end{equation}
For the cross-spectrum, the cosmic variance is only slightly more
complicated~\cite{1997PhRvD..55.1830Z,1997PhRvD..55.7368K}:
\begin{equation}
\mathrm{var}(\hat{C}_l^{TE} ) = \frac{1}{2l+1} \left[(C_l^{TE})^2
+ C_l^T C_l^E \right] \; .
\end{equation}
The second term arises from chance correlations between $\Delta T$ and $E$
in the single realisation of the sky
that we have available. It is the only term present in the cosmic
variance of estimators for the cross-spectra that vanish in the mean by
parity, $C_l^{TB}$ and $C_l^{EB}$.
Due to the presence of correlations between the temperature and the
electric polarization, estimates of $C_l^T$ and $C_l^E$ have non-vanishing
covariance, and each is also correlated with the estimator for $C_l^{TE}$;
see~\cite{1997PhRvD..55.1830Z,1997PhRvD..55.7368K} for details.

The form of the higher-order polarization spectra implied by rotational
invariance in non-Gaussian theories is similar to that for
temperature and is described in~\cite{2002PhRvD..66f3008O}.

\subsection{Thomson scattering and free-streaming}

Polarization is generated by scattering the quadrupole of the temperature
anisotropy around
recombination and reionization. Prior to recombination, on scales
larger than the mean-free path, Thomson scattering keeps the
CMB radiation very nearly isotropic in the rest-frame of the baryons,
and therefore unpolarized. It is only as the mean-free path grows
as recombination is approached that the radiation can free-stream over
an appreciable fraction of a wavelength of the perturbations between
scatterings and so generate a temperature quadrupole. Subsequent scattering
of this quadrupole generates linear polarization. After the CMB last scatters
around recombination, the polarization is preserved until further scattering
at reionization (see Sec.~\ref{subsec:pol_reion}).

Thomson scattering changes the linear polarization state of the radiation
as~\cite{1997PhRvD..56..596H}\footnote{%
When discussing the production and propagation of polarized radiation,
it is convenient to associate the radiation with the point $\ve$ on the sphere.
At this point, the propagation direction is outwards so a right-handed
polarization basis is formed by $\hat{\vtheta}$ and $\hat{\vphi}$ there;
this is equivalent to using $\hat{\vtheta},-\hat{\vphi}$ at $\vnhat = -\ve$
and so the Stokes parameters are the same in both descriptions.
However, $(Q+iU)(\ve)$ is then spin $+2$ and its appropriate expansion
in spin harmonics is
\begin{equation}
(Q\pm i U)(\ve) = \sum_{lm} (E_{lm} \pm i B_{lm}){}_{\pm 2}Y_{lm}(\ve)
\nonumber \; .
\label{eq:changepol}
\end{equation}
The multipoles $E_{lm}$ and $B_{lm}$ here are related to those in the 
line-of-sight description by factors of $(-1)^l$ and $(-1)^{l+1}$ respectively
and so the parity-invariant power spectra are unchanged.
}
\begin{equation}
\D (Q \pm i U) = \dot{\tau}\D\eta (Q\pm iU)(\ve)
- \frac{3}{5} \dot{\tau}\D\eta
\sum_{|m| \leq 2}\left(E_{2m} - \frac{1}{\sqrt{6}}
\Theta_{2m}\right){}_{\pm 2}Y_{2m}(\ve)
\label{eq:genpol}
\end{equation}
Here, $\Theta_{lm}$ are the multipoles of the fractional temperature
fluctuation, $\Theta(\eta,\vx,\ve) = \sum_{lm} \Theta_{lm}(\eta,\vx)
Y_{lm}(\ve)$, and, recall, $\dot{\tau} = -a n_{\mathrm{e}}\sigma_{\mathrm{T}}$.
The first term arises from scattering out of the beam and reduces the
polarization. The second term, arising from in-scattering,
involves the quadrupoles of the temperature
anisotropies and the $E$-mode polarization. In particular, scattering
of unpolarized radiation generates linear polarization
$\D (Q + i U) =  \dot{\tau} \D \eta \eth^2 \left(\sum_m
\Theta_{2m} Y_{2m}\right)/20$ which is an $E$-mode quadrupole.
However, the
polarization does not generally remain in this state in the presence of
spatial inhomogeneities in the polarization induced over the last scattering
surface. To see how this works, it is simplest to consider the important
cases of density perturbations and gravitational waves separately.

\subsubsection{Density Perturbations}

Density perturbations transform as scalars under reparameterisation
of spatial coordinates. In Fourier space, this means that every perturbed
tensor is azimuthally-symmetric about the wavevector $\vk$. The same
is true of the temperature fluctuation and so for the quadrupole
we must have
\begin{equation}
\sum_{|m| \leq 2} \Theta_{2m}(\eta,\vk) Y_{2m}(\ve) \propto
P_2(\vkhat \cdot \ve) = \frac{4\pi}{5} \sum_{|m| \leq 2}
Y_{2m}(\ve) Y_{2m}^*(\vkhat) \; ,
\end{equation}
where, in the second equality, we have used the addition theorem for
spherical harmonics. As a concrete example, consider scales which
are large compared to
the mean-free path $l_p$. Applying Eq.~(\ref{eq:17}) over one scattering time,
a quadrupole temperature anisotropy builds up most efficiently by
the Doppler term. 
If we locate ourselves at the origin, the temperature anisotropy is
\begin{equation}
\Theta(\eta,\vzero,\ve) \sim \ve \cdot \vvb(\eta-l_p,-l_p \ve)
\approx \ve \cdot \vvb(\eta-l_p,\vzero) - l_p \ve \cdot (\ve \cdot \vnabla
\vvb)(\eta-l_p,\vzero) \; .
\end{equation}
The quadrupole part of this comes from the last term which, for a single
Fourier mode, gives
\begin{equation}
\sum_{|m| \leq 2} \Theta_{2m}(\eta,\vk) Y_{2m}(\ve) \sim
\frac{2}{3} kl_p P_2(\vkhat \cdot \ve) \velb(\vk) \; .
\end{equation}
Note that this is $O(kl_p)$ whereas the monopole source terms in
Eq.~(\ref{eq:17}) can only produce a quadrupole of $O(k^2 l_p^2)$.
If we choose coordinates with the $z$-axis along $\vk$, the temperature
anisotropies are all in the $m=0$ modes for density perturbations.

Scattering of the quadrupole generates
$Q \pm i U \sim \sum_{|m| \leq 2} Y_{2m}^*(\vkhat) {}_{\pm 2} Y_{2m}(\ve)$
locally.
If we take $\vk$ along $z$, we have $Q \propto \sin^2 \theta$ and $U=0$.
How does the character
of the observed polarization change by free-streaming (ignoring
reionization for the moment)? What we observe from this single Fourier mode
still has $U=0$, but the angular dependence of $Q$ is further modulated by
the spatial dependence of the perturbation over the last-scattering surface.
This modulation
transfers polarization from $l=2$ to higher multipoles (with most
appreciable power appearing at $l = k \chi_*$), but preserves the
electric character of the polarization~\cite{1997PhRvD..55.1830Z,1997PhRvD..55.7368K}.
We can see that this is
reasonable by noting that the modulated polarization field has its
polarization direction either parallel or perpendicular to the direction in
which the polarization amplitude is changing~\cite{1997NewA....2..323H}.
In more detail, if we take $\hat{\vk}$ along the $z$-axis,
the observed polarization at the origin is
\begin{equation}
(Q\pm i U)(\ve) = (Q\pm i U)(\eta_*,-\chi_* \ve,\ve)
\propto (1-\mu^2)  \E^{-i k \chi_* \mu} \, ,
\label{eq:newpol1}
\end{equation}
where $\mu \equiv \hat{\vk} \cdot \ve = \cos\theta$. (Note that
$U$ is zero in this orientation.) Since the polarization is azimuthally
symmetric, the same will be true of any $E$ and $B$ modes.
Taking
\begin{equation}
Q+ iU = \eth^2 (\psi_E + i \psi_B) \, , \quad
Q- iU = \beth^2 (\psi_E - i \psi_B) \, ,
\end{equation}
where $\psi_E = \sum_{l\geq 2} \sqrt{(l-2)!/(l+2)!} E_{l0} Y_{l0}(\theta,\phi)$
and similarly for $\psi_B$, to satisfy Eq.~(\ref{eq:newpol1}) we must have
\begin{equation}
(1-\mu^2)(\psi_E'' \pm i \psi_B'') \propto (1-\mu^2) \E^{-i k\chi_* \mu} \, ,
\end{equation}
where primes denote differentiation with respect to $\mu$. It follows that
$\psi_B=0$ (the non-zero solutions of $\psi_B''=0$ are combinations of
$l=0$ and $l=1$ modes and cannot generate polarization) so there are no
$B$ modes. Since $E$ and $B$ modes transform irreducibly under
rotations, there will be no $B$ modes for a general $\vk$.
Dropping the proportionality constant, the
relevant solution of $\psi_E'' = \E^{-i k \chi_* \mu}$ is
$a_0(k\chi_*) + a_1 (k\chi_*) \mu -  \E^{-i k \chi_* \mu}/(k\chi_*)^2$
where $a_0$ and
$a_1$ should be chosen to remove the $l=0$ and $l=1$ modes. Expanding
the exponential gives
\begin{equation}
E_{l0} = - (-i)^l \sqrt{4\pi} \sqrt{2l+1} \sqrt{\frac{(l+2)!}{(l-2)!}}
\frac{j_l(k\chi_*)}{(k\chi_*)^2} \, .
\end{equation}
The multipoles for a general direction of $\hat{\vk}$ can be found
by the appropriate rotation. Note that the spatial-to-angular projection
of scalar polarization is controlled by the same function,
$j_l(k\chi_*)/(k\chi_*)^2$, as for the 
temperature anisotropies from gravitational waves (see Eq.~(\ref{eq:tensproj})).

The polarization power spectra produced by adiabatic density
perturbations are plotted in the left panel of
Fig.~\ref{fig:scalarplustensor}. They were computed
with the Boltzmann code CAMB~\cite{2000ApJ...538..473L}.
The $E$-mode power peaks around
$l \sim 1000$, corresponding to the angle subtended by the width of the
visibility function at recombination.
On larger scales
the polarization probes the electron-baryon velocity at last
scattering, as described above. Acoustic oscillations of the plasma
prior to last scattering imprint an oscillatory structure on the angular
power spectra of the CMB observables. Asymptotically, the photon-baryon
overdensity oscillates as a cosine, and the peculiar velocity of the
plasma as a sine, to satisfy the continuity equation. The peaks in the
$E$-mode spectrum therefore coincide with the troughs of the temperature
spectrum. Note also that the temperature and $E$-mode polarization are
correlated~\cite{1994PhRvL..73.2390C} and that modes caught at either the midpoint or
extrema of their oscillation at last scattering 
give zeroes in the temperature-polarization
cross-correlation.
Large-angle polarization from the last-scattering
surface is very small since the generation of a quadrupole in the temperature
anisotropy occurs via causal physics (photon diffusion) that is suppressed
outside the horizon.
The increase in polarization on large scales in Fig.~\ref{fig:scalarplustensor}
is due to reionization~\cite{1997PhRvD..55.1822Z} and is discussed
further in Sec.~\ref{subsec:pol_reion}.

\subsubsection{Gravitational Waves}

For gravitational waves, a temperature quadrupole builds up over a
scattering time due to the shear of the gravitational wave:
\begin{equation}
\Theta(\ve) \sim -\frac{1}{2} l_p \dot{h}_{ij} e^i e^j \, .
\end{equation}
Consider a circularly-polarized gravitational wave with $\vk$ along
the $z$-axis so that the temperature quadrupole is a $|m|=2$ mode.
Equations~(\ref{eq:shearsource}) and~(\ref{eq:genpol}) show
that Thomson scattering of this temperature quadrupole produce
polarization that locally has
\begin{equation}
Q \pm i U \sim  l_p \dot{h}^{(p)}(k{\hat{\vz}})\frac{1}{20} 
\sqrt{\frac{8\pi}{5}}{}_{\pm 2} Y_{2\, p}(\ve) \, ,
\label{eq:polgrav1}
\end{equation}
where $h^{(p)}(\vk)$ is the Fourier amplitude of the gravitational
wave with $p=\pm 2$ labelling the helicity.\footnote{A more careful
treatment of Eq.~(\ref{adc:eq43}),
balancing in- and out-scattering in the presence of the shear source,
and properly including the polarization dependence of Thomson scattering,
introduces
a prefactor of $10/3$ in these expressions; see~\cite{1997PhRvD..56..596H,2000CQGra..17..871C,2000PhRvD..62d3004C}.}
Unlike scalar perturbations, the local polarization has both $Q$ and
$U$ non-zero and modulating these with the plane wave $\E^{-i\chi_* \vk\cdot
\ve}$ will produce both $E$ and $B$-mode polarization~\cite{1997PhRvD..55.1830Z,1997PhRvD..55.7368K}.
In detail, the observed polarization is
\begin{eqnarray}
(Q \pm i U)(\ve) & \propto & {}_{\pm 2} Y_{2p}(\ve) \E^{-ik\chi_* \cos\theta}
\nonumber \\
&=& {}_{\pm 2} Y_{2p}(\ve) \sum_L \sqrt{4\pi} \sqrt{2L+1} (-i)^L j_L(k\chi_*)
Y_{L0}(\ve) \nonumber \\
&=& \sqrt{5} \sum_L \Biggl[ \sqrt{2L+1} (-i)^L j_L(k\chi_*) \nonumber \\
&&\mbox{} \times \sum_l \sqrt{2l+1} \threej{2}{L}{l}{-p}{0}{p}\threej{2}{L}{l}{\pm 2}{0}{\mp 2}
{}_{\pm 2}Y_{lp}(\ve) \Biggr] \nonumber \\
&=& -\sqrt{5} \sum_l (-i)^l \sqrt{2l+1} {}_{\pm 2} Y_{lp}(\ve)
\left[\epsilon_l(k\chi_*) \pm \frac{p}{2}i \beta_l(k\chi_*)\right] \, ,
\end{eqnarray}
where we have used an addition theorem for the spin spherical harmonics,
and the functions
\begin{eqnarray}
\epsilon_l(x) &\equiv & \frac{1}{4} \left[\frac{\D^2 j_l(x)}{\D x^2}
+ \frac{4}{x} \frac{\D j_l(x)}{\D x} + \left(\frac{2}{x^2}-1\right)
j_l(x) \right] \; ,\\
\beta_l(x) &\equiv & \frac{1}{2} \left[ \frac{\D j_l(x)}{\D x} +
\frac{2}{x} j_l(x) \right]
\end{eqnarray}
follow from substituting the explicit form of the $3j$ symbols and using
the recursion relations for the the spherical Bessel functions.
Reinstating numerical factors from Eq.~(\ref{eq:polgrav1}), and rotating to
a general $\vk$, the non-zero
$E$- and $B$-mode multipoles are
\begin{eqnarray}
E_{lm} &=& -\frac{\sqrt{2\pi^2}}{5} l_p (-i)^l \dot{h}^{(\pm 2)}(\vk)
\epsilon_l(k\chi_*){}_{\mp 2} Y_{lm}^*(\hat{\vk})  \,
\label{eq:gravpol2}\\
B_{lm} &=& \mp \frac{\sqrt{2\pi^2}}{5} l_p (-i)^l \dot{h}^{(\pm 2)}(\vk)
\beta_l(k\chi_*) {}_{\mp 2} Y_{lm}^*(\hat{\vk})  \, .
\label{eq:gravpol3}
\end{eqnarray}
We see that gravitational waves project onto both $E$- and $B$-mode
polarization. The projection from last scattering peaks at multipoles
$l \sim k \chi_*$ and divides power roughly equally between
$E$ and $B$~\cite{1997PhRvD..56..596H}. Note also that under a parity transformation,
$\dot{h}^{(\pm 2)}(\vk) \rightarrow \dot{h}^{(\mp 2)}(-\vk)$. Given
that ${}_s Y_{lm}(-\hat{\vk}) = (-1)^l {}_{-s}Y_{lm}(\hat{\vk})$,
Eqs~(\ref{eq:gravpol2}) and~(\ref{eq:gravpol3}) manifestly give
$E_{lm} \rightarrow (-1)^l E_{lm}$ and $B_{lm} \rightarrow
-(-1)^l B_{lm}$, as required, under parity.

The polarization power spectra from a scale-invariant background of
gravitational waves with tensor-to-scalar ratio $r=0.22$
is shown in the right-hand panel
of Fig.~\ref{fig:scalarplustensor}. As for temperature
anisotropies, the spectra peak on scales corresponding to the horizon size
at recombination. Note that the gravitational wave contribution to the
temperature anisotropies and electric polarization is sub-dominant to
the scalar perturbations, but gravitational waves dominate the $B$-mode
polarization on large angular scales. (On small scales, the non-linear
$B$-modes induced by the action of gravitational lensing on the largely
$E$-mode primary polarization are dominant.) 
The increase in polarization power on the largest scales in
Fig.~\ref{fig:scalarplustensor} is due to reionization, which we now
briefly discuss.

\subsection{Polarization from reionization}
\label{subsec:pol_reion}

For polarization, re-scattering suppresses the signal from last-scattering
by a factor $e^{-\tau_\mathrm{re}}$ but also produces a new large-angle
signal~\cite{1997PhRvD..55.1822Z} (see Fig.~\ref{fig:scalarplustensor}).
From Eq.~(\ref{eq:genpol}) we see that
the strength of this large-angle polarization varies almost linearly with
$\tau_\mathrm{re}$ and the size of the temperature quadrupole
at reionization. Since reionization appears to have completed before
dark energy came to dominate the expansion, the temperature quadrupole
at last scattering is sourced by the Sachs-Wolfe and Doppler effects
on the last scattering surface of the re-scattering electron.
Therefore, the quadrupole is dominated by modes with 
$k(\eta_{\mathrm{re}}-\eta_*) \sim 2$. After scattering, the newly-generated
$l=2$ $E$-mode polarization free-streams to give $E$-mode (and,
for gravitational waves, $B$-mode) polarization
that peaks at multipoles
$l \sim 2(\eta_0 - \eta_\mathrm{re})/(\eta_\mathrm{re}-\eta_*)$.
The position of the reionization feature is thus controlled by the
epoch of reionization. The detailed history of reionization does generate
some further small features in large-angle polarization but extracting this is rather limited by cosmic variance~\cite{2003ApJ...583...24K,2003PhRvD..68b3001H}.

Currently WMAP is the only polarized dataset with sufficient sky coverage to
extract $\tau_\mathrm{re}$. The latest
results give $\tau_\mathrm{re} = 0.087 \pm 0.017$~\cite{2009ApJS..180..306D}. For
instantaneous reionization, this limits the redshift $z_{\mathrm{re}} > 8.2$
at 95\% confidence, which, combined with quasar absorption spectra that
indicate reionization was complete by $z=6$
(e.g.~\cite{2006AJ....132..117F}), points to reionization being an
extended process. The determination of $\tau_\mathrm{re}$ required
aggressive cleaning of polarized Galactic foreground emission by the
WMAP team, but their result is stable to variations in the details
of the cleaning process~\cite{2009ApJS..180..306D}.

\subsection{Current measurements of polarization power spectra}

\begin{figure}[t!]
\centerline{\includegraphics[width=0.6\textwidth,angle=-90]{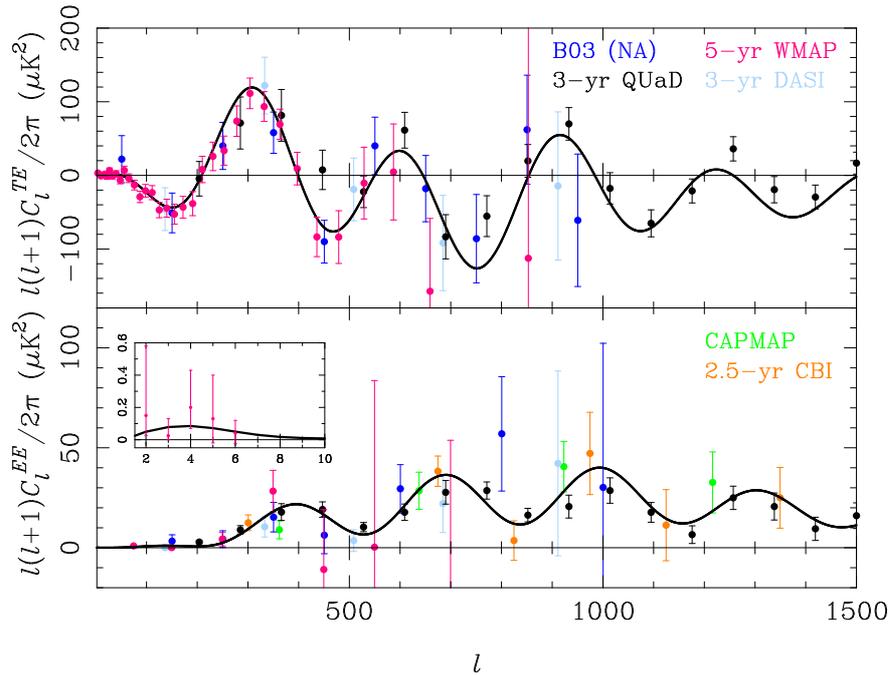}}
\caption{Current measurements of $C_l^{TE}$ (top) and $C_l^{E}$ (bottom)
from WMAP5 (magenta;~\cite{2009ApJS..180..296N}), QUaD  (black;~\cite{2009ApJ...692.1247P}),
BOOMERanG (blue;~\cite{2006ApJ...647..833P,2006ApJ...647..813M}), DASI (cyan;~\cite{2005ApJ...624...10L}), CAPMAP
(green;~\cite{2008ApJ...684...771B}) and CBI (orange;~\cite{2007ApJ...660..976S}).
The line is the best-fit
flat, $\Lambda$CDM model to the WMAP5 data.
}
\label{fig:Epoldata}
\end{figure}
%

In Fig.~\ref{fig:Epoldata},
we show all current measurements of the $E$-mode power spectrum and
the $T$-$E$ correlation. WMAP~\cite{2009ApJS..180..296N} provides the best measurements on
large scales while the recent QUaD data~\cite{2009ApJ...692.1247P} is most constraining
on intermediate and small scales. The data is still rather noisy but the
qualitative agreement of the polarization spectra to the best-fit
$\Lambda$CDM model (which is still driven by the higher signal-to-noise
temperature data) is striking. 
This is an important test of the structure
formation model: the polarization mainly reflects the
plasma bulk velocities around recombination and these are consistent,
via the continuity equation,
with the density fluctuations that mostly seed the temperature anisotropies.
In a more detailed analysis of consistency, carried out in~\cite{2009arXiv0901.0810Q}, a
mild tension is reported between the QUaD data (particularly for $C_l^{TE}$)
and the best-fitting model to the WMAP data. The amplitude of the
measured acoustic oscillations in $C_l^{TE}$ is rather higher for $l > 500$
than the best-fit model and this appears to be responsible for a high $\chi^2$
(probability to exceed of 7\%) between the QUaD polarization data and the
best-fit model. There will be significant improvements in measurements
of the $E$-mode power spectra in the next three years, most notably from
Planck. In addition, several of the new breed of polarimeters targeting
$B$-mode polarization have the angular resolution and sensitivity
to measure $E$ modes at high signal-to-noise to $l \sim 2000$, but their
spectra will likely not be competitive with Planck due to their limited
sky coverage. It will be interesting to see if the tension in the
QUaD polarization data persists in these future datasets.

\section{Cosmological parameters from the CMB}

In this section we briefly discuss the dependencies of the
CMB power spectra on the main cosmological parameters and the
current measurements of the parameters via this route. Since the CMB
observables are mostly a projection of conditions on the last-scattering
surface, parameters can have an influence either through the
acoustic physics of the pre-recombination plasma or through the
angular diameter distance to last scattering, $d_A$. The latter controls the projection
of linear distances at recombination to observed angular scales on the sky.
We defer discussion of the effect of the primordial power spectra to
Sec.~\ref{subsec:inflationandstructure}.

The parameters that influence the perturbations up to the
time of recombination are only the baryon and CDM densities -- given that
we know the CMB temperature rather precisely -- in models
with massless neutrinos. These densities set two length scales, the sound horizon
$r_{\mathrm{s}}$ and the damping scale $k_D^{-1}$, and influence
the relative heights of the acoustic peaks in the CMB power spectra (see
below). The modulation of the peak heights largely determines
the densities so that $r_{\mathrm{s}}$ and $k_D^{-1}$ become standard rulers
at the time of last scattering. The angular scale of the anisotropies
depends on the angles that these standard rulers subtend today, and hence
on $d_A$. The main
influence of the geometry of space, dark energy and light (but not
massless) neutrinos on the observed anisotropies is through their
effect on $d_A$. They also have a significant effect on the growth of structure at
late times, but this only influences the linear CMB anisotropies through the late-time ISW effect on large scales where cosmic variance is large.

To zero order, the primary CMB anisotropies and polarization determine
the physical baryon and CDM densities, $\Omega_{\mathrm{b}} h^2$ and $\Omega_{\mathrm{c}} h^2$,
and $d_A$. Large-angle polarization further determines the optical
depth to reionization.
Models with the same values for these parameters, and the same
primordial power spectra, produce essentially the same CMB power
spectra except on the largest scales~\cite{1999MNRAS.304...75E}.
This leads to a very precise \emph{geometric degeneracy} with CMB data alone
that can only be broken with external data, e.g.\ the Hubble constant,
luminosity distances from supernovae, and $d_A(z)$ from
the baryon acoustic oscillation (BAO) feature in galaxy clustering, 
or high-resolution observations of secondary effects in the CMB.
A consequence of the geometric degeneracy is that the
WMAP data \emph{alone} is consistent with a model with a closed geometry
and no dark energy, albeit with a very low Hubble constant
(e.g.~\cite{2003ApJS..148..175S}).
There are other \emph{approximate} degeneracies in current CMB data,
such as between the shape of the primordial
perturbation spectrum (i.e.\ its spectral index, $n_s$; see
Sec.~\ref{subsec:inflationandstructure})
and the matter density or the tensor-to-scalar ratio, $r$.

\subsection{Matter composition}

\begin{ltxfigure}[t!]
\begin{center}
\includegraphics[width=0.6\textwidth,angle=-90]{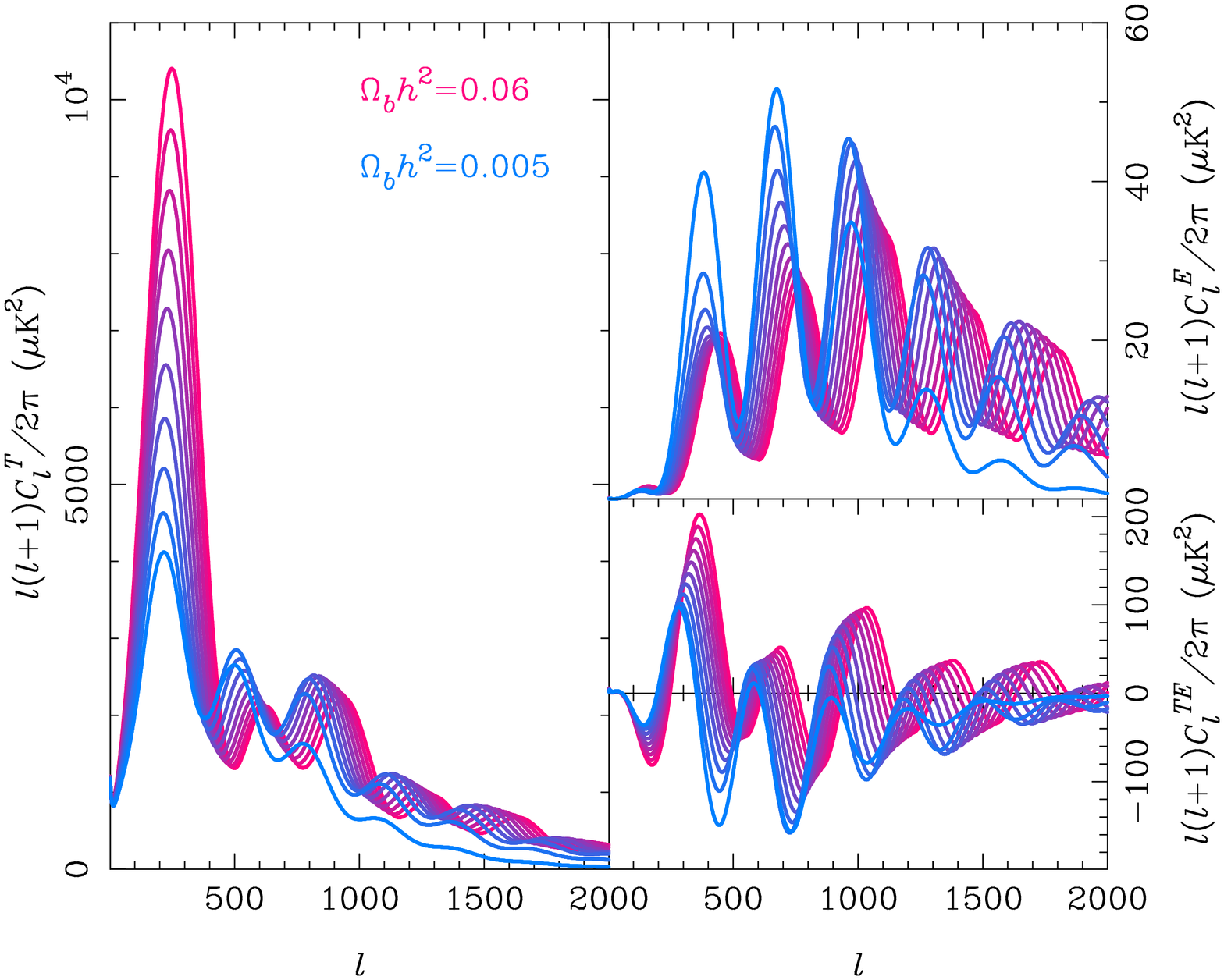}
\\
\includegraphics[width=0.6\textwidth,angle=-90]{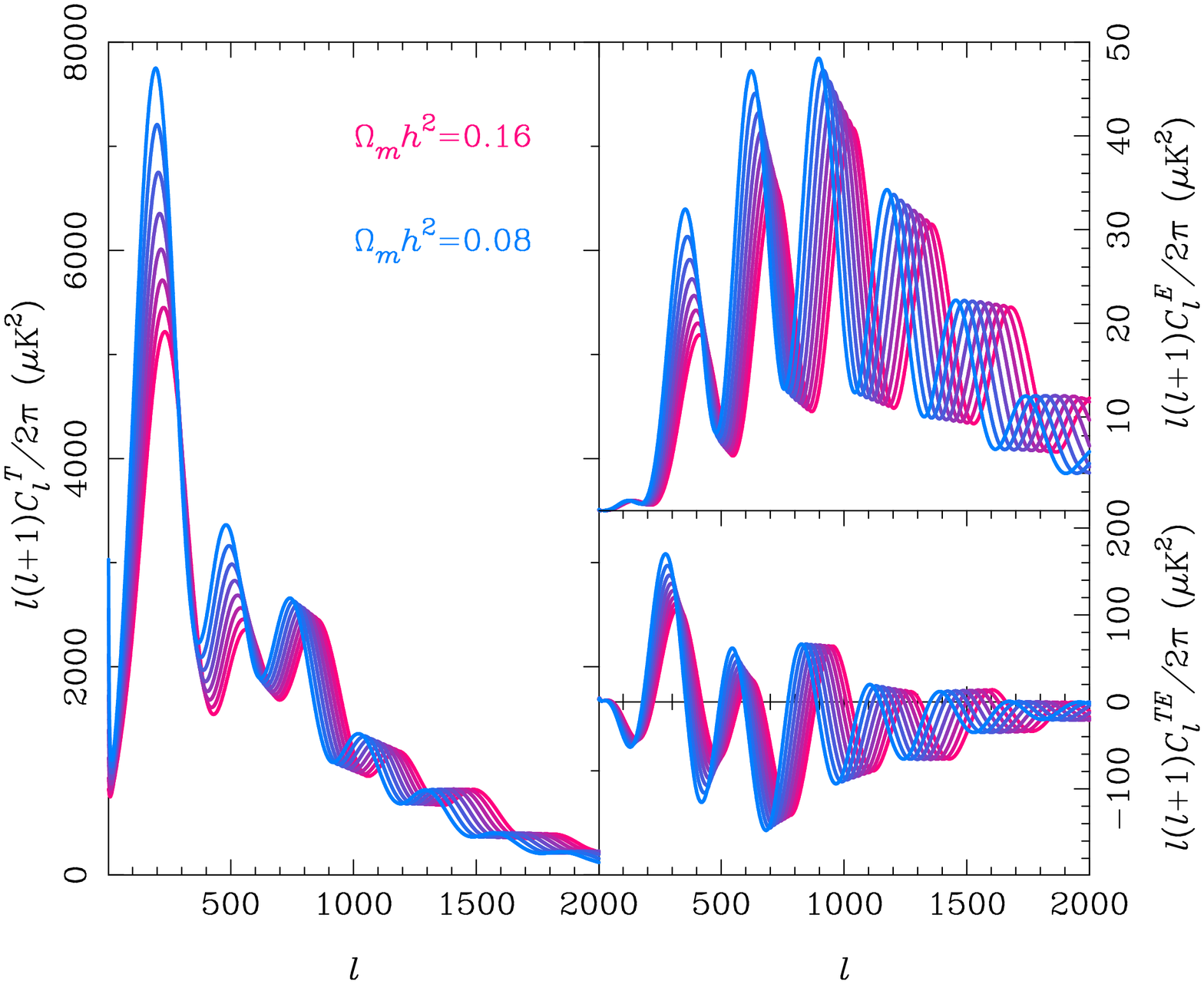}
\end{center}
\caption{Variation of $C_l^T$ (left), $C_l^E$ (top right) and $C_l^{TE}$
(bottom right) as $\Omega_{\mathrm{b}} h^2$ is varied with fixed $\Omega_{\mathrm{m}} h^2$ (top)
and as $\Omega_{\mathrm{m}} h^2$ is varied at fixed $\Omega_{\mathrm{b}} h^2$ (bottom). 
For the latter the angular diameter distance to last scattering has been
held fixed. All models are flat.
}
\label{fig:ombandomc_vars}
\end{ltxfigure}
%

The dependence of the temperature and $E$-mode polarization power spectra
on $\Omega_{\mathrm{b}} h^2$ and $\Omega_{\mathrm{c}} h^2$ is illustrated in
Fig.~\ref{fig:ombandomc_vars}. The main effects can be
understood as follows.
Consider increasing $\Omega_{\mathrm{b}} h^2$ at fixed
matter density $\Omega_{\mathrm{m}} h^2 \equiv \Omega_{\mathrm{b}} h^2 + \Omega_{\mathrm{c}} h^2$.
The increase in baryon inertia reduces the sound speed in the
pre-recombination plasma by reducing its bulk modulus. As explained
in Sec.~\ref{sec:acosutic}, this shifts the mid-point of the
acoustic oscillations to greater overdensities and enhances the
height of the compressional peaks ($n=1,3,\ldots$ for adiabatic
initial conditions) in $C_l^T$. The sound horizon is also reduced,
shifting the acoustic peaks in temperature and polarization
to smaller scales (larger $l$). The increase in the number density of
electrons in the plasma reduces the photon mean-free path, $l_p$, reducing the
amount of diffusion damping and so increasing power on small scales.
On larger scales, where tight-coupling still holds approximately,
the reduction in $l_p$ decreases the quadrupole
anisotropy around recombination (the reduction in the plasma bulk velocity
also contributes the same way) and the polarization is reduced.

Consider instead increasing $\Omega_{\mathrm{m}} h^2$ at fixed $\Omega_{\mathrm{b}} h^2$. At
fixed dark energy density and curvature radius, the expansion rate would
increase at all redshifts and $d_A$ would fall. We have factored this
effect out in Fig.~\ref{fig:ombandomc_vars} by reducing the dark energy density
to keep $d_A$ fixed. The main effects of increasing $\Omega_{\mathrm{m}} h^2$ are then to push
matter-radiation equality back further in time and reduce the sound horizon and 
diffusion scale (since the conformal age of the universe is reduced).
Since the residual radiation is then lessened at recombination, the
early-ISW effect and thus the first peak in $C_l^T$ are also reduced.
The resonant driving effect that enhances the amplitude of the
acoustic oscillations for modes
that enter the sound horizon during radiation domination is also limited to
higher peak order. Together, these effects lead to the relative enhancement
of the third and higher peaks over the first two that can be seen in
$C_l^T$ in
Fig.~\ref{fig:ombandomc_vars}. The reduction in the sound horizon shifts the
temperature and polarization peaks to larger $l$. Decreasing diffusion
reduces the polarization on large scales where tight-coupling holds,
as does the lessened impact of resonant driving.

These effects on the 
morphology of the acoustic peaks have allowed accurate measurements
of the matter and baryon densities from the CMB.
Five years of WMAP data alone give $\Omega_{\mathrm{b}} h^2 = 0.02273\pm 0.00062$
(i.e.\ 3\% uncertainty)
and $\Omega_{\mathrm{c}} h^2 = 0.1099\pm 0.0062$ (and so
$\Omega_{\mathrm{m}} h^2 = 0.1326\pm 0.0063$; 6\% uncertainty) in
flat $\Lambda$CDM models~\cite{2009ApJS..180..306D}.
These numbers should improve to sub-percent levels with
better measurements of the third and higher peaks
with the Planck
data in the future~\cite{2006astro.ph..4069T}.

Neutrino oscillations imply that at least two flavours are massive
(see~\cite{2006PhR...429..307L} for a review in the cosmological context). The minimum
mass of the heaviest eigenstate is $\approx 0.05\, \mathrm{eV}$. At this
minimum level, neutrinos are still relativistic at recombination and
their effect on the dynamics of the primordial plasma are indistinguishable
from massless neutrinos. A neutrino is non-relativistic at recombination
if its mass exceeds $\approx 0.6 \, \mathrm{eV}$. At this mass, the
eigenstates must be very nearly degenerate and the total (summed) neutrino
mass is therefore $\sum_{\nu} m_\nu \approx 1.8\,\mathrm{eV}$.
This sets a rough limit
on the determination of masses from the primary CMB anisotropies and is close
to the limit $\sum_{\nu} m_\nu < 1.3\,\mathrm{eV}$ from the five-year WMAP data
alone in flat, $\Lambda$ models~\cite{2009ApJS..180..330K}. Even with minimal masses,
two of the neutrino eigenstates are non-relativistic today and therefore
affect $d_A$ and suppress the late-time growth of structure on scales small
enough that the neutrinos cannot cluster. The former effect can
be used to improve constraints on $\sum_\nu m_\nu$ from the CMB
if external data is used to break the geometric degeneracy.
In WMAP5, the constraint on $\sum_\nu m_\nu$ improves to
$0.67\, \mathrm{eV}$ in flat, $\Lambda$ models by including supernovae and
BAO distance data. The effect of sub-eV neutrinos on structure
formation affects the lensing of the CMB by large-scale structure,
making CMB lensing a promising technique for determining masses (see
Sec.~\ref{subsec:lensing}).

\subsection{Geometry and dark energy}

\begin{figure}[t!]
\includegraphics[width=0.35\textwidth,angle=-90]{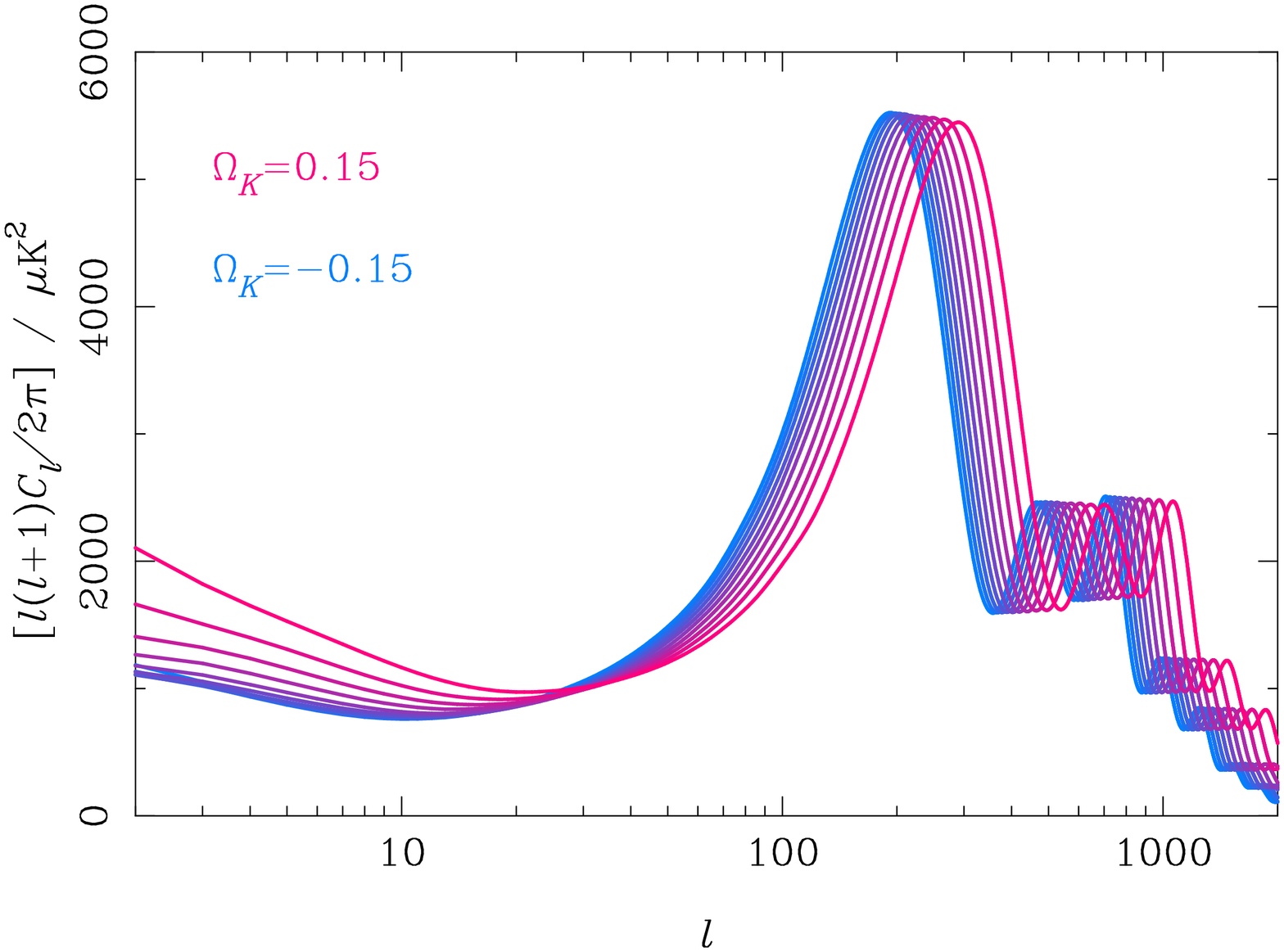}
\includegraphics[width=0.35\textwidth,angle=-90]{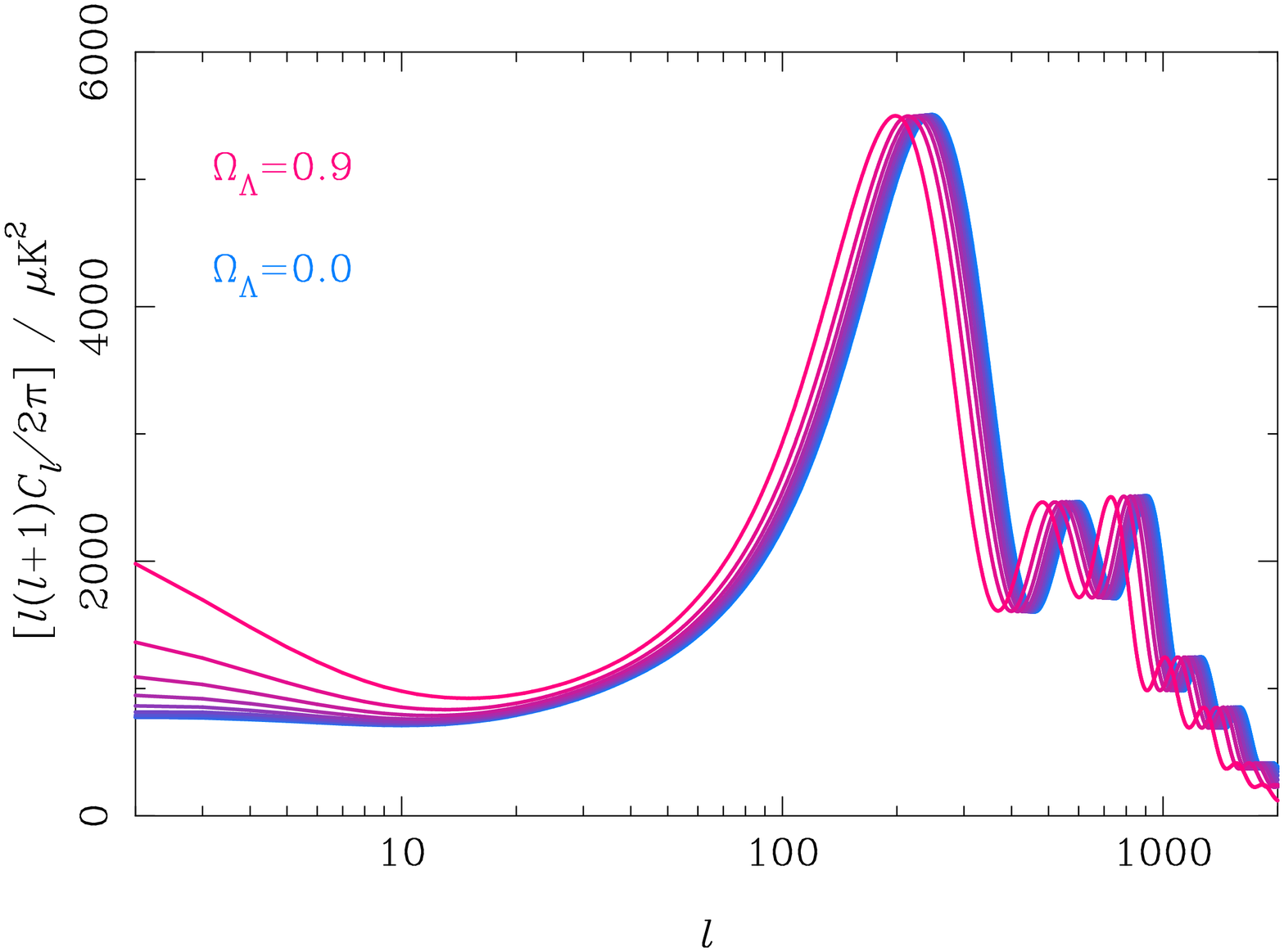}
\caption{Variation of $C_l^T$ with spatial curvature (left) and
dark energy density (right). In both cases, $\Omega_{\mathrm{b}} h^2$ and
$\Omega_{\mathrm{c}}h^2$ are fixed and the dark energy model is a cosmological
constant.
}
\label{fig:OmkandOmL}
\end{figure}
%

Generalising our discussion of flat models in Sec.~\ref{sec:acosutic}
to non-flat models, the positions of the acoustic peaks in
models with adiabatic initial conditions are asymptotically at
$l = n \pi d_A / r_{\mathrm{s}}(\eta_*)$ since curvature is
never dynamically important before recombination. In flat models, note that
$d_A$ reduces to $\chi_*$, the comoving distance to last scattering.
With the sound horizon determined from the peak morphology,
the CMB alone can be used to measure $d_A$: constraints from
WMAP5 give $d_A = 14.1 \pm 0.2 \, \mathrm{Gpc}$ in their most general
open models with dynamical dark energy~\cite{2009ApJS..180..330K}.

The dependence of $C_l^T$ on the curvature fraction, $\Omega_K$ (i.e.\
the ratio of the current Hubble radius to curvature radius and defined
to be positive/negative for open/closed models), and dark energy density
in $\Lambda$CDM models is illustrated in Fig.~\ref{fig:OmkandOmL}.
Both parameters principally affect the anisotropies through $d_A$ and
so simply shift the peaks. The remaining effect on large scales
is due to the late-time ISW effect and, in closed models, from 
mode quantisation. Disentangling the effects of dark energy and
curvature from the linear CMB anisotropies requires external data
to break the geometric degeneracy. Combining WMAP5 with BAO and
supernova data gives
$-0.0178 < \Omega_K < 0.0066$ in $\Lambda$ models, fully consistent with
flatness. (Only one
external dataset is required in this case; BAO is the most constraining.)
The geometric degeneracy also limits constraints from the CMB alone on
more complex models with dynamical dark energy, even for a flat universe;
a detailed discussion is given in~\cite{2009ApJS..180..330K}.
The current constraints combining WMAP with \emph{both} BAO and supernova data
are consistent with flatness and non-dynamical
dark energy: the equation of state parameter $w$ is
consistent with $-1$ at the
15\% level.

\section{Constraining early-universe physics with the CMB}

\subsection{Inflation and the origin of structure}
\label{subsec:inflationandstructure}

So far, we have not discussed the origin of the primordial perturbation
which provided the seeds for cosmological structure formation under the action of gravitational instability. The study of this question has the potential to expose deep connections between cosmology and physics at immensely high energies which are forever beyond the reach of earth-bound particle accelerators. 
Inflation \cite{Guth, Linde, Steinhardt,mfb92,star79,star85}, an epoch in which the expansion of the universe is accelerating, solves a number of puzzles associated with the standard big bang cosmology. During a phase of accelerated expansion, $H^{-1}$ (the physical Hubble radius) remains almost fixed but the
physical separation of particles initially in causal contact grows
exponentially. The result is that regions today separated by cosmological distances were actually in causal contact before/during inflation. At that time, these regions were given the necessary initial conditions, smoothness and {\em small perturbations about smoothness}, we observe today. 

The big bang puzzles (e.g.\ horizon problem, flatness problem, homogeneity problem and monopole problem) are not strict inconsistencies in the standard big bang theory, but must instead be {\it assumed} as extreme fine-tuning of the initial conditions, thus severely restricting the predictive power of the theory. Inflation provides a {\it dynamical} mechanism to solve the big bang problems. At the heart of the inflationary paradigm is an elegantly simple idea: {\it invert the behaviour of the comoving Hubble radius} i.e.\ make it {\it decrease} sufficiently in the very early universe.
It is easy to see from the second Friedmann equation that the three equivalent conditions necessary for inflation are decreasing comoving Hubble radius, accelerated expansion, and violation of the strong energy condition (i.e.\ negative pressure):
\begin{equation}
\frac{\D}{\D t} \left(\frac{H^{-1}}{a}\right) < 0 \quad \Rightarrow \quad
\frac{\D^2 a}{\D t^2} >0 \quad \Rightarrow \quad \rho+3p <0 \, .
\end{equation}
The requirement for negative pressure means that inflationary dynamics cannot be achieved via normal matter or radiation! Inflation can be phenomenologically described by a scalar field with special dynamics, evolving in a self-interaction potential. Although no fundamental scalar field has yet been detected in experiments, there are fortunately plenty of such fields in theories beyond the standard model of particle physics. 
Can a scalar field have $\rho + 3p <0$? The equation of state for a homogeneous scalar field $\Phi$, 
\begin{equation}
w  = \frac{p}{\rho} = \frac{(\partial_t \Phi)^2/2 -V}{(\partial_t \Phi)^2/2 +V}
\, ,
\end{equation}
shows that a scalar field can lead to accelerated expansion ($w<-1/3$) if the potential energy $V$ dominates over the kinetic energy $(\partial_t \Phi)^2/2$. The equation of motion of such a homogeneous
scalar field is
\begin{equation}
\partial_t^2 \Phi + 3 H \partial_t \Phi + V_{,\Phi} = 0 \, 
\label{eq:scalfield}
\end{equation}
where $V_{,\Phi} \equiv \D V / \D \Phi$.

In the following discussion, for notational convenience we will set the reduced
Planck mass to unity, i.e.\ $8\pi G \equiv M_{\rm Pl}^{-2} \equiv 1$.
As we have seen, inflation occurs if the field is evolving slow enough that the potential energy dominates over the kinetic energy, and in order to solve the classical big bang problems, the second time derivative of $\Phi$ is small enough to allow this slow-roll condition to be maintained for a sufficient period. Thus, successful inflation requires
\begin{eqnarray}
(\partial_t \Phi)^2 &\ll& V(\Phi)\, \\
|\partial_t^2 \Phi | &\ll& |3H \partial_t \Phi |, |V_{,\Phi}|\, ,
\end{eqnarray}
where the latter condition means the motion is friction dominated,
i.e.\ $3 H \partial_t \Phi \approx - V_{,\Phi}$ from Eq.~(\ref{eq:scalfield}).
Satisfying these conditions requires the smallness of two dimensionless quantities known as {\it potential slow-roll parameters}
\begin{eqnarray}
\epsilon_V(\Phi) &\equiv& \frac{1}{2} \left(\frac{V_{,\Phi}}{V}\right)^2\\
\eta_V(\Phi) &\equiv& \frac{V_{,\Phi\Phi}}{V}\, .
\end{eqnarray}
In the slow-roll regime, $\epsilon_V, |\eta_V| \ll 1$, with background evolution 
\begin{eqnarray}
H^2 &\approx& \frac{1}{3} V(\Phi) \approx \rm{const.}\, , \label{eq:sl1}\\
\frac{\D \Phi}{\D t} &\approx& -\frac{V_{,\Phi}}{3H}\, ,\label{eq:sl2}
\end{eqnarray}
and the
spacetime is approximately \emph{de Sitter}, with the scale factor evolving as $a(t) \sim e^{Ht}\,$ where the Hubble parameter  $H$ is approximately constant.

Inflation ends when the slow-roll conditions are violated: $\epsilon_V(\Phi_{\rm end}) \approx 1$. The number of $e$-folds before inflation ends is
\begin{equation}
N_e(\Phi) \equiv \ln \frac{a_{\rm end}}{a} = \int_t^{t_{\rm end}} H \D t  \approx \int_{\Phi_{\rm end}}^\Phi \frac{V}{V_{,\Phi'}} \D \Phi' \, .
\end{equation}
Roughly $60$ $e$-folds of inflation must occur for the classic big bang puzzles to be resolved. The universe must then \emph{reheat}, the energy density in the scalar field being converted to radiation to start off the  radiation-dominated era.

Besides solving the big bang puzzles, the decreasing comoving horizon during inflation is the key feature required for the quantum generation of cosmological perturbations. During inflation, quantum fluctuations are generated on sub-Hubble scales and are then stretched out of the Hubble radius by the accelerated
expansion. In other words, the superluminal expansion stretches the perturbations to apparently acausal distances. They become classical superhorizon density perturbations which re-enter the Hubble radius in the subsequent
non-accelerating evolution and then undergo gravitational collapse to form the large-scale structure in the universe. 

To analyse fluctuations during inflation, the inflaton field is split into
a homogeneous background $\bar \Phi(t)$ and a spatially varying perturbation $\delta \Phi(t, \vx)$:
\begin{equation}
\Phi(t, \vx) = \bar \Phi(t) + \delta \Phi(t,\vx)\, .
\end{equation}
Perturbations of the inflaton field value $\delta \Phi$ satisfy the equation of motion of a harmonic oscillator with time-dependent mass. The quantum treatment of inflaton perturbations therefore parallels the quantum treatment of a collection of one-dimensional harmonic oscillators.  Just as zero-point fluctuations of a harmonic oscillator induce a non-zero variance for the oscillation amplitude $\langle x^2 \rangle$, the quantum fluctuations of a light scalar field\footnote{Such a field has (effective) mass $V_{,\Phi\Phi} \ll H^2$ which
is equivalent to $|\eta_V| \ll 1$.} during inflation induce a non-zero variance for the inflaton perturbations \cite{Mukhanov}. The variance
of these fluctuations in Fourier space, i.e.\ the power spectrum, is
\begin{equation}
\label{equ:zero}
\mathcal{P}_{\delta\Phi}(k) = \left(\frac{H_k}{2\pi}\right)^2 \, ,
\end{equation} 
where the right-hand side is evaluated when the mode exits the Hubble radius.
Our Fourier conventions follow those in Sec.~\ref{subsubsection:spatang}.
Technically, the field fluctuation here is defined on hypersurfaces
with zero intrinsic curvature and, in this gauge, the result is the
same as if metric perturbations (i.e.\ the back-reaction of
$\delta \Phi$ on the spacetime geometry) were ignored.

A pedagogical review of the generation of density (scalar) perturbations is beyond the scope of this article and we direct the interested reader to the comprehensive treatment in Ref.~\cite{Kinney:2009vz}. The quantity we wish to
compute is the comoving curvature perturbation $\mathcal{R}$, since this
is conserved from Hubble exit during inflation to Hubble re-entry during the
standard radiation or matter-dominated epochs. With $\mathcal{R}$ we can
reliably compute the primordial fluctuation (in single-field models of inflation) in the radiation era without needing to model the dynamics of the reheating
process. During inflation, comoving hypersurfaces have the property that
they coincide with the hypersurfaces over which the (total) inflaton
$\Phi$ is homogeneous. These hypersurfaces are not the same as the
zero-curvature surfaces on which Eq.~(\ref{equ:zero}) holds -- there
is a time delay between them, $\delta t = - \delta \Phi / \partial_t
\bar{\Phi}$, such that the evolution of the background
$\bar{\Phi}$ in this time compensates for the perturbation $\delta \Phi$
to give a smooth total $\Phi$. The differential background expansion during
this time delay means that the intrinsic curvature of the comoving hypersurfaces is simply $\mathcal{R} = - H \delta \Phi / \partial_t \bar{\Phi}$.
In this simple class of inflation models, the power spectrum of $\mathcal{R}$ is thus~\cite{LiddleLyth}
\begin{equation}
\label{equ:SRPs}
\mathcal{P}_{\mathcal{R}}(k) = \left( \frac{H^2}{2\pi \partial_t \bar{\Phi}}
\right)^2_{k=aH} \approx \left(\frac{V^3}{12\pi^2 V_{,\Phi}^2} \right)_{k=a H} \, ,
\end{equation}
where the second equality uses the slow-roll approximations~(\ref{eq:sl1})
and (\ref{eq:sl2}). The efficiency of this mechanism for producing
cosmological curvature perturbations depends on both the height of
the potential, which determines the expansion rate and hence size
of $\delta \Phi$, and its slope, which enters through the conversion of
inflaton fluctuations to time delays and so curvature.

For gravitational waves, the two polarization modes of the metric perturbations $h_{ij}$ (see Eq.~\ref{adc:eq40}) satisfy the same equation of motion as the
perturbations of a massless scalar field. The power spectrum of $h_{ij}$ is
therefore~\cite{star79, star85, LiddleLyth}
\begin{equation}
\label{equ:SRPt}
\mathcal{P}_h(k) = 8\left(\frac{H_k}{2\pi}\right)^2
\approx \left. \frac{2}{3\pi^2}  V \right|_{k=aH}\, ,
\end{equation}
where the second equality uses the slow-roll approximation. Our Fourier
conventions for $h_{ij}$ are such that the real-space variance of
the metric fluctuation is
\begin{equation}
\langle h_{ij} h^{ij} \rangle = \int \mathcal{P}_h(k) \D \ln k \, .  
\end{equation}
Note that the tensor power spectrum depends only on the expansion rate
during inflation. The amplitude of the tensor power spectrum
relative to the curvature spectrum defines the tensor-to-scalar ratio
\begin{equation}
\label{equ:r}
r \equiv \frac{\mathcal{P}_h(k_\star)}{\mathcal{P}_\mathcal{R}(k_\star)}
\approx  8 \left( \frac{V_{,\Phi}}{V} \right)^2 =16 \epsilon_V \, .
\end{equation}
Here, $k_\star$ is some (arbitrary) reference scale.
As we will explain below, the value of $r$ is of fundamental importance in the quest to understand the  microscopic origin of the inflationary dynamics.

The power spectra in Eqs.~(\ref{equ:SRPs}) and (\ref{equ:SRPt}) are to be evaluated when a fluctuation with (physical) wavenumber $k/a$ exits the Hubble radius $H^{-1}$.  Different scales exit at different times when the inflationary potential $V(\Phi)$ has slightly different values.  This leads to a small (but computable) scale-dependence of the primordial power spectra. In terms of an empirical parameterisation of this weak scale-dependence, the scalar power spectrum can be written in terms of an amplitude $A_s$ and a spectral index $n_s$ as 
\begin{equation}
\mathcal{P}_\mathcal{R}(k) = A_s (k/k_\star)^{n_s -1}\, ,
\label{eq:pk_scalar}
\end{equation}
both measured at the reference scale $k_\star$. Similarly, its tensor counterpart can be written as 
\begin{equation}
\mathcal{P}_h(k) = A_t (k/k_\star)^{n_t}\, .
\label{eq:pk_tensor}
\end{equation}
The deviation from scale invariance ($n_s=1$, $n_t=0$) of these power spectra is sensitive to the \emph{shape} of the inflaton potential:
\begin{equation}
n_s - 1 = 2  \frac{V_{,\Phi\Phi}}{V}  - 3 \left(\frac{V_{,\Phi}}{V}\right)^2\, ,
\end{equation}
and
\begin{equation}
n_t = - 4 \left( \frac{V_{,\Phi}}{V}\right)^2\, .
\end{equation}
The primordial perturbation spectra are rather directly imprinted in
the observable CMB angular power spectra. For example, for
scalar perturbations, the dominant Sachs-Wolfe contribution to the
anisotropies gives $C_l \propto \mathcal{P}_{\mathcal{R}}(l/d_A)$, where
the $l$-dependent proportionality depends on the processing of the
primordial perturbations by gravity and acoustic physics.
Measurements of the scale-dependence of the primordial power spectra therefore have the power to probe the shape of the inflaton potential $V(\Phi)$. Note the consistency relation, $r=-8 n_t$, in the simple single-field models we are considering so far.

Two important pieces of information about inflation would follow from a detection of $A_t$: (1) the energy scale of inflation; and (2) the field variation of the inflaton. The energy scale of inflation $V^{1/4}$ is proportional to
$A_t^{1/4}$. Using the measured value for $A_s$, this can be expressed in terms
of the tensor-to-scalar ratio on CMB scales, $r$, as
\begin{equation} V^{1/4} = 1.06 \times 10^{16} \, {\rm GeV} \left(\frac{r}{0.01}\right)^{1/4}\, . \end{equation} 
To date, high-energy physicists only have two indirect clues about physics at this scale: the apparent unification of gauge couplings, and experimental lower bounds on the proton lifetime, and these energy scales are forever beyond the reach of earth-bound particle accelerators. Thus, if a primordial tensor mode were to be detected, we would be presented with a unique opportunity to use the universe as a high-energy physics laboratory. 

From manipulations of Eq.~(\ref{equ:r}) one may derive \cite{Lyth:1996im} the following relation between the tensor-to-scalar ratio $r$ and the distance in inflaton field space between the end of inflation and the point when the scales of CMB fluctuations were created:
\begin{equation}
\frac{\Delta \Phi}{M_{\rm pl}} \gtrsim \Bigl( \frac{r}{0.01} \Bigr)^{1/2}\, .
\end{equation}
Here, we have reinstated the reduced Planck mass.
A large tensor amplitude, $r>0.01$, therefore correlates with a super-Planckian field variation during inflation. Super-Planckian field excursions have interesting theoretical implications \cite{WhitePaper}:
to control the shape of the inflaton potential over a super-Planckian range requires the existence of an approximate shift symmetry in the ultraviolet (UV) limit of the underlying particle theory for the inflaton, $\Phi \to \Phi + {\rm const}$. In string theory it has only recently become possible to construct controlled large-field inflation models with approximate shift symmetries in the UV \cite{EvaAlex, EvaLiamAlex, Baumann:2009ni}.

\subsection{Observable predictions and current observational constraints}

The inflationary proposal requires a huge extrapolation of the known laws of physics. In the absence of a complete theory, a phenomenological approach has been commonly employed, where an effective potential $V(\Phi)$ is postulated. Ultimately, $V(\Phi)$ has to be derived from a fundamental theory, and significant progress in implementing inflation in string theory has been made in recent years \cite{Baumann:2009ni}. However, while it is challenging to understand the origin of inflation from a particle physics point of view, it is also a great opportunity to learn about ultra-high-energy physics from cosmological observations.
 
The simplest inflationary scenarios consist of a single light scalar field
with a canonical kinetic term, $(\nabla \Phi)^2/2$, in its action.
They predict the following observable characteristics.

\begin{enumerate}
\item {\it Flat geometry}, i.e.\ the observable universe should have no spatial curvature. As we have seen, flatness has been verified at the 1\% level by the location, or, better, separation, of the CMB acoustic peaks combined with some low-redshift distance information.
\item {\it Gaussianity}, i.e.\ the primordial perturbations should correspond to Gaussian random variables to a very high precision. 
\item {\it Scale-invariance}, i.e.\ to a first approximation, there should be equal power at all length-scales in the perturbation spectrum, without being skewed towards high or low wavenumbers. 
In terms of the parameterisation (\ref{eq:pk_scalar}) and (\ref{eq:pk_tensor}) this corresponds to $n_s =1$ and $n_t= 0$. However, small deviations from scale-invariance are also a typical signature of inflationary models and tell us about the dynamics of inflation.
\item {\it Adiabaticity}, i.e.\ after reheating, there are no perturbations in the relative number densities of different species on super-Hubble scales (so no isocurvature modes). This follows from the assumption that only a single field is important during inflation. Constraints on isocurvature modes were discussed in Sec.~\ref{subsubsec:iso} where we noted that current data shows no evidence for a non-adiabatic component of the primordial perturbation.
\item {\it Super-Hubble fluctuations}, i.e.\ there exist correlations between anisotropies on scales larger than the apparent causal horizon, beyond which two points could not have exchanged information at light-speed during the history of a non-inflationary universe. This corresponds to angular separations on the sky larger than $\sim 2^\circ$.
\item {\it Primordial gravitational waves}, which give rise to temperature and polarization anisotropies as described above. These tensor modes must exist; however, their predicted amplitude can vary by many orders of magnitude depending on the underlying microphysical mechanism implementing inflation. 
\end{enumerate}

It is beyond the scope of these lecture notes to describe in algorithmic detail how CMB observations can be used to constrain inflationary physics; the interested reader is referred to Ref.~\cite{Baumann:2008bn} for a beginner-level introduction and a guide to the primary literature. The tightest constraints on inflationary physics currently come from the WMAP data in combination with complementary cosmological data. We now summarise some of the main consequences of the five-year data for inflation where these have not been covered elsewhere in these notes.

\subsubsection{Super-horizon correlations}

The WMAP detection of an (anti-)correlation between CMB temperature
and polarization fluctuations at angular separations $5^\circ > \theta >
1^\circ$ (corresponding to the $TE$ anti-correlation seen for multipoles
$l \sim 50$--$150$ in Fig.~\ref{fig:Epoldata})
is a distinctive
signature of adiabatic fluctuations on super-horizon scales at the
epoch of recombination, confirming a fundamental prediction of the
inflationary paradigm~\cite{1994PhRvL..73.2390C,spergel/zaldarriaga:1997,peirisetal03}.
Inflation, in which microscopic quantum fluctuations are stretched to super-Hubble scales, is the most compelling causal mechanism for generating such apparently-acausal perturbations.
The observed $TE$ anti-correlation on large scales is remarkable qualitative evidence for this 
basic mechanism.

\subsubsection{Measurements of the scalar spectrum}

\begin{table}[t]
\begin{tabular}{ccc}
\hline
&
\tablehead{1}{c}{c}{5-year WMAP}
&
\tablehead{1}{c}{c}{WMAP+BAO+SN} \\
\hline
 $n_s$ & $0.963_{-0.015}^{+0.014}$ & $0.960\pm 0.013$ \\
\hline
$n_s$ & $0.986 \pm 0.022$ & $0.970 \pm 0.015$ \\
 $r$ & $<0.43$ & $<0.22$ \\
\hline
 $n_s$ & $1.031_{-0.055}^{+0.054}$ & $1.017_{-0.043}^{+0.042}$ \\
 $\alpha_s$ & $-0.037 \pm 0.028$ & $-0.028_\pm 0.020$ \\
\hline
$n_s$ & $1.087_{-0.073}^{+0.072}$ & $1.089_{-0.068}^{+0.070}$ \\
 $r$ & $<0.58$ & $<0.55$ \\
$\alpha_s$ & $-0.050 \pm 0.034$ & $-0.053 \pm 0.028$ \\
\hline
\end{tabular}
\caption{ \label{tab:param} Five-year WMAP constraints on the primordial power spectra in the power-law parameterisation~\cite{2009ApJS..180..330K}.}
\end{table}

Komatsu et al.~\cite{2009ApJS..180..330K} recently used the WMAP five-year temperature and polarization
data, combined with the luminosity distance data of Type-Ia supernovae (SN) at redshifts $z\le 1.7$~\cite{unionsn} and BAO data at redshifts $z=0.2$ and $0.35$~\cite{bao}, to
put constraints on the shape of the primordial power spectra (see Table \ref{tab:param}).

The WMAP analysis employed the standard power-law parameterisation of the scalar power spectrum given in Eq.~(\ref{eq:pk_scalar}).
The amplitude of scalar fluctuations at $k_\star = 0.002\, {\rm Mpc}^{-1}$ is measured to be
$A_s(k_\star) = (2.445\pm 0.096) \times 10^{-9}$.
From both numerical simulations and analytical estimates one finds that with this initial amplitude of density fluctuations, there is sufficient time
for gravity to form the large-scale structures that we observe today.
Assuming {\it no} tensor perturbations ($r\equiv 0$) the scale-dependence of the power spectrum is
\begin{equation}
n_s = 0.960 \pm 0.013 \quad (r\equiv 0)\, .
\end{equation}
The scale-invariant Harrison-Zel'dovich-Peebles spectrum, $n_s = 1$, is
$3.1\sigma$ away from the mean of the likelihood. Including the possibility of a non-zero $r$ in the parameter estimation, the marginalised constraint on $n_s$ becomes
\begin{equation}
n_s = 0.970 \pm 0.015 \quad (r \ne 0)\, .
\end{equation}
The (slight) worsening of the constraint simply means that there is a degeneracy between $n_s$ and $r$, and the current data cannot simultaneously measure them independently. Since adding tensor perturbations enhances the large-scale power, it can be offset by increasing $n_s$.

\subsubsection{Constraints on the spectrum of tensor perturbations}

At current sensitivities, constraints on the amplitude of tensor perturbations
are driven by the temperature data. In the future, this situation should change
with weight shifting to the $B$-mode of polarization, as discussed later in
Sec.~\ref{subsec:Bmodesandtensors}.
With five years of WMAP data alone, one finds the following upper limit on the tensor-to-scalar ratio:
\begin{equation}
\label{equ:rlimit}
r < 0.43 \quad (95 \%\, {\rm C.L.})\, .
\end{equation}
This improves to $r < 0.22$ (95\% C.L.) if SN and BAO data are included.
The latter datasets add only geometric information and so are not influenced
by tensor perturbations. However, what they do bring is a better determination
of the matter density which, through degeneracies, leads to an improvement in
$n_s$ and hence $r$. The limit $r<0.22$ is close to the fundamental limit,
set by cosmic variance, that can be achieved with temperature data alone.
Upcoming polarization experiments with sensitivity to $B$-modes
should improve on this limit by an order of magnitude or more.

\subsubsection{Constraints on inflationary models}

\begin{figure}[t!]
    \centering
        \includegraphics[width=.8\textwidth]{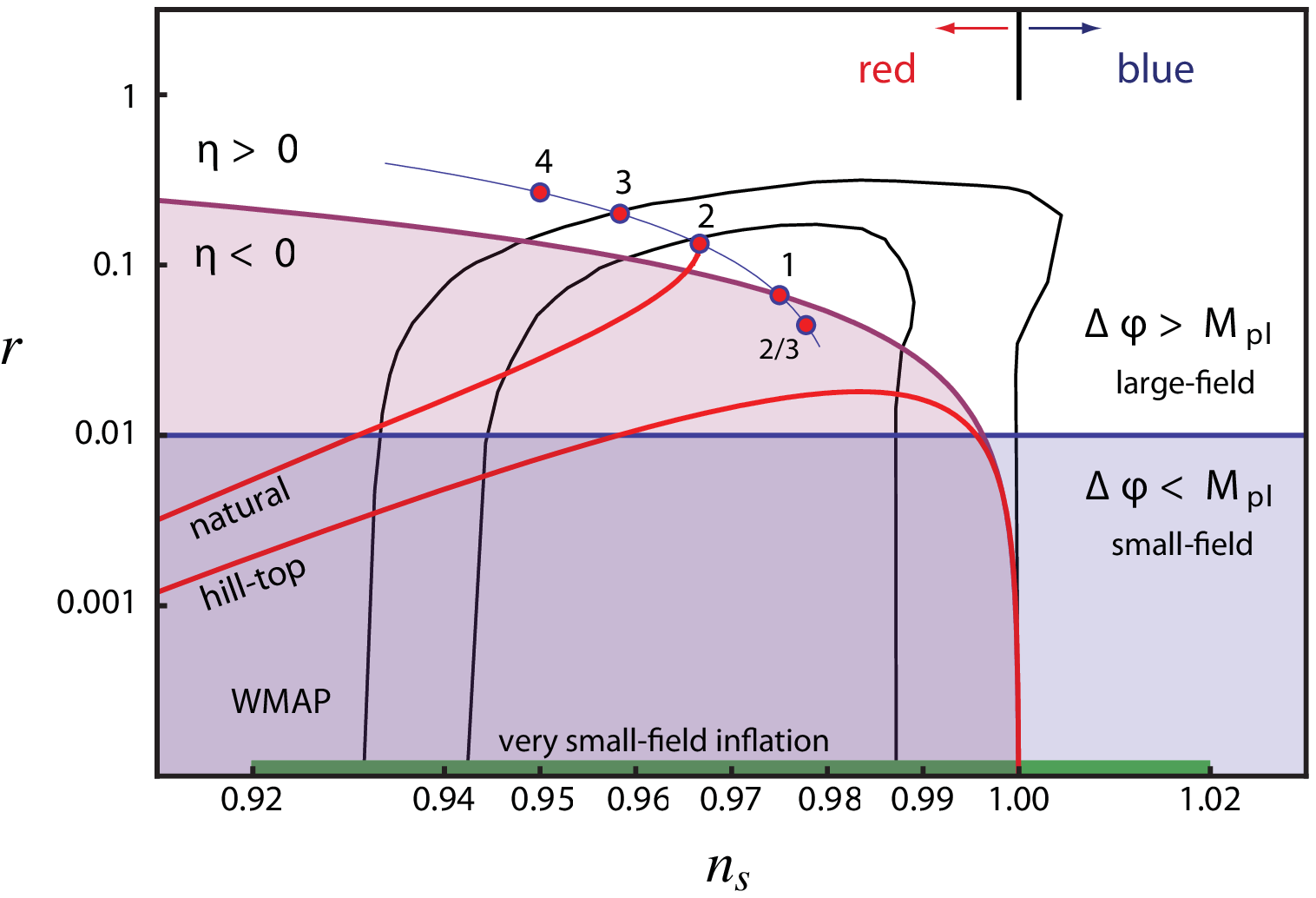}
          \caption{Constraints on single-field slow-roll models in the $n_s$-$r$ plane~\cite{WhitePaper}. The value of $r$ determines whether the model involves large or small inflaton field variations.  The sign of $n_s-1$ classifies the scalar spectrum as red or blue. The combination $2\eta_V = n_s -1 + 3r/8$
determines whether the curvature of the potential was positive ($\eta_V > 0$) or negative ($\eta_V < 0$) when the observable universe exited the Hubble radius. Shown also are the WMAP five-year constraints on $n_s$ and $r$~\cite{2009ApJS..180..330K}, as well as
the predictions of a few representative models of single-field slow-roll inflation: {\it chaotic inflation} -- $\lambda_p\, \Phi^p$, for general $p$ (thin solid line) and for $p=4, 3, 2, 1, 2/3$ (filled circles) -- models of which with $p=1$~\cite{MSW} and $p=2/3$~\cite{EvaAlex} have recently been obtained in string theory; 
{\it natural inflation} -- $V_0 [1-\cos(\Phi/\mu)]$ (solid line); {\it hill-top inflation} -- $V_0 [1-(\Phi/\mu)^2] + \dots$ (solid line); 
{\it very small-field inflation} -- models of inflation with a very small tensor amplitude, $r \ll 10^{-4}$ (green bar), e.g.\ brane inflation \cite{KKLMMT, delicate1, BDKM, Holographic}, K\"ahler inflation \cite{Conlon:2005jm}, and racetrack inflation \cite{BlancoPillado:2006he}, which often arise in string
theory.}
    \label{fig:nsr}
\end{figure}

At the present stage we are still testing basic aspects of the inflationary {\it mechanism} rather than details of its specific implementation. However, constraints on specific inflationary models have been obtained (see Fig.~\ref{fig:nsr}): inflationary models predicting a blue spectrum ($n_s > 1$) are now virtually ruled out; this includes models of hybrid inflation like $V(\Phi) =V_0[1 + m^2 \Phi^2]$. Assuming that the tensor amplitude is small, the measurement $n_s < 1$ implies a constraint on the curvature of the inflaton potential, $V'' < 0$.
Finally, models predicting a very large tensor amplitude are ruled out, e.g.\ $V(\Phi) = \lambda \Phi^4$. In summary, many popular models are still allowed by the data, but an increasing number of models are on the verge of being tested seriously.

\subsection{Future prospects for constraining early-universe physics with the CMB}
\label{sec:future}

There are exciting prospects for improving our understanding of early-universe physics with new CMB data that will be available in the next decade. The Planck satellite~\cite{2006astro.ph..4069T} will measure the temperature power spectrum over a large range of scales with unprecedented accuracy and resolution. In addition, it will provide greatly improved constraints on $E$- and $B$-mode polarization. The Planck data will be supplemented by many sub-orbital experiments with a special focus on measurements of the small-scale temperature fluctuations and/or the polarization power spectra. Finally, plans are being made for a next-generation satellite dedicated to the measurement of CMB polarization~\cite{WhitePaper,2008ExA...tmp...45D}. Such an experiment proposes to improve the sensitivity to $B$-modes by almost two orders of magnitude over the current situation.
As we now describe, the combination of this wealth of data will allow detailed tests of the physics of the early universe.

\subsubsection{Scalar perturbation spectrum on small scales}

The shape of the spectrum of primordial density fluctuations can be an important diagnostic of the inflationary dynamics.
Equation~(\ref{eq:pk_scalar}) is, of course, only a simple parameterisation of the power spectrum in terms of an amplitude $A_s$ and a spectral index $n_s$, both defined at the pivot scale $k_\star$.
This power-law parameterisation may be refined by allowing a non-zero
running of the spectral index, $\alpha_s \equiv \D n_s/\D\ln k = \D^2 \ln
\mathcal{P}_{\mathcal{R}} / \D (\ln k)^2$, i.e. by defining
\begin{equation}
\mathcal{P}_{\mathcal{R}}(k) = A_s(k_\star) \left(\frac{k}{k_\star} \right)^{n_s(k_\star)-1 + \alpha_s(k_\star) \ln(k/k_\star)/2} \, .
\end{equation}
Generically, slow-roll inflation predicts that the running should be a small
effect, second order in the slow-roll parameters, and so
$|\alpha_s| \sim O(0.001)$.  At present, the data is insufficiently precise to detect $\alpha_s$ at that level.
The current constraints are rather dependent on the combination of CMB data
used, and the details of the treatment of contaminants at high multipoles
such as emission from extra-Galactic sources and the Sunyaev-Zel'dovich effect
in galaxy clusters. Full details can be found in~\cite{2008arXiv0801.1491R,
2009ApJS..180..330K}; the
basic result is that no running ($\alpha_s = 0$) is consistent with various
data combinations at the 95\% confidence level but the mean values of
$\alpha_s$ are persistently negative.
The key to improving constraints on running is to increase the lever arm in $k$. Future small-scale CMB measurements, possibly combined with large-scale structure surveys, should
improve constraints considerably though probably not to the level where
the prediction of minimal inflation becomes detectable.
Forecasts for Planck suggest $1\sigma$ (marginalised) errors of $0.005$ for $\alpha_s$~\cite{2006astro.ph..4069T}.

In the case of slow-roll inflation, a definitive measurement of runnning of
the spectral index, $\alpha_s\neq 0$, is a signal that $\xi_H$, the third {\it Hubble slow-roll} parameter (defined in analogy to the first two {\it potential slow-roll} parameters discussed previously),\footnote{The primes
here denote derivatives with respect to $\Phi$.}
\begin{equation}
\xi_H \equiv 4 M_{\rm pl}^4\left[\frac{H'(\Phi) H'''(\Phi)}{H^2(\Phi)}\right],
\label{eq:xiH}
\end{equation}
played a significant role in the dynamics of the inflaton \cite{Chongchitnan:2005pf} as the CMB scales exited the Hubble radius. In the context of slow-roll inflation it would be hard to explain a value of $\alpha_s$ that is either negative or positive but much larger in magnitude than $0.001$.
The consequences for the physics of inflation differ depending on whether the running is negative or positive, and both options would dramatically complicate the theoretical understanding of inflation.

A large negative running implies that  $\xi_H$ was (relatively) large and positive  as the cosmological perturbations were laid down ($\alpha_s \approx
- 2 \xi_H$ in this limit).
It can be shown that   $\xi_H >0$ generally hastens the end of inflation (relative to    $\xi_H =0$), provided the higher order slow-roll parameters can be ignored.  With these assumptions, we find a tight constraint on   $\xi_H$  if we are to avoid a premature end to slow roll, with inflation terminating soon after the observable scales exit the Hubble radius \cite{Chung:2003iu, Malquarti:2003ia, Easther:2006tv, Peiris:2006sj}. Thus, a definitive observation of a large negative running would imply that any inflationary phase requires higher order slow-roll parameters to become important after the observable scales leave the Hubble radius \cite{Chung:2003iu, Malquarti:2003ia, Ballesteros:2005eg, Adshead:2008vn}, multiple fields yielding complicated spectra \cite{Sasaki:1995aw}, two or more bursts of inflation \cite{Silk:1986vc, Holman:1991ht, Polarski:1992dq, Lyth:1995ka, Burgess:2005sb}, or a temporary breakdown of slow roll (such as a feature in the potential) \cite{peirisetal03,Starobinsky:1992ts, Adams:2001vc, Covi:2006ci, Hamann:2007pa, Hunt:2004vt, Hunt:2007dn}. There have also been proposals for explicit models where a large, transient negative running occurs as a ``stringy'' signature in multi-throat brane inflation \cite{Chen:2004gc,Chen:2005ad,Bean:2007eh}.

The current cosmological data disfavour inflationary models with a blue tilt, $n_s>1$ \cite{2009ApJS..180..306D,2009ApJS..180..330K}; however, a significant parameter space is still allowed where $n_s<1$ but with a large positive running (implying a large negative $\xi_H$), which would lead to a strongly blue-tilted spectrum on scales below those of cosmological interest~\cite{Peiris:2008be}. Again under the hypothesis that this parameterisation can be extrapolated to the end of inflation, we find a class of solutions where $\epsilon \rightarrow 0$ as $H$ remains finite, and the field rolls towards a minimum with a substantial vacuum energy. The  perturbation spectrum grows at small scales, possibly diverges, and can lead to over-production of primordial black holes \cite{Peiris:2008be,Hawking:1971ei, Carr:1974nx, Carr:1994ar, Kim:1996hr, Green:1997sz, Covi:1998mb, Yokoyama:1999xi, Leach:2000ea, Zaballa:2006kh, Chongchitnan:2006wx, Kohri:2007qn}, or even the onset of eternal inflation \cite{Peiris:2008be,Linde:1986fd, Kohri:2007qn}.  Consequently, if we measure a large positive running we will again conclude that any inflationary phase is not described within the single field, slow-roll formalism, or that higher order terms in slow roll are important.

\subsubsection{Tensor perturbations}
\label{subsec:Bmodesandtensors}

The WMAP upper limit on the amplitude of primordial tensor fluctuations, $r < 0.22$, is only beginning to be a serious constraint on inflationary model-building. As noted above, current constraints are driven by the large-angle temperature data, but this route is fundamentally limited by the cosmic variance of the dominant scalar perturbations. The same is true for $E$-mode polarization but not for $B$-mode polarization since this is not generated by scalar perturbations in linear theory. Improved constraints on $B$-modes of CMB polarization will therefore drive future limits on the tensor amplitude.
The quest for a $B$-mode detection is one of the most exciting developments in observational cosmology.

\begin{figure}[t!]
\centerline{\includegraphics[width=0.6\textwidth,angle=-90]{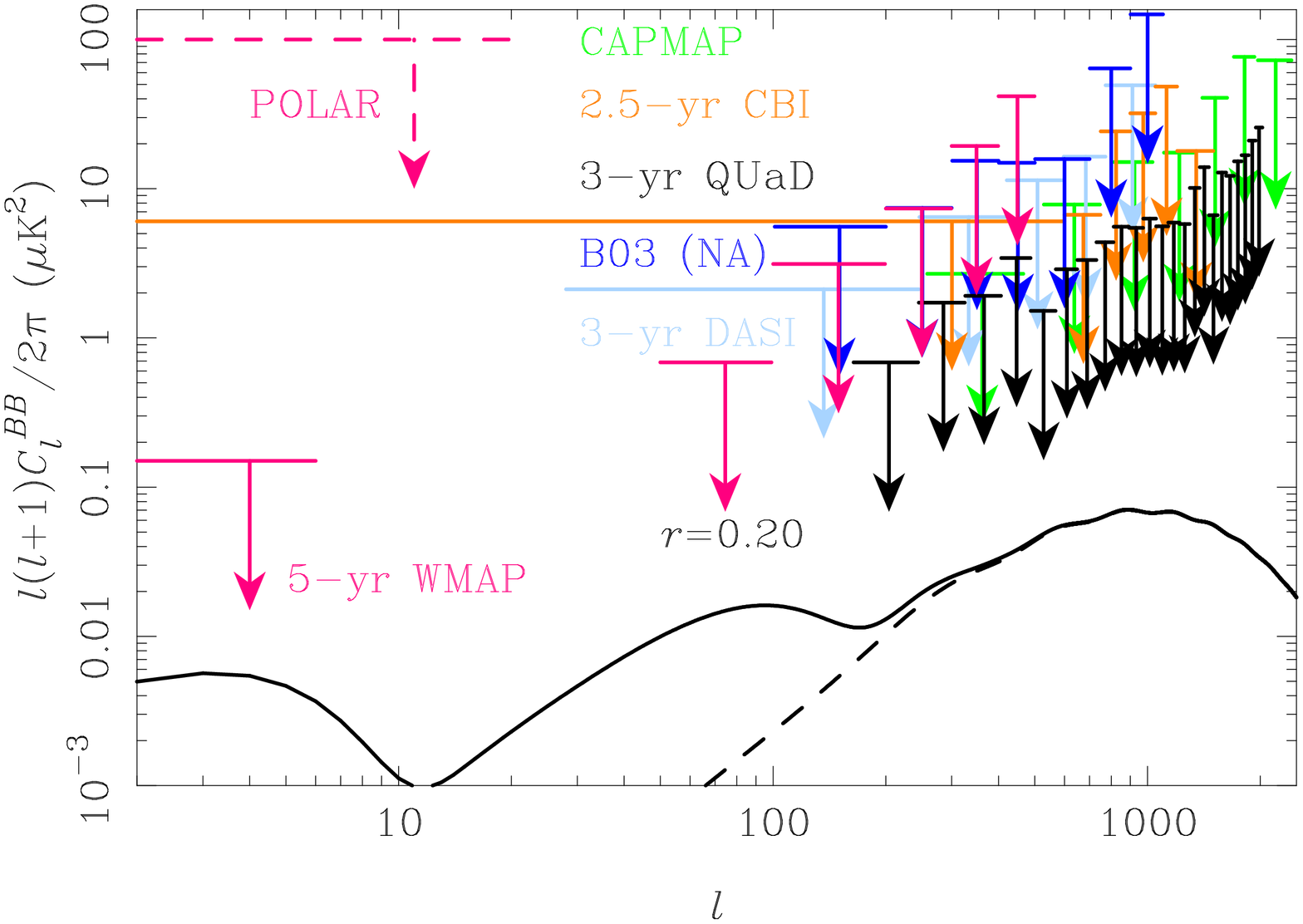}}
\caption{Current 95\% upper limits on $C_l^{B}$ from WMAP5
(magenta;~\cite{2009ApJS..180..296N}), QUaD  (black;~\cite{2009ApJ...692.1247P}),
BOOMERanG (blue;~\cite{2006ApJ...647..833P,2006ApJ...647..813M}), DASI (cyan;~\cite{2005ApJ...624...10L}), CAPMAP
(green;~\cite{2008ApJ...684...771B}), CBI (orange;~\cite{2007ApJ...660..976S}) and
POLAR (dashed magenta;~\cite{2001ApJ...560L...1K}).
The solid line is the contribution from gravitational waves for
a tensor-to-scalar ratio $r=0.2$. The dashed line is the $B$-mode
power induced by weak gravitational lensing.
}
\label{fig:Bpoldata}
\end{figure}

Current limits on $B$-mode power are compared in Fig.~\ref{fig:Bpoldata}
to the power expected for $r=0.2$. (The secondary contribution from
conversion of $E$ modes to $B$ by weak gravitational lensing is also
shown; see Sec.~\ref{subsec:lensing}.) The primordial signal is very weak,
with r.m.s.\ $=0.18 \sqrt{r/0.2} \,\mu\mathrm{K}$, but there are no
fundamental technical or cosmological obstacles to achieving
$r \sim 0.001$, although it calls for arrays of thousands of detectors.
However, it is still unclear whether polarized Galactic emission, particularly
thermal dust emission, is sufficiently well behaved to allow its removal
down to this level on the basis of multi-frequency data.
There are two characteristic scales for attempting detection. The reionization
signal at $l<10$, and the signal from recombination at $l < 100$.
Where most information lies depends on the value of $r$, the instrument
sensitivity, the scale-dependence of Galactic foregrounds, and the
size of the survey (i.e.\ whether the large-scale reionization modes
are accessible). Realistic forecasts for Planck suggest
$r=0.05$ may be achievable,
and the constraint is driven by the reionization signal~\cite{2009arXiv0902.4803E}.
In contrast, most of the next generation of ground-based and balloon
experiments will survey smaller regions of the sky in areas of (known) low
foreground contamination, and their target is the signal from
recombination~\cite{2006SPIE.6275E..51Y,2008arXiv0805.3690N,2008arXiv0806.4334S,2008arXiv0805.4527C,2004SPIE.5543..320O,2008SPIE.7010E..79C}.
Tensor-to-scalar ratios $r \sim 0.01$ may be achievable
with these surveys. Looking further ahead, $r \sim 10^{-3}$
with a full-sky survey from a dedicated polarization satellite seems
achievable with current technology.
Measuring tensors at that level of sensitivity would mark a qualitative shift in our ability to test the inflationary paradigm.  If tensors are not seen at that level, all large-field models of inflation are ruled out.
On the other hand,
a $B$-mode detection would be an extraordinary discovery and a ``smoking gun'' for inflation.

A natural target sensitivity for most future $B$-mode experiments
is $\sim 5\, \mu\mathrm{K}$-$\mathrm{arcmin}$ in the polarization maps.
At this level, the ``noise''
due to the $B$ modes produced by gravitational lensing is comparable
to the instrument noise~\cite{2002PhRvD..65b3505L,2002PhRvD..65b3003H}. For reference, this is 10 times the
sensitivity of the nominal Planck survey.
Pushing for further improvements in sensitivity is
then only worthwhile if it improves rejection of foreground contamination,
or if the lensing contamination can be removed coherently
(i.e.\ at the level of the maps), for example with quadratic reconstruction
techniques~\cite{2002ApJ...574..566H}. Without reconstruction, the
($3\sigma$) upper limits on
$r$ set by lensing, assuming $r=0$, are $1\times 10^{-4}$ from the reionization
signal and $8\times 10^{-4}$ from recombination alone (i.e.\ $l > 10$).
In practice, astrophysical foregrounds and instrumental systematic effects
will surely be more of a limitation than lensing noise. Foregrounds will
be particularly troubling for the signal from reionization, with
estimates suggesting they may limit detections to $r \sim 0.01$ (99\% C.L.)
from $l < 10$ data
alone given the frequency coverage and sensitivity being considered for
a future satellite mission~\cite{2006JCAP...01..019V,2008arXiv0811.3915D}.

\subsubsection{Non-Gaussianity}
\label{subsubsec:nonGauss}

So far, we have assumed that the primordial fluctuations have a Gaussian distribution. Indeed, this is a fundamental prediction of slow-roll inflation and the current observational limits confirm this at the 0.1\% level.\footnote{By this measure, non-Gaussianity has been constrained more accurately than curvature.}
However, as we will now discuss, a small degree of primordial non-Gaussianity can be a crucial probe of the inflationary dynamics.

Non-Gaussianity is a sensitive probe of inflationary physics that is difficult to access by any other means. Specifically, it is a measure of {\it inflaton interactions}. To allow for slow-roll inflation the inflaton field is necessarily weakly interacting [$V(\Phi)$ is very flat] and the non-Gaussianity is predicted to be small \cite{2003JHEP...05..013M}. However, going beyond the single-field slow-roll paradigm, non-trivial kinetic terms (derivative self-interactions in the inflaton action), the presence of more than one light field during inflation, the (temporary) violation of slow roll, and a non-adiabatic initial vacuum state for the inflaton may lead to large, observationally distinct non-Gaussianity (e.g.\ \cite{2004PhR...402..103B, Chen:2006nt}).

Gaussian fluctuations are characterised completely by their two-point correlation function, or, equivalently, their power spectrum, $\mathcal{P}_{\mathcal{R}}(k)$. Non-Gaussianity is therefore measured by considering the connected
higher-order correlation functions, i.e.\ the part that remains after all
possible contractions are subtracted. The connected $n$-point functions
with $n>2$ vanish for a Gaussian random field and so
the leading non-Gaussian effect is usually given by the three-point function. To calculate
the non-Gaussianity from inflation, generally one must go beyond linear
perturbation theory, expanding the action to third order to capture all
cubic interactions between perturbed quantities~\cite{2003JHEP...05..013M}.
The primordial
3-point function from inflation is conveniently calculated for quantities
$\zeta$ and $\gamma_{ij}$ which are the non-linear generalisations of
$\mathcal{R}$, the linear-theory comoving-gauge curvature perturbation, and
$h_{ij}$, the linear-theory gravitational wave amplitude, respectively.\footnote{%
These quantities are defined so that the induced line element on comoving
hypersurfaces, over which $\delta \Phi = 0$, is~\cite{2003JHEP...05..013M}
\begin{equation}
{}^{3}g_{ij} \D x^i \D x^j \equiv a^2 e^{2\zeta} e^{\gamma_{ij}} \, .
\end{equation}
Here, $\gamma_{ij}$ is trace-free and transverse,
$\partial_i \gamma^i{}_j = 0$, so that $\det e^{\gamma_{ij}} = 1$.
Note that the true curvature perturbation
of comoving hypersurfaces differs from $\zeta$ by terms quadratic in
$\zeta$ and $\gamma_{ij}$.}
Both $\zeta$ and $\gamma_{ij}$ are conserved on super-Hubble scales
in the simplest inflation models.
The bispectrum for $\zeta$ is defined
by
\begin{equation}
\langle \zeta(\vk_1) \zeta(\vk_2) \zeta(\vk_3) \rangle =
B_\zeta (k_1, k_2, k_3) \delta(\vk_1 + \vk_2 + \vk_3) \, .
\end{equation}
Three other bispectra with different combinations of
$\zeta$ and $\gamma_{ij}$ can also be constructed.
Note that the bispectrum depends only on the magnitudes of the three
wavevectors as a consequence of translational, rotational and
parity invariance.

The bispectrum is measured by sampling {\it triangles} in Fourier space. Much physical information is contained in the momentum dependence or the shape of the bispectrum. Ref. \cite{Fergusson:2008ra} presents visualisations of the full scale/shape dependence of the bispectra for a variety of inflationary scenarios in which significant (observable) amounts of non-Gaussianity could be potentially produced. While these have complex shapes, fortuitously, different physical mechanisms that produce significant non-Gaussianity result in bispectrum signals that peak on different triangular configurations (e.g.\ \cite{2003PhRvD..67b3503L,Chen:2006nt,Linde:1996gt, Babich:2004gb, Holman:2007na,Chen:2006xjb}):
\begin{itemize}
\item multi-field models peak on squeezed triangles ($k_3 \ll k_2 \sim k_1$);
\item models with non-canonical kinetic terms peak on
equilateral triangles ($k_1 \sim k_2 \sim k_3$);
\item models with non-adiabatic initial vacua peak on flattened/folded triangles ($k_3 \sim k_2 \sim 2 k_1$); and
\item non-slow-roll models peak on more complex (model dependent) configurations requiring a matched template for their analysis.
\end{itemize}
Thus, configuration-dependent studies of non-Gaussianity may become a powerful probe of ultra-high-energy physics and inflation.

Given a primordial bispectrum, how do we compute the observable bispectrum,
$B_{l_1 l_2 l_3}$, in the CMB (see Eq.~\ref{eq:5a})? In theories with weak
primordial non-Gaussianity, such as single-field inflation,
this is a difficult task since the non-linear perturbation evolution and
radiative transfer will induce additional non-Gaussianity in the
CMB that exceeds (and distorts) the primordial contribution.
Signatures of specific non-linear effects have been computed, but a full
calculation is still lacking
(see~\cite{2006JCAP...05..010B,2006JCAP...06..024B,2006astro.ph.10110B,2008PhRvD..78f3526P,2008arXiv0812.3652S,2009CQGra..26f5006P,2009arXiv0903.0894N} for some recent progress towards this goal). Things are more straightforward, though, if the
initial level of non-Gaussianity is large enough, since then one can continue to employ linear
perturbation theory to compute the higher-order statistical properties of the CMB, replacing the bispectrum of the linear-theory $\mathcal{R}$ with
$B_\zeta(k_1,k_2,k_3)$.

Generalising Eq.~(\ref{eq:19}) to include all sources of anisotropy,
the multipoles of the temperature anisotropy, $T_{lm}$, can be written in
the form
\begin{equation}
T_{lm} = 4\pi i^l \int \frac{\D^3 k}{(2\pi)^{3/2}} g_l(k) \mathcal{R}(\vk)
Y_{lm}^*(\hat{\vk}) e^{i \vk \cdot \vx_R} \end{equation}
in linear theory with adiabatic perturbations.
Here, $g_l(k)$ is the \emph{temperature transfer function} which linearly
relates the observed CMB anisotropy to the primordial curvature perturbation.
The observed CMB bispectrum then becomes
\begin{equation}
B_{l_1 l_2 l_3} = \sqrt{\frac{(2l_1+1)(2l_2+1)(2l_3+1)}{4\pi}}
\left( \begin{array}{ccc} l_1 & l_2 & l_3 \\
			    0 & 0 & 0 \end{array}
\right) b_{l_1 l_2 l_3} \; ,
\end{equation}
where the reduced bispectrum $b_{l_1 l_2 l_3}$ is~\cite{2000PhRvD..61f3504W}
\begin{equation}
b_{l_1 l_2 l_3} = \frac{64}{(2\pi)^{3/2}} \int_0^\infty r^2 \D r \int 
\prod_{i=1}^3 \left[ k_i^2 \D k_i\, g_{l_i}(k_i)j_{l_i}(k_i r)\right]
B_\zeta(k_1,k_2,k_3) \; .
\label{eq:nongauss1}
\end{equation}

As a simple and well-studied example, we consider the local model of primordial
non-Gaussianity in which $\zeta$ is a sum of a Gaussian piece and
the square of a Gaussian in real-space:
\begin{equation}
\zeta(\vx) = \zeta_{\mathrm{G}}(\vx) + \frac{3}{5}f_{\mathrm{NL}} \left[
\zeta_{\mathrm{G}}^2(\vx) - \langle \zeta_{\mathrm{G}}^2 \rangle \right] \; .
\end{equation}
The factor of $3/5$ is conventional and is chosen so that $-f_{\mathrm{NL}}$
parameterises the quadratic contribution to the gravitational potential
$\phi$ in matter domination (for $|f_{\mathrm{NL}}| \gg 1$ so
further non-linear evolution can be ignored).\footnote{Our sign
choices and normalisation are consistent with~\cite{2009ApJS..180..330K}.
Note that on large scales, the temperature anisotropies are
skewed \emph{negative} by a positive $f_{\mathrm{NL}}$.}
Single-field, slow-roll inflation
with canonical kinetic term approximates to the local model,
with a very small primordial $f_{\mathrm{NL}}$ of the
order of the slow-roll parameters, i.e.\ $\sim O(0.01)$~\cite{2003JHEP...05..013M} (non-linear effects
 -- e.g.\ in the relation between the gravitational potential and $\zeta$ --
increase the effective observable $f_{\mathrm{NL}}$ to $O(1)$). Large local
non-Gaussianity can be generated in models with multiple fields, such
as the curvaton model. The bispectrum in the local model evaluates to
\begin{equation}
B_\zeta(k_1,k_2,k_3) = \frac{3f_{\mathrm{NL}}}{5(2\pi)^{3/2}}
\left(\frac{2\pi^2}{k_1^3}\clp_\clr(k_1) \frac{2\pi^2}{k_2^3}\clp_\clr(k_2)
+ \mbox{perms}\right) \; .
\end{equation}
Significantly, this is factorisable which makes the integration in
Eq.~(\ref{eq:nongauss1}) easily tractable.
Numerical examples of the reduced bispectrum in the local model can be found
in Ref.~\cite{2001PhRvD..63f3002K}; it displays similar acoustic oscillations
to the power spectrum.
The best observational constraints on the local form of non-Gaussianity are
from the 3-point function of the WMAP5 data. With an optimal weighting
of the data, Ref.~\cite{2009arXiv0901.2572S} find no evidence for non-zero $f_{\mathrm{NL}}$:
$-4 < f_{\mathrm{NL}} < 80$ (95\% C.L.).
Planck should improve this constraint with its better measurement of the
anisotropies on smaller scales; a detection limit of 
$f_{\mathrm{NL}} \sim 5$ is forecasted~\cite{2001PhRvD..63f3002K}.

\section{Secondary anisotropies}

Secondary anisotropies are generated after recombination. They can arise
from gravitational effects (the integrated Sachs-Wolfe effect or
gravitational lensing) and from scattering events during or after the
epoch of reionization. We have already met
some examples, such as the large-angle polarization generated from
scattering at reionization (Sec.~\ref{subsec:pol_reion}). In this
section we briefly review the other main sources of secondary anisotropy
and their cosmological potential. For a recent, thorough review of
the subject, see~\cite{2008RPPh...71f6902A}.
Fully exploiting the
primary anisotropies on multipoles $l \sim 2000$, for example to
tighten constraints on $n_s$, requires careful
modelling of secondary effects. Such modelling is also important
since the secondary signals themselves
contain valuable cosmological information about structure formation
at late times and reionization. A number of ongoing ground-based surveys
with arcmin resolution are seeking to exploit the secondary anisotropies
in this way~\cite{2008MNRAS.391.1545Z,2007ApJ...663..708M,2004SPIE.5498...11R,2006NewAR..50..969K}.

\begin{figure}[t!]
\centerline{\includegraphics[width=0.99\textwidth,angle=0]{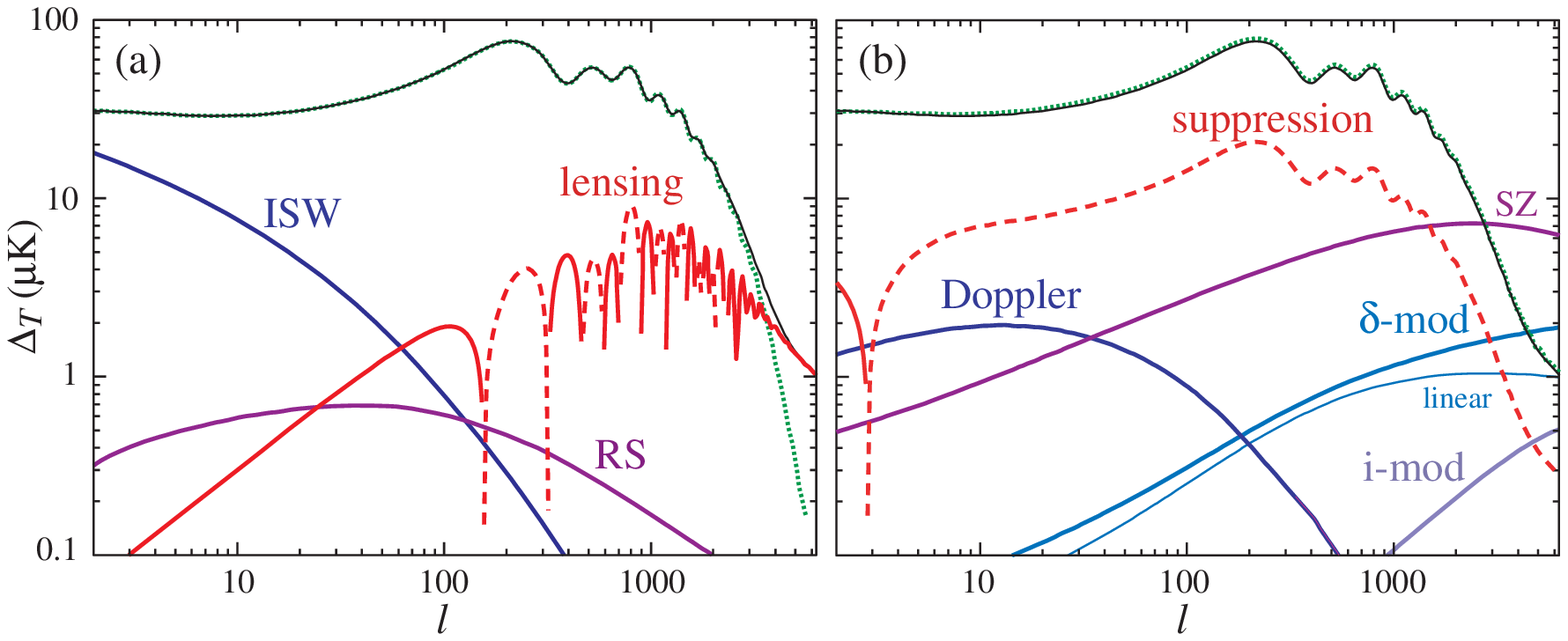}}
\caption{Predicted secondary anisotropies from gravitational effects (left) and
scattering effects (right) from~\cite{2002ARA&A..40..171H}.
(a) The ISW (blue), Rees-Sciama (purple) and
lensing (red; negative where dashed) effects compared to the primary anisotropies from
recombination (green) and the sum of primary and gravitational secondaries
(black). (b) The $\approx
- \tau_{\mathrm{re}}$ fractional suppression of the primary anisotropies (red;
discussed in Sec.~\ref{subsubsec:Treion}),
linear Doppler effect (blue), thermal Sunyaev-Zel'dovich effect
(purple; calculated in the Rayleigh-Jeans limit), the density-modulated Doppler
effect with linear densities (thin cyan) and the total including
non-linear modulation e.g.\ from clusters of galaxies (thick cyan), and
the Doppler effect modulated by the patchiness of reionization (``i-mod'';
purple) compared with the primary anisotropies (green).
The predictions for the SZ effect, non-linear Doppler-modulation and
patchy reionization are based on simplified halo models;
see~\cite{2002ARA&A..40..171H} for details.
The quantity $\Delta T \equiv \sqrt{l(l+1)C_l/2\pi}$.
}
\label{fig:secondaries}
\end{figure}

\subsection{Gravitational secondary effects}

\subsubsection{Integrated Sachs-Wolfe effect}

The integrated Sachs-Wolfe (ISW) effect is described by the last term on
the right of Eq.~(\ref{eq:11}). 
It is an additional source of anisotropy due
to the temporal variation of the gravitational potentials along the line of
sight. If a potential well were to deepen as a photon crossed it, the
photon would receive a net redshift, and there would be a decrement to
the CMB temperature along the line of sight. Conversely,
a decaying potential well gives a temperature increment.
At late times, the potential evolves only when
dark energy (or curvature) start to dominate the background dynamics -- a
linear-theory effect -- or as non-linear structures form.

The linear effect from evolving potentials contributes to the
temperature anisotropies only on large scales (see Fig.~\ref{fig:secondaries}).
This is because there is little power in the gravitational potentials
today on scales that entered the Hubble radius during radiation domination,
and, furthermore, as an integrated effect, the contribution of
small-scale fluctuations is suppressed.
The physics of dark energy affects the decay of the potential
through
the background expansion rate (via its energy density and equation of state)
and through its clustering properties (i.e.\ sound speed). Both
impact on the amplitude and detailed scale-dependence of the ISW
contribution~\cite{2003MNRAS.346..987W}.
The late-time ISW effect is the only way to probe
late-time structure growth with linear CMB anisotropies alone,
but this is hampered by cosmic variance on large angular scales. A more
promising route is to cross-correlate the CMB with a tracer of large-scale
structure~\cite{1996PhRvL..76..575C}. A positive correlation is expected from decaying
potentials, and was first detected by correlating the first-year WMAP data
with the X-ray background and the distribution of
radio galaxies~\cite{2004Natur.427...45B}. Since the initial detection,
the correlation has been detected with a number of large-scale
structure tracers. The most complete analyses to date~\cite{2008PhRvD..77l3520G,2008PhRvD..78d3519H} find $\sim 4.5\sigma$ detections. The power of the ISW effect in
constraining dark energy models is rather limited by chance correlations
between the primary CMB and large-scale structure; the total signal-to-noise
can never exceed $\sim 7$, even for a tracer perfectly correlated with the
ISW signal. Current detections provide independent evidence for dark energy,
but no evidence for departures from $\Lambda$CDM (i.e.\ dynamical
dark energy).

The non-linear late-time effect is sometimes called
the Rees-Sciama effect~\cite{1968Natur.217..511R}. It arises from
the non-linear growth of structures and, more generally, from the
bulk motion of clustered matter. The power spectrum of the Rees-Sciama
effect is broad but is a sub-dominant effect on all scales~\cite{1996ApJ...460..549S}; see Fig.~\ref{fig:secondaries}.

\subsubsection{Weak gravitational lensing}
\label{subsec:lensing}

Weak gravitational lensing of CMB photons by the large-scale matter
distribution at $z < 10$ has several important effects on
the CMB; see~\cite{2006PhR...429....1L} for a recent review.
The deflection arises from the cumulative effect of large-scale
structure along the line of sight. The total deflections are predicted to
have an r.m.s.\ $\sim 3$~arcmin, and be coherent over scales
$\approx 1^\circ$ (the angle subtended by the typical coherence size of the
gravitational potential, $\sim 10^2 \,\mathrm{Mpc}$, at a comoving radial
distance $\chi_*/2$ where lensing is most efficient).
Lensing conserves brightness, simply re-mapping points according to
the deflection field. For the temperature
anisotropies, this re-mapping leads to a smoothing of the acoustic peaks
resulting in fractional changes in $C_l^T$ of around 10\% at the troughs by
$l \sim 2000$. Moreover, it transfers large-scale power to small scales
so that it dominates the
primary anisotropies for $l > 4000$; see Fig.~\ref{fig:secondaries}.
The effects in polarization are
similar, but with the important addition that $B$-mode polarization is
generated from $E$~\cite{1998PhRvD..58b3003Z}. This arises since
the re-mapping by lensing does
not preserve the geometrical relation between polarization direction and
the angular variation of the amplitude which defines $E$-mode polarization.
The resultant $B$ modes have an almost white spectrum
for $l \ll 1000$ with $C_l^B \sim 2\times 10^{-6}\,\mu\mathrm{K}^2$;
see~Fig.~\ref{fig:scalarplustensor}.
Characterising the $B$-mode polarization from lensing is a secondary
science goal of upcoming polarization experiments that have gravitational
waves as their primary target~\cite{2008arXiv0811.3916S}.

As the angular resolution and sensitivities of CMB experiments have improved,
it has become
important to include the effect of lensing when deriving parameter
constraints from CMB temperature data. For example, in the recent
analysis of ACBAR data~\cite{2008arXiv0801.1491R}, a significantly better fit
of $\Lambda$CDM models to the data is found when lensing is included.
Including the amplitude of the lensing correction as a parameter in the
analysis (with a predicted value $q=1$), gives $q = 1.23^{+ 0.83}_{-0.76}$
(95\% C.L.).

An additional feature of lensing is that it introduces non-Gaussian
signatures in the CMB. Roughly, sub-degree scale CMB features behind a given
lens are sheared coherently producing what locally looks like a
statistically-anisotropic pattern of Gaussian CMB fluctuations. Averaged
over lenses, statistical-isotropy is restored but the lensed CMB
becomes non-Gaussian. This non-Gaussianity, manifesting itself as
a non-zero trispectrum~\cite{2001PhRvD..64h3005H}, may be used to reconstruct a noisy estimate
of the deflection field (e.g.~\cite{2001ApJ...557L..79H}).
The noise arises from both instrument noise and
chance alignments in the primary (random) CMB field that mimic the
effect of a lens. The more small-scale features in the field being lensed,
the better the reconstruction. 
Current data is too noisy to allow a direct detection of lensing via
the reconstruction route. The signal-to-noise can be improved by
cross-correlating the reconstruction with a tracer of large-scale structure.
[For a high density of sources, the lower (Poisson) noise compensates for
the reduced level of correlation.]
The lensing effect in the CMB temperature has now been detected
at the $3\sigma$ level in cross-correlation with
galaxy surveys~\cite{2007PhRvD..76d3510S,2008PhRvD..78d3520H}.

Moving beyond these first detections, the interest in CMB lensing
will turn to exploiting it as a probe of the growth of
structure; see~\cite{2008arXiv0811.3916S} for a recent overview.
In particular, CMB lensing has the potential to provide constraints on
dark energy, modifications of gravity, and massive neutrinos that are
inaccessible with the primary CMB anisotropies alone. CMB nicely
compliments cosmic shear surveys~\cite{2008PhR...462...67M}
as it probes structure at higher redshift and with quite different
systematic effects.

Note finally that since the effect of gravity on photons is achromatic,
the frequency dependence of all gravitational secondary anisotropies follows
that of the primary anisotropies.

\subsection{Scattering secondary effects}
\label{subsec:secondary_scattering}

\subsubsection{Thermal Sunyaev-Zel'dovich effect}

The thermal Sunyaev-Zel'dovich (SZ) effect arises from Compton up-scattering
of CMB photons by electrons in hot, ionized gas such as 
in galaxy clusters~\cite{1972CoASP...4..173S}; see
also~\cite{1999PhR...310...97B} for a review. The CMB temperature
at the redshift of the cluster is much less than the gas temperature,
$k_{\mathrm{B}} T \sim \mathrm{few}\, \mathrm{keV}$, so photons
gain energy, on average, by scattering. The net fractional
change in the effective temperature of the CMB is
\begin{equation}
\Theta(\vnhat;\nu) = y(\vnhat) \left[\left(\frac{h \nu}{k_{\mathrm{B}}T_{\mathrm{CMB}}}\right) \coth  \left(\frac{h \nu}{2k_{\mathrm{B}}T_{\mathrm{CMB}}}\right)
- 4 \right] \, ,
\label{eq:sz1}
\end{equation}
where the Compton $y$ parameter is determined by the integral of the
gas pressure along the line of sight: $y(\vnhat)
\equiv (\sigT / m_\mathrm{e}c^2)
\int p_\mathrm{e} \, \D l$ where $m_\mathrm{e}$ is the electron mass.
Unlike the gravitational secondaries, the thermal SZ effect has a
characteristic frequency dependence -- negative below the null
at $217\,\mathrm{GHz}$ and positive above. For a typical galaxy cluster,
$\Theta \sim 10^{-4}$ around $100\,\mathrm{GHz}$ giving a decrement of
$O(100)\,\mu\mathrm{K}$. In the Rayleigh-Jeans limit, $h \nu \ll
k_{\mathrm{B}} T_{\mathrm{CMB}}$, the decrement is
independent of frequency: $\theta(\vnhat;\nu) \approx
- 2 y(\vnhat)$.

The SZ effect is now routinely detected in targeted observations towards
known clusters. A significant feature of the SZ effect is that
\emph{the observed temperature change is independent of the redshift of the cluster}
making SZ also a promising route to search for high-redshift clusters.
Indeed, a number of blind surveys
are now underway, e.g.\
AMI~\cite{2008MNRAS.391.1545Z}, SZA~\cite{2007ApJ...663..708M},
SPT~\cite{2004SPIE.5498...11R} and ACT~\cite{2006NewAR..50..969K},
to do just this. These ground-based instruments, covering a broad spread of
frequencies, all have the high resolution ($\sim 1\,\mathrm{arcmin}$) required
to resolve clusters, and, together, expect to find thousands of clusters.
The Planck satellite, despite its poorer
resolution, will produce a full-sky catalogue with $O(1000)$ SZ clusters.
Recently, the detection of three previously unknown clusters was reported
by the SPT team~\cite{2008arXiv0810.1578S}; these are the first clusters
to be found in a blind SZ survey.

Once detected, clusters will be
followed-up with redshift measurements to determine cluster counts as
a function of redshift, $\D^2 N / \D z\D\Omega$.
This is related to the
underlying cosmology through the expansion rate $H(z)$,
the cluster mass function, $\D n(M,z) /\D M$
giving the number density of clusters per mass $M$ as a function
of redshift, and the survey selection function. The
mass function depends sensitively on the growth of structure
making cluster counts a promising route to improving
our understanding of dark energy~\cite{2002PhRvL..88w1301W}.
However, to achieve this goal, the survey selection function must
be very well understood and this depends on the 
relation between the SZ observables and
the cluster mass. Fortunately, the 
total SZ flux is expected to be a good proxy for the cluster mass,
with the two being related by scaling relations. We can see that this is
reasonable by noting that the SZ flux is
proportional to the volume integral of the electron pressure,
by Eq.~(\ref{eq:sz1}), and hence electron thermal energy. For a relaxed
cluster, the virial theorem relates the thermal energy, hence SZ flux,
to the virial mass. For a flux-limited SZ survey, the selection function
should be rather uniform in mass.
Better understanding the relation between mass and SZ observables,
both with further simulation work and X-ray and optical follow-up,
is an important aspect of SZ cluster research.

The thermal-SZ effect is an important foreground for studies of the
primary CMB temperature anisotropies on small scales. The power
spectrum is shown in Fig.~\ref{fig:secondaries} in the Rayleigh-Jeans
limit.
On large scales it is essentially white noise, but turns over
on scales where clusters are resolved.
Since the frequency spectrum is known, and differs from the primary
CMB anisotropies, the thermal SZ effect can be removed with multi-frequency
observations or by observing close to the null at $217\,\mathrm{GHz}$.
Moreover, the signal is very non-Gaussian. However, for a survey like
Planck, masking clusters detected at high significance
in the CMB maps themselves is not sufficient since the residual contribution
from high-redshift, low-mass clusters below the detection threshold is
still sizeable~\cite{2009MNRAS.392.1153T}.
The power spectrum of residual thermal SZ emission may already have been
seen by CBI~\cite{2009arXiv0901.4540S} and BIMA~\cite{2006ApJ...647...13D}
as an excess of observed small-scale power over that expected from the
primary anisotropies. However, to reconcile the observed excess with
SZ requires around twice the number density of clusters than
predicted for standard cosmological parameters. 
Interestingly, recent data at the same frequency ($30\,\mathrm{GHz}$)
from the Sunyaev-Zel'dovich
Array~\cite{2009arXiv0901.4342S} find no evidence for excess power after
correction for the contribution of residual point sources.
The origin of the discrepancy between these results is still unclear.

\subsubsection{Doppler effects}

Thomson scattering of CMB photons off
free electrons in bulk flows after reionization
generate frequency-independent temperature anisotropies
\begin{equation}
\Theta(\vnhat) = - \int a n_e \sigma_{\mathrm{T}} \E^{-\tau} \vnhat \cdot
\vv_{\mathrm{b}} \, \D \eta,
\end{equation}
via the Doppler effect. For an optical depth, $\tau_{\mathrm{re}} \sim
0.1$, and r.m.s.\ peculiar velocities $\sim 10^{-3}$, the linear Doppler
effect would swap the primary anisotropies were it not for
a rather precise cancellation from integrating along the line of sight.
The cancellation follows since $\vnhat \cdot \vv_{\mathrm{b}}$ is the
radial gradient of a potential in linear theory. The small net effect
is shown in Fig.~\ref{fig:secondaries} and is at most two orders
of magnitude below the power spectrum of the primary anisotropies.

Any spatial modulation of the electron density will tend to avoid
the line-of-sight cancellation. This can arise from patchiness
of the ionization fraction during reionization~\cite{1998ApJ...508..435G},
or from the
clumpiness of the (almost) fully-ionized intergalactic medium
after reionization is complete~\cite{1972CoASP...4..173S,1986ApJ...306L..51O,1987ApJ...322..597V}.
Both effects
produce anisotropies on scales of a few arcmin and below with the same thermal
spectrum as the primary anisotropies.
Computations of the power spectra of the modulated Doppler signals are
shown in Fig.~\ref{fig:secondaries}. 
The details of the patchy signal depend on the morphology
of the ionized fraction (which affects where the power peaks) and
the duration of reionization. Since the 
physics of reionization is still rather uncertain,
and calculating the anisotropies from patchy reionization properly
requires large numerical simulations incorporating radiative transfer
for the ionizing radiation, there remain significant modelling uncertainties
in the predicted power spectrum.\footnote{The computation in
Fig.~\ref{fig:secondaries} assumes a simple semi-analytic prescription
for the growth and clustering of the ionized regions~\cite{1998ApJ...508..435G}.} However, the patchy signal is expected to be sub-dominant to the
density-modulated Doppler signal. There are contributions to the
latter from the linear density field (often called the Ostriker-Vishniac
effect~\cite{1986ApJ...306L..51O,1987ApJ...322..597V}),
which is easily calculated
analytically, and from non-linear structures such
as clusters of galaxies and filaments at low redshift (the kinetic
Sunyaev-Zel'dovich effect~\cite{1972CoASP...4..173S}).
The density-modulated Doppler signal is dominated by the non-linear effect
on small scales ($l > 1000$).

Observations of the small-scale temperature anisotropies can
provide complementary information on reionization to that from large-angle
polarization. In particular, the latter is mostly sensitive to the integrated
optical depth $\tau_{\mathrm{re}}$,
with only weak sensitivity to the duration of
reionization~\cite{2003ApJ...583...24K}, and none to the morphology.
By having one observing channel close to the null of the
thermal Sunyaev-Zel'dovich effect, experiments like the
South-Pole Telescope and the Atacama Cosmology Telescope should
detect the power spectrum of the Doppler signals at very high
significance~\cite{2005ApJ...630..657Z}.

Small-angle polarization generated at reionization should be around
two orders of magnitude below the temperature signal generated there
and will likely remain unobservable~\cite{2000ApJ...529...12H}.
Note, however, that it is a further source of $B$-mode polarization.




\begin{theacknowledgments}
AC thanks the organisers for the invitation to participate in what was
an excellent summer school in a beautiful location.
HVP was supported in part by Marie Curie grant MIRG-CT-2007-203314  
from the European
Commission, and by a STFC Advanced Fellowship. HVP thanks the Galileo  
Galilei Institute
for Theoretical Physics for the hospitality and the INFN for partial  
support during the
completion of this work.

\end{theacknowledgments}



\bibliographystyle{aipproc}   


\IfFileExists{\jobname.bbl}{}
 {\typeout{}
  \typeout{******************************************}
  \typeout{** Please run "bibtex \jobname" to optain}
  \typeout{** the bibliography and then re-run LaTeX}
  \typeout{** twice to fix the references!}
  \typeout{******************************************}
  \typeout{}
 }


\end{document}